\documentclass [letter, 12pt]{article}
\usepackage{a4wide, amsmath, amsfonts, amssymb, geometry}

\usepackage{bm}

\usepackage{fancyhdr, dsfont}
\usepackage[latin1]{inputenc}
\usepackage[pdftex]{graphicx}  % uncomment for latex-use
\usepackage{graphicx}
\usepackage{float}
\restylefloat{figure}

\usepackage{epstopdf}

\newcommand{\req}[1]{(\ref{#1})}
\def\vev#1{\langle #1 \rangle}
\def\ket#1{[ #1 ]}
\def\bet{\beta}
\def\fc#1#2{\frac{#1}{#2}}
\def\h{\frac{1}{2}}
\newcommand{\nwc}{\newcommand}
\nwc{\ba}  {\begin{array}}
\nwc{\ea}  {\end{array}}
\nwc{\bdm} {\begin{displaymath}}
\nwc{\edm} {\end{displaymath}}

\nwc{\bea} {\begin{equation}\ba{lcl}}
\nwc{\eea} {\ea\end{equation}}

\nwc{\bda} {\bdm\ba{lcl}}
\nwc{\eda} {\ea\edm}

\nwc{\bc}  {\begin{center}}
\nwc{\ec}  {\end{center}}

\nwc{\ds}  {\displaystyle}

\nwc{\nn} {\nonumber}
\nwc{\nnn} {\nonumber \vspace{.2cm} \\ }
\nwc{\ra}{\rightarrow}
\nwc{\lra}{\longrightarrow}
\def\lf{\left}\def\ri{\right}

\nwc{\p} {\partial}
\nwc{\Tr}{{\rm Tr}}

\def\ap{\alpha'}

\def\Pc{{\cal P}}\def\Mc{{\cal M}}
\def\Qc{{\cal Q}}
\def\Rc{{\cal R}}
\def\ov{\overline}
\def\th{\theta}

\def\FF#1#2{{_#1F_#2}}

\def\s{{\hat s}}
\def\gpp#1{{G''}^{\rho_1}\hskip-0.5cm_{#1}\hskip0.3cm}
\def\gp#1{{G'}^{\rho_1}\hskip-0.4cm_{#1}\hskip0.2cm}
\def\gpps#1{{G''}^{\rho_2}\hskip-0.5cm_{#1}\hskip0.3cm}
\def\gps#1{{G'}^{\rho_2}\hskip-0.4cm_{#1}\hskip0.2cm}
\newtheorem{pre-defi}[thm]{Definition}

\newtheorem{pre-rem}[thm]{Remark}

\newtheorem{pre-exple}[thm]{Example}

\newtheorem{pre-ue}[thm]{Exercise}

{\begin{sloppypar}\noindent{\bf Proof.}}%
{\hspace*{\fill}$\square$\bigskip\end{sloppypar}}
\def\beq{\begin{equation}}
\def\eeq{\end{equation}}

\def\Re{{\rm Re\,}}

\newcommand{\vecb}{\left(\begin{array}{c}}
\newcommand{\vece}{\end{array}\right)}
\newcommand{\ccb}{\left(\begin{array}{cc}}
\newcommand{\cce}{\end{array}\right)}
\newcommand{\cccb}{\left(\begin{array}{ccc}}
\newcommand{\ccce}{\end{array}\right)}
\newcommand{\ccccb}{\left(\begin{array}{cccc}}
\newcommand{\cccce}{\end{array}\right)}
\newcommand{\cccccb}{\left(\begin{array}{ccccc}}
\newcommand{\ccccce}{\end{array}\right)}
\newcommand{\pa}{\partial}

\newcommand{\al}{\alpha}
\newcommand{\be}{\beta}
\newcommand{\ga}{\gamma}
\newcommand{\de}{\delta}

\newcommand{\vep}{\varepsilon}
\newcommand{\si}{\sigma}
\newcommand{\la}{\lambda}

\newcommand{\Ga}{\Gamma}

\newcommand{\mto}{\rightarrow}
\newcommand{\te}{\textrm}
\newcommand{\eq}{ \ \ = \ \ }

\newcommand{\co}{\ , \ \ \ \ \ \ }

\newcommand{\dd}{d}

\newcommand{\dbe}{\dot{\beta}}

\newcommand{\dde}{\dot{\delta}}

\newcommand{\RR}{{\bf R}}

\newcommand{\ZZ}{{\bf Z}}

\allowdisplaybreaks

\begin{document}

\title{\textbf{\hskip-1cm The~LHC~String~Hunter's~Companion~(II):} \\
\textbf{Five--Particle Amplitudes and Universal Properties} \\ }
\author{\large D. L\"ust$^{\te{a,b}}$,  O. Schlotterer$^{\te{a}}$,
S. Stieberger$^{\te{a}}$  and  T.R. Taylor$^{\te{a,b,c}}$\\[2cm]}
\date{}
%\bigskip
\maketitle
\vskip-2cm
\centerline{\it $^{\te{a}}$ Max--Planck--Institut f\"ur Physik,
Werner--Heisenberg--Institut,}
\centerline{\it 80805 M\"unchen, Germany}
\bigskip
\centerline{\it $^{\te{b}}$ Arnold--Sommerfeld--Center for Theoretical Physics,}
\centerline{\it Ludwigs--Maximilians--Universit\"at, 80333 M\"unchen, Germany}
\bigskip
\centerline{\it $^{\te{c}}$ Department of Physics, Northeastern University,
Boston MA 02115, USA}

\medskip\bigskip
\abstract{\noindent We extend the study of scattering amplitudes presented in
``The LHC String Hunter's Companion'' \cite{LHC} to the case of five-point processes that
may reveal the signals of low mass strings at the LHC and are potentially useful
for detailed investigations of fundamental Regge excitations.
In particular, we compute the full--fledged string disk amplitudes describing all
$2\to 3$ parton scattering subprocesses leading to the production of three hadronic
jets. We cast our results in a form suitable for the implementation of stringy partonic
cross sections in the LHC data analysis. We discuss the universal, model--independent
features of multi-parton processes and point out the existence of even stronger
universality relating $N$-gluon amplitudes to the amplitudes involving $N{-}2$
gluons and one quark-antiquark pair. We construct a particularly simple basis
of two functions describing all universal five--point amplitudes. We also discuss
model--dependent amplitudes involving four fermions and one gauge boson that may be
relevant for studying jets associated to Drell--Yan pairs and other processes depending
on the spectrum of Kaluza--Klein particles, thus on the geometry of compact dimensions.}

\begin{flushright}
{\small  MPP--2009--139}\\
\small LMU--ASC 31/09
\end{flushright}

\thispagestyle{empty}

\newpage
\tableofcontents
\newpage
\numberwithin{equation}{section}
% Falls auf Kapitel- oder Abschnittebene Gleichungen nummeriert werden sollen
%________________________________________________________________________________

\section{Introduction}

As it is well known for a long time \cite{Lerche:1986cx},
string theory contains a huge number of ground states, a problem which is often referred to as
the string landscape problem. This observation raises the question about
the predictive power of string theory or, respectively, if string theory is testable.
In particular one likes to understand if all or at least some four-dimensional string vacua share some common,
model independent features, which are true in large region of the string landscape.
In a statistical approach \cite{Douglas:2004kp,Blumenhagen:2004xx}
to the string landscape one would also search
for correlations or anti-correlations between low energy observables.

One way to obtain model independent statements about four-dimensional string compactifications
is to completely decouple gravity.
Decoupling gravity, namely  performing the limit $M_{\rm Planck}/M_{\rm FT}\rightarrow\infty$
($M_{\rm FT}$ is the field theory scale in question, e.g. the GUT scale $M_{\rm GUT}$ or the
scale of the Standard Model $M_{\rm SM}$), basically means that one keeps the field theory degrees
of freedom of the low energy effective gauge theory, given by a GUT theory or by the Standard Model (SM), and
throws away all extra heavy string modes.
It is then interesting to see if the effective field theory possesses some universal features that
are independent from the details of the compactification, i.e. from the gravity theory at high energies.
This strategy was recently pushed forward in F-theory compactifications \cite{Beasley:2008dc},  where it was claimed
that F-theory on Calabi-Yau four--folds, which are  fibrations over del-Pezzo surfaces, lead to model
independent predictions.

Warped compactifications, or in the extreme case the AdS/CFT correspondence provide concrete realizations
of decoupling gravity and heavy string excitations. Due to the warping of space-time, the excited string states
are blue-shifted and therefore become very heavy. In the extreme case of exponential warping
(the AdS geometry) the excited string modes become infinitely massive and therefore completely decouple
from the field theory (gauge) degrees of freedom.

Undoubtedly,  decoupling of all string degrees of freedom might be a good test to see if the effective field theory
exhibits testable, new and perhaps universal features. However this procedure is certainly not a real
test of string theory itself. A much more powerful, interesting
and in some sense orthogonal strategy is to search for the excited string states in high energy experiments. Of course, up to now this can be done only if these states are not too heavy, i.e.
accessible to collider (LHC) energies. This is precisely what we discussed in a previous article \cite{LHC}.
In fact, the heavy string excitations, called Regge states, are comparably light in scenarios of large
extra dimensions that are a very appealing solution to the hierarchy problem \cite{ADD,Nima}, see \cite{Ignatios} for a review.
The gravitational and gauge interactions are unified at the electroweak scale
and the observed weakness of gravity at lower energies is due to the existence
of large extra dimensions. Gravitons may scatter into the extra space and
by this the gravitational coupling constant is decreased to its observed value.
Extra dimensions arise naturally in string theory. Hence, one obvious question is how
to embed the above scenario into string theory and how to compute
cross sections.

The basic relation between the Planck mass, the string scale (the mass of the Regge states) and the volume $V_6$
of the compact (unwarped) internal space is well known, see e.g. \cite{LMRS}:
\begin{equation}
M_{\rm Planck}^2 \ \ \sim \ \ M_{\rm string}^8 ~V_6 \, .
\end{equation}
As it is clear from this relation, $M_{\rm Planck}$ can be kept large in two extreme ways, namely (i)
choosing the string scale $M_{\rm string}$ high, of order of the Planck mass, and $V_6\sim {\cal O}(1)$, or (ii)
to choose the string scale low, $M_{\rm string}\sim {\cal O}(1~{\rm TeV})$, in which case the volume
of the extra dimensions is large, $V_6\sim {\cal O}(10^{32})$.
It is of course not clear if (ii) -- the large extra dimension scenario -- is realized in nature, but if so it would imply dramatic
consequences for near future collider physics: around or even a bit below  the mass thresholds of the excited string Regge excitations
SM-processes would receive large stringy corrections, which could be seen at the LHC.
This program was initiated in Refs. \cite{Anchordoqui:2007da,Anchordoqui:2008ac}, where high transverse momentum photons associated with  hadronic jets were proposed as probes of low mass strings. The underlying parton subprocess is dominantly the gluon fusion into gluon plus photon. The corresponding amplitude was evaluated at the string disk level, and it was found completely universal, i.e.\  common to all D--brane models and independent of the compactification details. Actually, this universality was known \cite{ST,STi} to hold for all amplitudes involving gauge bosons (without matter fields) although its phenomenological importance was not so widely appreciated.
In \cite{LHC} we computed all tree--level cross sections for external SM particles at the
four parton level, and in Refs. \cite{LHC1,Anchordoqui:2009mm} we discussed possible signatures of these string corrected
cross sections in dijet events. The most important result of the analysis done in Refs. \cite{LHC,LHC1,Anchordoqui:2009mm} can be summarized as follows.
We could show that
in a large class of intersecting brane models (type IIA/B orientifolds -- see e.g.
\cite{Blumenhagen:2006ci}) not only the purely gluonic amplitudes but also
the amplitudes involving one quark-antiquark pair plus any number of gluons
are completely model independent.
So computing these effects for the four gluon and two gluon -- two fermion amplitudes at the LHC, one obtains universal tree level answers irrespective of the details of the type II landscape. In this sense model independent, stringy predictions
are possible, and the problem of the string landscape is nullified at the LHC.
On the other hand, four--fermion amplitudes do depend on the details of the internal geometry, namely on the masses
of the Kaluza--Klein (KK) and/or winding modes.
Therefore to compute the four--fermion amplitudes and to measure their
effects at the LHC would allow investigating some properties of the internal compact space.

In this paper we generalize the previous computations  to five-point amplitudes. These are again very relevant for
LHC physics, since five-parton amplitudes describe in particular $2\to 3$ ``hard'' parton scattering underlying the production of three hadronic jets. They are also relevant for many other processes important for studying string effects, including photon-dijet production, one jet plus lepton-antilepton pair creation {\it etc}.
Here again we find that five-point amplitudes with at most two fermions depend on the
Regge modes only and not on the internal geometry. We also discuss universal properties of higher-point amplitudes describing multi-jet production and other multi-particle processes.

The paper is organized as follows. In Section 2, we discuss some general properties of five--point amplitudes and explain how they are affected by the presence of Regge excitations.
We discuss the universality issue and set up the formalism for computing full-fledged string disk amplitudes  in D--brane constructions of the SM. In Section 3, we summarize previous results on five-gluon amplitudes. In Section 4, we compute the amplitudes involving three gluons and two fermions and we recast them in a form very similar to five--gluon amplitudes. We explain  the origin of the relation between these amplitudes.
In Section 5, we discuss further relations that allow expressing all partial amplitudes in terms of two independent functions  combining hypergeometric functions and kinematic variables (in general, in terms of $(N-3)!$
functions for $N$-point amplitudes), which can be very useful for assembling partial amplitudes into the full amplitude. We find a very simple basis common to $ggggg$ and $gggq\bar q$ processes. In Section \ref{GQQQQ}, we compute the amplitudes with one gluon and four fermions.
In Section 7, we discuss relations between the partial amplitudes involving one vector boson $g$ and four fermions,
$g\chi\bar\chi\chi\bar\chi$ ($\chi$ are gauginos) and  $gq\bar q q\bar q$.
In Section 8, we present all amplitudes squared, summed over helicities and color indices,
in a form suitable for applications to collider physics. Some phenomenologically-oriented readers will certainly find out that Sections 2 and 8 are the ones most relevant to the LHC applications.

\section{String Regge excitations and five parton amplitudes}

\subsection{Massive string states}

Each SM particle is accompanied  by an infinite number of massive string  excitations.
Generically, in any string compactification of this kind we expect, among other fields
(neutral moduli, bulk KK modes), the following open string excitations of the SM fields, which will
be important for our discussion:

\vskip0.3cm
\noindent
{\it (i) String Regge excitations:}

\vskip0.2cm
\noindent
The most direct way to see stringy effects is to  measure the open string oscillator excitations, the so-called Regge
modes. There are infinitely many open string Regge modes, and their mass-squares are multiples of the string scale $M_{\rm string}^2$:
\begin{equation}\label{reggemasses}
M_{\rm Regge}^2 \eq n \,M_{\rm string}^2\ \ \ ,\ \ \ n \in {\bf N} \ .
\end{equation}
This part of the mass spectrum follows the well known Regge trajectories, since these particles are the oscillator excitations of
the SM particles with in general higher spins, where the highest possible spin at each level ($n+1$) is limited
by the oscillator number $n$. Moreover,
these stringy Regge excitations carry SM quantum numbers, i.e. they may be produced by $pp$-collisions.

\vskip0.3cm
\noindent
{\it (ii) Small (longitudinal) cycle KK particles:}

\vskip0.2cm
\noindent
Since the D-branes are wrapped around internal, p-dimensional cycles of the compact space, one always gets the open string KK excitations
of every SM particle.
The masses of the KK particles are related to the  volumes of the internal cycles, and   one generically obtains:
\begin{equation}
M_{\te{KK}} \eq {m \over V_p^{1/p}}\ \ \ ,\ \ \ m \in {\bf N}\ .
\end{equation}
Since the volumes $V_p$ are of the order of $M_{\rm string}^{-p}$ to get the correct values for the gauge coupling constants,  the masses of the KK modes
are roughly also of the order of the string scale.
The integer $m$  denotes internal KK charge, i.e. the internal momentum of the KK modes along the internal
directions of the wrapped D-branes. This KK charge has to be preserved in any scattering process of the light SM fields.
In addition, the KK particles carry SM quantum numbers,  and hence they can be
directly produced at the LHC.
In addition to the KK particles, one may also have winding strings, if the $p$-cycles allow for non-contractible circles as submanifolds.
Again their masses will be of order $M_{\rm string}$.

\vskip0.3cm
\noindent
{\it (iii) Mini black holes:}

Let us also briefly comment on the possible creation  of (mini) black holes in string scattering processes.
In contrast to the string Regge excitations and the KK (winding) modes, the black hole states are of non-perturbative nature, and
their masses are basically determined by the scale of gravity in the higher dimensional bulk theory. More precisely,
the energy threshold for black hole production is given by\footnote{This estimate is derived
 by comparing the black hole entropy, ${\cal S}_{\rm b.h.}\sim1/G_N\sim1/g_{\rm string}^2$ with the string entropy,\\ ${\cal S}_{\rm string}\sim\sqrt{n}$  \cite{Horowitz:1996nw}.}:
\begin{equation}
E_{\rm b.h.}\ \simeq\  \fc{M_{\rm string}}{g_{\rm string}^2}\ .
\end{equation}
Comparing this with the masses of the perturbative string excitations in eq.(\ref{reggemasses}), one recognizes
that one first produces $n^*\sim 1/g_{\rm string}^4$ string states before one reaches $E_{\rm b.h}$.
In case of weak string coupling, $n^*$ is a quite large number, e.g. for the string coupling constant being of the same order as the gauge
coupling, $g_{\rm string}\approx 0.2$, $n^*\approx 625$. Hence black hole production will not be of any relevance for our further
discussion.

In this work  we compute five--point scattering amplitudes among the massless SM model
open string fields as external particles. These amplitudes will be dominated first by the exchange of the light SM fields
themselves, second by the exchange
of the string Regge excitations and third possibly also by the exchange of KK and/or winding modes. The heavy Regge modes
together with the KK/winding modes will therefore constitute the stringy corrections to any SM scattering process. In the limit
of $\alpha'\rightarrow 0$, i.e. $M_{\rm string}\rightarrow\infty$, one should always recover the known field theoretic SM scattering amplitudes.
As we will discuss in the following, there exist entire classes of string amplitudes, for which only the string Regge modes but
not the KK/winding modes contribute. Hence this class of string amplitudes is completely universal and not affected
by any detail (geometry, topology) of the internal compact space.

String Regge resonances in models with low string scale are also discussed
in \cite{CullenEF,AAB,Anchordoqui:2007da,Anchordoqui:2008ac,LHC,LHC1,
CHEM,Anchordoqui:2009mm,more}, while KK graviton exchange  into the bulk, which
appears at the next order in perturbation theory, is discussed in
\cite{CullenEF,Dudas} at the level of four--point amplitudes.

\subsection{Five--particle kinematics}

We begin by summarizing our notation, conventions and some basic facts about
five--particle kinematics.
The kinematic variables commonly used in data analysis are mass dimension two invariants
\begin{equation}\label{sidim}
 s_{ij}=(k_i+k_j)^2\ ,
\end{equation}
where $k_i$ denotes the momentum of $i$-th particle.
Note that the momenta are always subject to the momentum conservation constraint,
$\sum\limits_{i=1}^{5}k_i=0$, and all external particles are on-shell, $k_i^2=0$.
By using momentum conservation, the scalar products $s_{ij}$ can be  expressed in terms of
5 independent invariants $s_i,~i=1,\dots, 5$
\begin{equation}\label{sinodim}
 s_1=(k_1+k_2)^2~,~ s_2=(k_2+k_3)^2~,~ s_3=(k_3+k_4)^2~,~
 s_4=(k_4+k_5)^2~,~  s_5=(k_5+k_1)^2~,
\end{equation}
as written in Table 1.

\vskip0.5cm
\begin{center}
\begin{tabular}{p{1cm}p{1cm}p{1cm}p{1cm}p{1cm}p{1cm}}
 \cline{2-5}
  \multicolumn{1}{c}{} &\multicolumn{1}{|c}{$\,\;\;\qquad
 2\qquad\;\;\,$}&\multicolumn{1}{|c}{3}&\multicolumn{1}{|c}{4}&
 \multicolumn{1}{|c}{5}&\multicolumn{1}{|c}{}\\ \hline
 \multicolumn{1}{|c}{1} &
\multicolumn{1}{|c}{$ s_1$}&\multicolumn{1}{|c}{$ - s_1 -  s_2 +  s_4$}&  \multicolumn{1}{|c}{$ s_2 -  s_4 -  s_5$}& \multicolumn{1}{|c}{$ s_5$}
&  \multicolumn{1}{|c|}{1} \\  \hline
\multicolumn{1}{|c}{2} &  \multicolumn{1}{|c}{} &
 \multicolumn{1}{|c}{$ s_2$}&\multicolumn{1}{|c}{$ - s_2 - s_3 + s_5$}& \multicolumn{1}{|c}{$- s_1 + s_3
 -s_5$}
& \multicolumn{1}{|c|}{2} \\
\hline
 \multicolumn{1}{|c}{3} &\multicolumn{2}{|c}{} &
 \multicolumn{1}{|c}{$ s_3$}&\multicolumn{1}{|c}{$ s_1 - s_3 -s_4$}
&  \multicolumn{1}{|c|}{3}  \\
 \hline
\multicolumn{1}{|c}{4} & \multicolumn{3}{|c}{}&  \multicolumn{1}{|c}{$ s_4$} &  \multicolumn{1}{|c|}{4} \\ \hline
\end{tabular}\\[5mm]\begin{flushleft}
{\bf Table 1.\ } {\it
The scalar products
$s_{ij}\equiv 2 k_ik_j$, with $i$ and $j$ labeling  rows and columns, expressed in terms of the kinematic invariants} $ s_1,\dots,s_5$ \cite{STi}.\end{flushleft}
\end{center}

\noindent
In order to write down the amplitudes in a concise way, it is convenient to introduce
dimensionless variables
\begin{eqnarray}\label{sijnodim} \hat s_{ij}
\hskip -6mm &&= \alpha'\ s_{ij}~,\qquad \hat s_i=\alpha'\ s_i\\[1mm]
&&\hskip -1.4cm\epsilon(i,j,m,n)\;=\, \ap^2\ \epsilon_{\alpha\beta\mu\nu}\
k_i^{\alpha}\ k_j^{\beta}\ k_m^{\mu}\ k_n^{\nu}\ ,  \label{pseudo}\end{eqnarray}
where $\epsilon_{\alpha\beta\mu\nu}$ is the four-dimensional Levi-Civita symbol.
The factors of $\alpha'$ render the above invariants dimensionless.

\subsection{Regge excitations and contact interactions}

Due to the extended nature of strings, the  string amplitudes
are generically non--trivial functions of $\ap$ in addition to the
usual dependence on the kinematic invariants and degrees of freedom of the
external states.
In the effective field theory description this $\ap$--dependence gives rise to a
series of infinite many resonance channels\footnote{In addition, there may be
additional resonance channels
due to the exchange of KK and winding states, as it is the case for amplitudes
involving at least four quarks or leptons, c.f. also Subsection \ref{UNI}.}
due to Regge excitations and/or new contact interactions.

Generically, as we shall see in the next Section, tree--level string amplitudes involving
five gluons or amplitudes with three gluons and two fermions
are specified  by a basis of two Gaussian hypergeometric functions $f_1,f_2$
depending on the kinematic
invariants $\hat s_1,\dots, \hat s_5$.
More concretely, each disk amplitude $\Mc(k_1,k_2,k_3,k_4,k_5,\ap)$ with five external
open string states with momenta $k_i$ can be expressed in terms of two hypergeometric functions $f_1$ and $f_2$:
\bea
\ds{f_1}&=&\ds{\int^1 \limits_{0} \dd x \int^{1} \limits_{0} \dd y \
x^{\hat s_2-1} \,  y^{\hat s_5-1} \, (1-x)^{\hat s_3} \, (1-y)^{\hat s_4} \,
(1-xy)^{\hat s_{1}-\hat s_{3}-\hat s_{4}}\ ,} \label{BASIS} \\
\ds{f_2}&=&\ds{\int^1 \limits_{0} \dd x \int^{1} \limits_{0} \dd y \
x^{\hat s_2} \,  y^{\hat s_5} \, (1-x)^{\hat s_3} \, (1-y)^{\hat s_4} \,
(1-xy)^{\hat s_{1}-\hat s_{3}-\hat s_{4}-1}\ ,}
\eea
multiplying some kinematic factors. The $\alpha'$-dependence of the
two functions \req{BASIS} is implicitly contained in the invariants $\s_{ij}$ with $\ap$ setting the mass units for kinematic variables.

\subsubsection{Exchanges of string Regge excitations}

The five--point amplitudes $\Mc(k_1,k_2,k_3,k_4,k_5;\ap)$ may be understood as an
infinite sum over poles with intermediate string states $|k;n\rangle$ exchanged.
In the following we present two descriptions. For both descriptions we first rewrite
the two functions \req{BASIS} as:
\bea
\ds{f_1}&=&\ds{\sum_{n=0}^\infty \gamma(\s_{35},n)\
B(\s_{23}+n,\s_{34}+1)\
B(\s_{45}+1,\s_{51}+n)\ ,}\\[5mm]
\ds{f_2}&=&\ds{\sum_{n=1}^\infty \gamma(\s_{35}-1,n-1)\ B(\s_{23}+n,\s_{34}+1)\
B(\s_{45}+1,\s_{51}+n)\ .}
\label{rewrite}
\eea
Above we have introduced the residuum
\beq
\gamma(\ap s,n) :=\fc{(-\ap s,n)}{(1,n)} \eq \fc{1}{n!}\
\fc{\Gamma(-\ap s+n)}{\Gamma(-\ap s)}
= \fc{1}{n!} \prod_{j=1}^n(- \ap s-1+j) \sim (\ap s)^n\ ,
\label{residua}
\eeq
which determines the three--point coupling of the intermediate string
state $|k;n\rangle$ to the external particles.
Furthermore, the Euler Beta function:
\beq
B(a,b) = \int^{1}_{0} \dd x \ x^{a-1}\ (1-x)^{b-1}=\fc{\Ga(a)\ \Ga(b)}{\Ga(a+b)}
\label{Euler}\ .
\eeq
Now let us present two rewritings of the expression \req{rewrite} for $f_1,f_2$.
This manipulation will make manifest the exchange of string Regge excitations.

\begin{figure}[H]
\centering
\includegraphics[width=0.5\textwidth]{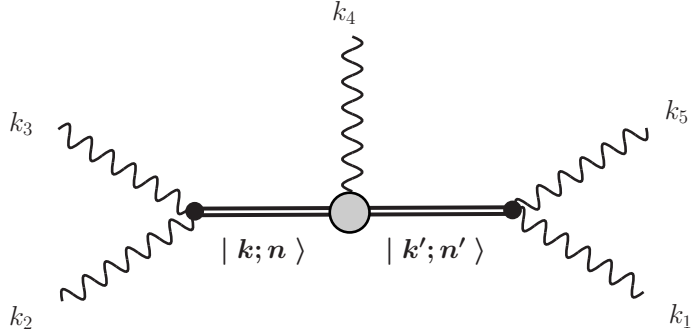}
\caption{Exchange of string Regge excitations.}
\label{regge1}
\end{figure}

In the first case the two functions $f_1,f_2$ may be written
as two infinite sums over $s_{23}$ and $s_{51}$--channel poles with the intermediate
string states $|k;j\rangle$ and $|k';j'\rangle$ exchanged.
More precisely, \req{rewrite} may be cast into:
\bea
\ds{f_1}&=&\ds{\sum_{n=0}^\infty\sum_{n'=0}^\infty\sum_{j=0}^{{\rm min}(n,n')}
\gamma(\s_{35},j)\ \fc{\gamma(\s_{34},n-j)}{\s_{23}+n}\
\fc{\gamma(\s_{45},n'-j)}{\s_{51}+n'}\ ,}\\[5mm]
\ds{f_2}&=&\ds{\sum_{n=1}^\infty\sum_{n'=1}^\infty\sum_{j=1}^{{\rm min}(n,n')}
\gamma(\s_{35}-1,j-1)\ \fc{\gamma(\s_{34},n-j)}{\s_{23}+n}\
\fc{\gamma(\s_{45},n'-j)}{\s_{51}+n'}\ ,}
\label{Regge1}
\eea
Since $k^2:=(k_2+k_3)^2= \s_{23}/\ap$ and $k'^2:=(k_1+k_5)^2=\s_{51}/\ap$
the expression \req{Regge1} describes the two intermediate exchanges of an
infinite number of string states $|k;n\rangle$ and $|k';n'\rangle$
with highest spins $n+1$ and $n'+1$, respectively and the sum of three--point couplings
$\gamma(s_{34},n-j)$ and $\gamma(s_{45},n'-j)$, respectively, c.f.
Figure~\ref{regge1}.
\noindent
Clearly, in the sum of $f_1$ the term $j=0$ encodes the canonical exchanges
$\fc{\gamma(\s_{34},n)}{\s_{23}+n}\ \fc{\gamma(\s_{45},n')}{\s_{51}+n'}$,~encountered
in~\cite{LHC}.

\begin{figure}[H]
\centering
\includegraphics[width=0.5\textwidth]{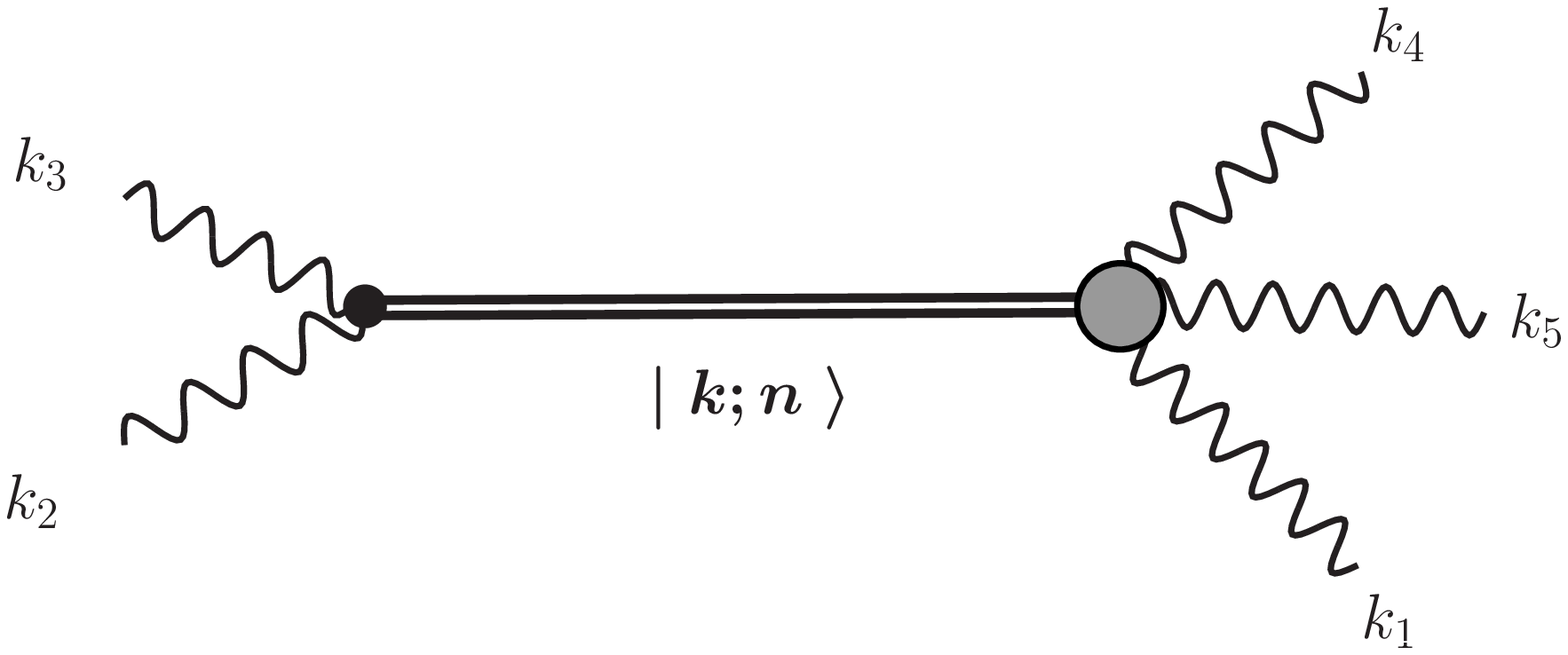}
\caption{Exchange of string Regge excitations.}
\label{regge}
\end{figure}

In the second case the functions $f_1,f_2$ may be written as one infinite sum
over $s_2$--channel poles with the intermediate string states $|k;n\rangle$ exchanged.
More precisely, \req{rewrite} may be cast into:
\bea
\ds{f_1}&=&\ds{\sum_{n=0}^\infty\sum_{j=0}^{n}
\gamma(\s_{35},j)\ \fc{\gamma(\s_{34},n-j)}{\s_{23}+n}\ B(\s_{45}+1,\s_{51}+j)\ ,}\\[5mm]
\ds{f_2}&=&\ds{\sum_{n=1}^\infty\sum_{j=1}^{n}
\gamma(\s_{35}-1,j-1)\ \fc{\gamma(\s_{34},n-j)}{\s_{23}+n}\ B(\s_{45}+1,\s_{51}+j)\ .}
\label{Regge}
\eea
Since $k^2 :=(k_2+k_3)^2=\s_{23}/\ap$ the expression \req{Regge} describes
the intermediate exchange of infinite string states $|k;n\rangle$
with highest spin $n+1$ and the sum of three--point
$\gamma(\s_{34},n-j)$ and four--point $B(\s_{45}+1,\s_{51}+j)$
couplings, c.f. Figure \ref{regge}.

\subsubsection{String contact interactions}

Another way of looking at the expressions \req{Regge1} and \req{Regge}
appears when we express each
term in the  sum as a power series expansion in $\ap$. For \req{Regge1}
we may rearrange the triple sum and expand each term by $\ap$:
\bea
\ds{f_1}&=&\ds{\sum_{n=0}^\infty\sum_{n'=0}^\infty\sum_{j=0}^{{\rm min}(n,n')}\ldots
\ \ \ =\ \ \ \ap^{-2}\ \underbrace{\sum_{n=n'=j=0}\ldots}_{=\fc{1}{s_2s_5}}
\ \ \ +\ \ \
\ap^0\ \biggl(\ \underbrace{\sum_{n'=j=0}\sum_{n=1}^\infty\ldots}_{=-\zeta(2)\ \fc{s_3}{s_5}}+
\underbrace{\sum_{n=j=0}\sum_{n'=1}^\infty\ldots}_{=-\zeta(2)\ \fc{s_4}{s_2}}\ \biggr)}\\[8mm]
&+&\ds{\ap^1\ \biggl(\ \underbrace{\sum_{n'=j=0}\sum_{n=1}^\infty\ldots}_{=\zeta(3)\
\fc{s_3\ (s_2+s_3)}{s_5}}+
\underbrace{\sum_{n=j=0}\sum_{n'=1}^\infty\ldots}_{=\zeta(3)\
\fc{s_4\ (s_4+s_5)}{s_2}}+\underbrace{\sum_{n=n'=j=1}^\infty\ldots}_{=\zeta(3)\
\lf(-s_1+s_3+s_4\ri)}\ \biggr)\ \ \ +\ \ \ \ldots\ ,}\\[8mm]
\ds{f_2}&=&\ds{\sum_{n=1}^\infty\sum_{n'=1}^\infty\sum_{j=1}^{{\rm min}(n,n')}\ldots
\ \ \ =\ \ \ \ap^0\ \underbrace{\sum_{n=n'=j=1}^\infty\ldots}_{=\zeta(2)}}\\[8mm]
&+&\ds{\ap^1\ \biggl(\hskip-0.5cm\underbrace{\sum_{n=n'=j=1}^\infty\ldots}_{=-\zeta(3)\ \lf(s_1+s_2-s_3-s_4+s_5\ri)}\ \ \ +\ \ \ \underbrace{\sum_{n=1}^\infty\sum_{n'>n}^\infty\sum_{j=n}\ldots}_{=-s_4\ W(1,1,1)}\ \ \ +\ \ \ \underbrace{\sum_{n'=1}^\infty\sum_{n>n'}^\infty\sum_{j=n'}\ldots}_{=-s_3\ W(1,1,1)}\ \biggr)\ +\ldots\ ,}
\label{decomp}
\eea
with $W(1,1,1)=2\zeta(3)$ \cite{6GG}.
In total, we have:
\bea
\ds{f_1}&=&\ds{\fc{1}{\s_2\s_5}-\zeta(2)\
\lf(\fc{\s_3}{\s_5} +\fc{\s_4}{\s_2}\ri)+ \zeta(3)\
\lf(-\s_1+\s_3+\s_4+\fc{\s_4^2+\s_4\s_5}{\s_2}+\fc{\s_2\s_3+\s_3^2}{\s_5}\ri) \ + \ {\cal O}(\ap^2) \ ,}\\[5mm]
\ds{f_2}&=&\ds{
\zeta(2) \ - \zeta(3)\ \lf(\s_1+\s_2+\s_3+\s_4+\s_5\ri) \ + \ {\cal O}(\ap^2) \ .}
\label{EXPANSIONS}
\eea
In \req{decomp} the first term of $f_1$ stemming from $n=n'=j=0$ gives the field--theoretical contribution\footnote{To obtain the correct mass dimensions these expressions are multiplied by $\ap^2$.}. On the other hand, in the next two brackets of $f_1$ and in $f_2$ infinite many SR states are summed up and give rise to  the first string corrections. Furthermore, in \req{decomp} string contact terms arise from terms without poles.
E.g. at the order $\ap^3$ the infinite sum over SR states comprises into a contact term  with five external states, c.f. the next Figure \ref{contact}.
\begin{figure}[H]
\centering
\includegraphics[width=0.35\textwidth]{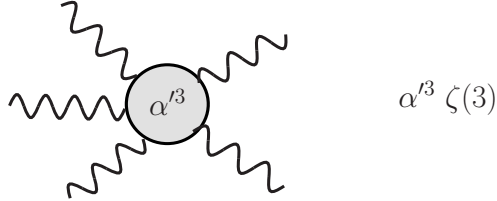}
\caption{New string contact interaction.}
\label{contact}
\end{figure}

\subsection{Universal properties of parton amplitudes}
\label{UNI}

\begin{figure}[H]
\centering
\includegraphics[width=0.8\textwidth]{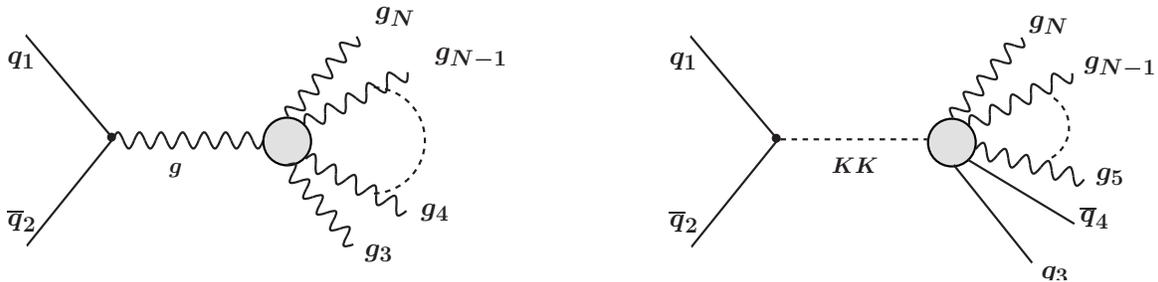}
\caption{Exchanges of gluons and KK states in $N$--parton amplitudes.}
\label{universal}
\end{figure}

Parton amplitudes are important for collider phenomenology since multijet production
is dominated by tree-level QCD scattering.
Therefore those parton amplitudes that are  generic to any string compactification are especially important, as they  may give rise to
universal string signals independent on any compactification details.
Amplitudes involving an arbitrary number of gluons $g$ (or gauginos $\chi$)
but only two quarks $\psi$ or
squarks $\phi$ are of this kind. More precisely, the following  $N$--point amplitudes:
\beq\ba{lcl}
{\cal M}(g^{a_1}\ldots g^{a_N})\ ,&&
{\cal M}(\psi^{a_1}\overline\psi^{a_2}g^{a_3}\ldots g^{a_{N}})\ ,\\
{\cal M}(\chi^{a_1}\overline\chi^{a_2}g^{a_3}\ldots g^{a_{N}})\ ,&&
{\cal M}(\phi^{a_1}\overline\phi^{a_2}g^{a_3}\ldots g^{a_{N}})\ ,
\ea\label{UNIVERSAL}
\eeq
are completely universal to any string compactification, i.e.\  they
do not depend on the compactification details. Using space--time supersymmetry we may also replace gluons $g$
by gauginos $\chi$.
The reason for the universality of the above amplitudes
can be explained in terms of the diagrams shown in Figure \ref{universal}.
\noindent
A possible model dependence would originate from an exchange of a KK or winding state between
the two quarks or squarks and the remaining $N-2$ gluons (or gauginos).
However, in contrast to gluons or gauginos the KK or winding state carries an internal charge.
Hence such an exchange would violate charge conservation, unless there is at least one quark--antiquark pair on the right hand side of the diagram.
Hence amplitudes involving four and more quarks are model dependent due to possible
KK or winding exchange, but the amplitudes \req{UNIVERSAL}
involving no more than one quark--antiquark pair share universal properties
insensitive to the compactification.

The  two amplitudes in the first column of \req{UNIVERSAL} are related through supersymmetric Ward identities \cite{MHV}, cf. also \cite{6G}. 
Similarly, the  last two amplitudes of the second column of \req{UNIVERSAL} are related through
supersymmetry. Since the first set of amplitudes involves only members of a vector multiplet,
while the second set also involves chiral multiplets one does not expect {\em a priori\/} a relation between
those two sets.
However, four--point amplitudes involving quarks and gluons exhibit a higher degree of universality which becomes obvious after writing the following partial amplitudes \cite{STi,LHC} side-by-side:
\begin{eqnarray}\label{side1}
{\cal M}(g^+_1,g^-_2, g^-_3, g^+_4)&=& 4 g^2\  V^{(4)}(\hat s_i)\
\frac{[14]^4}{[12][23][34][41]}\ \Tr(T^{a_1}T^{a_2}T^{a_3}T^{a_4})\\ \label{side2}
{\cal M}(g^+_1,g^-_2, q^-_3, \bar{q}^+_4)&=& 2g^2\  V^{(4)}(\hat s_i)\
\frac{[13][14]^3}{[12][23][34][41]}\ (T^{a_1}T^{a_2})^q_{\bar q}\ ,
\end{eqnarray}
which makes it clear that the same string form factor $V^{(4)}(\hat s_i)=\hat s\hat u\hat  t^{-1}B(\hat s,\hat u)$ (of Mandelstam variables normalized in string units) describe the effects of string Regge excitations in both amplitudes.  We will find a similar result for
five-point functions, where the effects of string Regge excitations are described by two universal, ``generalized'' string form factors. We  discuss it in more detail in Sections 4 and 5. In general, all universal $N$--point amplitudes \req{UNIVERSAL},
can be expressed in terms of $(N{-}3)!$ model--independent hypergeometric functions
\cite{STi}. The amplitudes involving more than one quark-antiquark pair can be also expressed in terms of $(N{-}3)!$ functions, but these functions are sensitive to the spectrum of KK excitations, thus to the geometry of extra dimensions, see Sections 6 and 7.

\subsection{Five--point string amplitudes of gauge and matter SM fields}

\subsubsection{D-brane set up for SM open strings and their excitations}

\begin{figure}[H]
\centering
\includegraphics[width=0.6\textwidth]{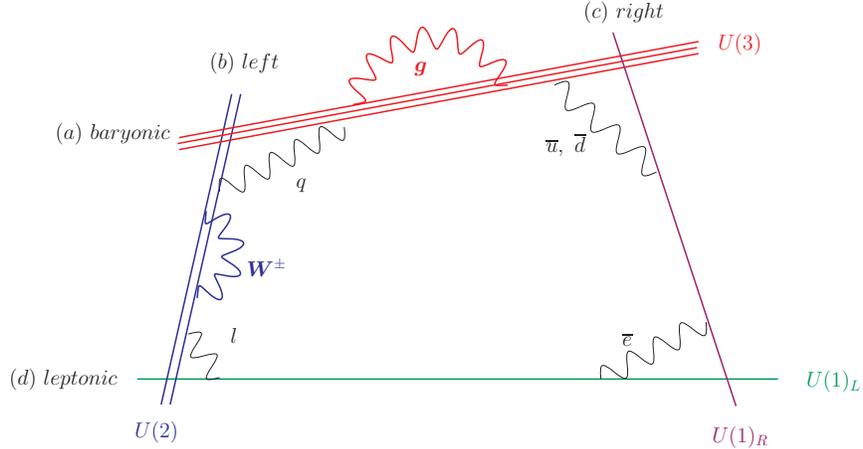}
\caption{The realization of the SM by four stacks of intersecting D-branes.}
\label{SM1}
\end{figure}

In this Section we will give a brief summary about how the SM is realized by
intersecting D-branes. More details can be found, e.g., in reference \cite{Blumenhagen:2006ci,LHC}.
The following set-up applies for type IIA and type IIB orientifolds with intersecting D6-
or D7-branes, respectively. The D-branes are space-time filling, and they are
wrapped around certain p-dimensional cycles ($p=3,4$) inside the compact space. This set-up is in principal also valid for F-theory,
which provides the non-perturbative uplift of the type IIB orientifolds with intersecting D7-branes.
Therefore the considered D-brane quiver locally describes a large class of four-dimensional string vacua.
As it is shown in Figure \ref{SM1}, the SM particles can be locally realized as massless open string  excitations that live
on a local quiver of four different stacks of intersecting D-branes.
The corresponding quiver SM gauge group is given by
\begin{equation}
{G} \eq {G}_a\times {G}_b\times {G}_c\times {G}_d \eq U(3)_a\times
U(2)_b\times U(1)_c\times U(1)_d\ .
\end{equation}

\noindent
Note that there are four different $U(1)$ gauge group factors. In a specific compactification, most of them will be anomalous,
such that the corresponding gauge bosons become massive by the Green-Schwarz mechanism.
However one has to ensure that the gauge boson of the linear combination, which determines the weak hyper charge, stays massive.
The $U(1)$ gauge symmetries, which are anomalous or which correspond to massive gauge bosons, nevertheless
remain as global symmetries in all perturbative scattering amplitudes. In particular,
the symmetry associated to $U(1)_a$ is the baryon number conservation. Hence all perturbative processes respect
baryon number conservation.
Of course there might be other dangerous processes in string perturbation theory, such as
flavor changing neutral currents, which are in general not prohibited by the global $U(1)$ symmetries of this D-brane
quiver. However this is a model-dependent issue, which we do not address here.

Specifically, the SM gauge bosons in the adjoint
representations of the gauge group ${G}$ such as the gluons, the weak gauge bosons and
the hyper charge gauge boson (being associated to  linear combination of the various $U(1)$
factors), correspond to open strings with two endpoints on the same stack of D-branes.
On the other hands, the SM matter fields such as quarks and leptons are open strings located at the various intersection points of the
four different D-branes (and their orientifold images).
They transform under bi-fundamental representations of the  four gauge group factors, and
they can also be in the anti-symmetric representation ${\bf 3}_A$ of $SU(3)_a$, in case the color stack of D-branes is intersected
by its orientifold image.
As it turns out the four stack D-brane quiver reproduces all quantum numbers of the SM particles
in a straightforward and natural way.
In fact, no GUT embedding is necessary to explain the gauge quantum numbers of the SM particles.
The family replication is explained by multiple intersections of the D-branes inside the compact space.
Moreover it is shown in \cite{Gmeiner:2008xq} that one can construct
consistent  type II string  compactifications
on the $\ZZ_6'$ orientifold which reproduce the spectrum of the SM with three
generations of quarks and leptons and without chiral exotics.

\subsubsection{Five open string amplitudes}

This Section deals with the superconformal field theory on the worldsheet
boundary. Basic ingredients for stringy scattering amplitudes are the SCFT
correlation functions involving the dimension 1 vertex operators of the
partons and gauge fields.

Disk amplitudes involving open strings $\Phi_i$ as external states comprise into
a sum over all the possible orderings of the corresponding
vertex operators along the disk boundary. For each cyclically inequivalent
$S_{5}$ permutation $\rho$, there is a partial amplitude ${\cal M}_{\rho}$
obtained from appropriate integration over worldsheet
positions. Schematically, this means
\begin{subequations}
\begin{align}
{\cal M}(\Phi_{1},\Phi_{2},\Phi_{3},\Phi_{4},\Phi_{5}) \ \ &= \ \
\sum_{\rho \in S_{5} / \ZZ_{5}} {\cal
 M}_{\rho}(\Phi_{1},\Phi_{2},\Phi_{3},\Phi_{4},\Phi_{5})
\label{0,1} \\
{\cal M}_{\rho}(\Phi_{1},\Phi_{2},\Phi_{3},\Phi_{4},\Phi_{5}) \ \ &= \ \
{V}_{\te{CKG}}^{-1}\ \int \limits _{{\cal I}_{\rho}}  \lf(\prod_{k=1}^{5}
\dd z_{k}\ri) \
\langle V_{\Phi_{1}}(z_{1}) \, ... \, V_{\Phi_{5}}(z_{5}) \rangle \label{0,2}
\end{align}
\end{subequations}
where the $\Phi_{i}$ represent SM particles, $V_{\Phi_{i}}$ their corresponding
vertex operators and ${\cal I}_{\rho}, {V}_{\te{CKG}}$ some details about
the integration over the vertex positions $z_k$ to be specified later.

The world sheet integrals relevant to any superstring disk amplitudes involving
five external partons are more involved to handle than the four particle analogue.
The latter basically boils down to the Euler Beta
function \req{Euler}. On the other hand, the five--point case is described by
the Gaussian hypergeometric function $_{3}F_{2}$, c.f. \cite{6GG} for a more detailed discussion:
\beq
\int^{1} \limits_{0} \dd x \int^{1} \limits_{0} \dd y \ x^{a-1} \, y^{b-1} \,
(1-x)^{c-1} \, (1-y)^{d-1} \, (1-xy)^{e} \eq \frac{\Ga(a) \, \Ga(b) \, \Ga(c)
 \, \Ga(d)}
{\Ga(a+c) \, \Ga(b+d)} \; _{3}F_{2} \biggl[ \begin{array}{c} a, \, b, \, -e
   \\ a+c,\, b+d \end{array} ; 1 \biggr]\ .
\label{2,7}
\eeq

\subsubsection{Open string vertex operators}
\label{VOPs}

The vertex operators of the gauge vectors are given by
\begin{subequations}
\begin{align}
V_{A^{a}}^{(-1)}(z,\xi,k) \ \ &= \ \ (2\al')^{1/2} \ g_{\te{D}p_{a}} \  \bigl(T^{a} \bigr)^{\al_{1}}_{\al_{2}} \  \xi_{\mu} \ \psi^{\mu}(z) \ e^{-\phi(z)}\ e^{ik X(z)}\ , \label{1,1a} \\
V_{A^{a}}^{(0)}(z,\xi,k) \ \ &= \ \ g_{\te{D}p_{a}} \  \bigl(T^{a} \bigr)^{\al_{1}}_{\al_{2}} \  \xi_{\mu} \  [\  i \pa X^{\mu}(z) \ + \ 2\al' \ k  \psi(z)\ \psi^{\mu}(z) \ ] \  e^{ik  X(z)} \ .
\label{1,1b}
\end{align}
\end{subequations}
in the $(-1)$ and $0$--ghost picture, respectively.
They are specified by the polarization vector $\xi_{\mu}$, the spacetime
four--momentum $k_{\mu}$ and the worldsheet position $z$ on the boundary of the disk. The coupling\footnote{In this work we refer to stack $a$ as the QCD stack and identify the respective
D$p$--brane coupling $g_{Dp_{a}}$ with the QCD coupling constant $g$, i.e.:  $g_{Dp_{a}}= g$.} $g_{\te{D}p_{a}}$ as well as the Chan Paton factors $T^{a}$ depend on the stack of D$p$--branes to which the gluon in question couples, c.f.  \cite{LHC} for details.

Furthermore, the chiral fermion vertex operators describing quarks and leptons are given by \cite{LMRS,LHC}
\begin{subequations}
\begin{align}
V_{\psi^{\al}_{\be}}^{(-1/2)}(z,u,k) \ \ &= \ \ \sqrt{2} \  \al'^{3/4} \  e^{\phi_{10}/2} \  \bigl(T^{\al}_{\be} \bigr)^{\be_{1}}_{\al_{1}} \  u^{\al} \  S_{\al}(z) \  e^{-\phi(z)/2} \  \Xi^{a \cap b}(z) \  e^{ik  X(z)}\ , \label{1,2a} \\
V_{\bar{\psi}^{\be}_{\al}}^{(-1/2)}(z,\bar{u},k) \ \ &= \ \ \sqrt{2} \  \al'^{3/4} \  e^{\phi_{10}/2} \  \bigl(T^{\be}_{\al} \bigr)^{\al_{1}}_{\be_{1}} \  \bar{u}_{\dbe} \  S^{\dbe}(z) \  e^{-\phi(z)/2} \  \bar{\Xi}^{a \cap b}(z) \  e^{ik  X(z)}\
\label{1,2b}
\end{align}
\end{subequations}
in the $(-1/2)$--ghost picture.
The helicity information is specified by the  (massless) Dirac four--spinors $u,\bar{u}$, which are contracted with the $D=4$ Ramond spin fields $S_{\al},S^{\dbe}$ of conformal weight $1/4$. The boundary changing operators $\Xi^{a \cap b}$, $\bar{\Xi}^{a \cap b}$ carry the internal degrees of freedom  of the Ramond sector associated to the internal part of the SCFT. Their two--point correlator is given by \cite{LHC}:
\beq
\langle \Xi^{a \cap b}(z_{1}) \, \bar{\Xi}^{a \cap b}(z_{2}) \rangle \ \ =
\ \ \fc{1}{(z_1-z_2)^{3/4}}\ .
\label{1,3b}
\eeq
The normalization of this correlator is fixed by unitarity.

\section{Five gluon amplitudes $\bm{ggggg}$}
\label{GGGGG}

\begin{figure}[H]
\centering
\includegraphics[width=0.35\textwidth]{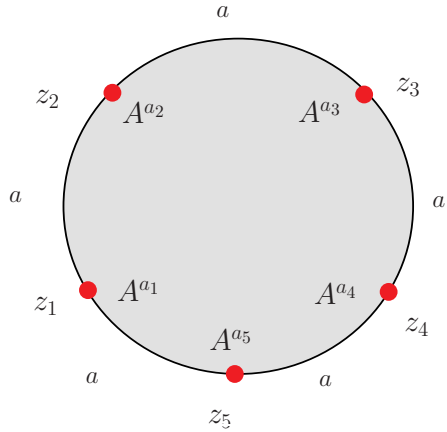}
\caption{Gauge boson vertex operators along the boundary of the disk world--sheet.}
\label{ggggg}
\end{figure}

Five gluon amplitudes have been studied in Refs. \cite{Barreiro:2005hv,6GG,ST,STi}.
The corresponding string disk diagram is shown in Figure  \ref{ggggg}.
\noindent
The complete amplitude can be generated from the maximally helicity violating
MHV amplitudes \cite{ST,STi}. There are $24$ possible orderings of the five
gluon vertex operators along the boundary of the disk, cf. Figure \ref{ggggg}.
Usually only one amplitude is written
explicitly -- a {\it partial\/} amplitude associated to one specific
Chan--Paton factor -- although all of them are necessary for collider applications.
We start from one specific ``mostly minus''
helicity amplitude with the Chan-Paton factor corresponding to the order
$\rho=(1,2,3,4,5)$.
In Refs. \cite{ST,STi}, we introduced\footnote{In Sections 3--7 and in the
Appendix we use dimensionless variables (\ref{sijnodim}). However, in order
to streamline notation, we omit carets. We will switch to standard
kinematic variables (\ref{sidim}) and (\ref{sinodim}) in Section 8.}
two  basis functions, which are given in Eq.~\req{BASIS} and
allow to write the amplitude in a very simple way
\begin{equation}\label{m5}
{\cal M}_{(12345)}(g^+_1,g^-_2, g^-_3, g^-_4, g^+_5) =
[\ {V^{(5)}(s_j)} \ - \ 2i\,
 P^{(5)}(s_j)\, \epsilon(1,2,3,4)\ ]\ {\cal M}_{(12345)}^{\te{QCD}}\!(g^+_1,g^-_2, g^-_3, g^-_4, g^+_5) ~,\end{equation}
where the QCD amplitude is
\begin{equation}
{\cal M}_{(12345)}^{\te{QCD}}\!(g^+_1,g^-_2, g^-_3, g^-_4, g^+_5) =
4\sqrt{2} \,g^3 \; \frac{[15]^4}{[12][23][34][45][51]}~\Tr ( T^{a_{1}} \, T^{a_{2}} \, T^{a_{3}} \, T^{a_{4}} \, T^{a_{5}}).\end{equation}
and the prefactor contains\begin{eqnarray}\label{vfac}
V^{(5)}(s_j) &\equiv&  s_2s_5\, f_1 \ + \ \frac{1}{2}\; (s_2s_3+s_4s_5-s_1s_2-s_3s_4-s_1s_5)\, f_2 \\
P^{(5)}(s_j)&\equiv& f_2 \label{pfac}
\end{eqnarray}
The result Eq. (\ref{m5}) has a nice factorized form, with the string effects succinctly extracted in a single factor multiplying the field-theoretical QCD amplitude. In order to discuss other orderings, it is convenient to introduce the kinematic function
\begin{equation}\label{cfun}
{\cal C}(k_1,k_2,k_3,k_4,k_5) \ \ \equiv \ \ \frac{  {V^{(5)}(s_j)} \ - \ 2i\,
 P^{(5)}(s_j)\, \epsilon(1,2,3,4)  }{[12][23][34][45][51]}~,\end{equation}
so that
\begin{equation}\label{mother}
{\cal M}_{(12345)}(g^+_1,g^-_2, g^-_3, g^-_4, g^+_5)= 4\sqrt{2} \  g^3 \,[15]^4 \
{\cal C}(k_1,k_2,k_3,k_4,k_5)\ \Tr ( T^{a_{1}} \, T^{a_{2}} \, T^{a_{3}} \, T^{a_{4}} \, T^{a_{5}})
\end{equation}
Note that the function $\cal C$ is even under cyclic permutations of the momenta and odd under mirror reflections $(1,2,3,4,5)\to (5,4,3,2,1)$.

The amplitudes associated to other orderings (for the same helicity configuration) can be \linebreak obtained from Eq. (\ref{mother}) by simultaneous permutations of the gauge group generators,
\linebreak
$(a_{1},a_{2},a_{3},a_{4},a_{5})\to (a_{1_{\rho}},a_{2_{\rho}},a_{3_{\rho}},a_{4_{\rho}},a_{5_{\rho}})$
 and momenta inside the ${\cal C}$-function,
 $(k_{1},k_{2},k_{3},k_{4},k_{5})\to (k_{1_{\rho}},k_{2_{\rho}},k_{3_{\rho}},k_{4_{\rho}},k_{5_{\rho}})$, where $\rho$ is one of the 24 permutations in $S_5/ \ZZ_5$ and $i_{\rho}\equiv\rho(i)$.
Note that the  ``helicity factor'' $[15]^4$ remains intact. Due to antisymmetry of $\cal C$-functions under mirror reflections, the relevant color factors are
\begin{eqnarray}\label{chanp5}
t^{a_{1}a_{2}a_{3}a_{4}a_{5}}&=&\Tr ( T^{a_{1}} \, T^{a_{2}} \, T^{a_{3}} \, T^{a_{4}} \, T^{a_{5}})-\Tr ( T^{a_{5}} \, T^{a_{4}} \, T^{a_{3}} \, T^{a_{2}} \, T^{a_{1}})\\ &=&
i \,  f^{a_{1} a_{2}n}  \big( d^{a_{3}a_{4}a_{5}n}  -  \frac{1}{12}  f^{a_{3}a_{4}m} f^{a_{5}nm} \big) +  i \,  f^{a_{1} a_{3}n}  \big( d^{a_{2}a_{4}a_{5}n}  - \frac{1}{12}  f^{a_{2}a_{4}m}  f^{a_{5}nm} \big) \nonumber \\
&+& \!\!i \,  f^{a_{2} a_{3}n}  \big( d^{a_{1}a_{4}a_{5}n} -  \frac{1}{12}  f^{a_{1}a_{5}m} \, f^{a_{4}nm} \big)+ i \,  f^{a_{4} a_{5}n}  \big( d^{a_{1}a_{2}a_{3}n}  +  \frac{1}{12} \; f^{a_{2}a_{3}m} f^{a_{1}nm} \big),\nonumber
\end{eqnarray}
where $d$--symbols denote symmetrized traces and $f$'s are the structure constants.
Now the full amplitude can be written as a sum of twelve terms
\begin{equation}\label{gmhv5}
{\cal M}(g^+_1,g^-_2, g^-_3, g^-_4, g^+_5) \eq 4\sqrt{2} \, g^3 \, [15]^4\sum_{\rho\in \Pi_5}\,
{\cal C}(k_{1_{\rho}},k_{2_{\rho}},k_{3_{\rho}},k_{4_{\rho}},k_{5})~
t^{a_{1_{\rho}}a_{2_{\rho}}a_{3_{\rho}}a_{4_{\rho}}a_{5}}\ ,
\end{equation}
where:
\begin{eqnarray}\label{p5set}
\Pi_5&\equiv&\Big\{\,(1,2,3,4,5),~(1,2,4,3,5),~(1,3,4,2,5),~(1,3,2,4,5),~(1,4,2,3,5),
~(1,4,3,2,5),~~~~~~~\nonumber\\
&& (2,1,3,4,5),~(2,1,4,3,5),~(2,3,1,4,5),~(2,4,1,3,5),~(3,1,2,4,5),~(3,2,1,4,5)\,\Big\}.
\end{eqnarray}
All other ``mostly minus'' amplitudes, with the two positive helicity gluons labeled by arbitrary $i$ and $j$ instead of 1 and 5,  can be obtained from Eq. (\ref{gmhv5}) by simply replacing the helicity factor $[15]^4\to [i^+j^+]^4$. ``Mostly plus'' amplitudes are obtained by complex conjugation.

\section{Three gluon and two fermion amplitudes $\bm{gggq\bar q}$}

In this Section we compute the full string
amplitude ${\cal M}$ for two quarks and three gluons. The SCFT correlation
function involved is the following, c.f. \req{0,2}:
\beq
\langle V_{A^{x}}^{(0)}(z_{1},\xi _{1},k _{1}) \, V_{A^{y}}^{(0)}(z_{2},\xi _{2},
k _{2}) \, V_{A^{z}}^{(-1)}(z_{3},\xi _{3},k _{3}) \,
V_{\psi^{\al}_{\be}}^{(-1/2)}(z_{4},u,k_4) \,
V_{\bar{\psi}^{\be}_{\al}}^{(-1/2)}(z_{5},\bar{v} ,k_{5}) \rangle_{\rho}\ .
\label{start}
\eeq
In \req{start} the ghost pictures are adjusted such that the background charge of $-2$
is compensated by the total ghost number. The $\rho$ subscript only affects
the Chan Paton degrees of freedom subject to the ordering $\rho$ of the five vertex operators to be specified later.

With all the ingredients presented in Section 2 and in the Appendix \ref{appA}  the full SCFT correlator \req{start} involving the five parton vertex operators  is evaluated in the following.

\subsection{Kinematic structure of the  amplitude}
\label{KINEMATICS}

After performing all Wick contractions the correlator \req{start} gives rise to many
kinematic terms.
For an easier bookkeeping, we will introduce shorthands for the index structures in which the momenta $k_i$, spinors $u,\bar{v}$ and polarization vectors $\xi_i$ occur. They are ordered according to the number of $\si$ matrices and reduced with the help of momentum conservation, the transverse gauge condition $k_i \xi_i = 0$ and the massless Dirac equation $k_{4\mu}  u\si^{\mu} \bar{v} = k_{5\mu}  u\si^{\mu} \bar{v} = 0$.

Let us first of all list the thirty terms where the Ramond spinors contract with a single Dirac matrix:
\begin{alignat*}{3}
{\cal X}^{1} \ \ &= \ \ \tfrac{1}{2\al'} \; \bigl( \xi  _{1}  \xi _{2} \bigr) \, \xi_{3\mu} \bigl( u \, \si^{\mu} \, \bar{v} \bigr)  \ , \ \ \ \ \  & {\cal X}^{16} \ \ &= \ \ \bigl( \xi  _{1}  k _{3} \bigr) \, \bigl( \xi _{3}  k _{2} \bigr) \, \xi _{2\mu} \, \bigl( u \, \si^{\mu} \, \bar{v} \bigr)\ , \\
{\cal X}^{2} \ \ &= \ \ \bigl( \xi  _{1}  k _{2} \bigr) \, \bigl( \xi _{2}  k _{1} \bigr) \, \xi _{3\mu} \, \bigl( u \, \si^{\mu} \, \bar{v} \bigr)  \ , \ \ \ \ \  & {\cal X}^{17} \ \ &= \ \ \bigl( \xi  _{2}  \xi _{3} \bigr) \, \bigl( \xi _{1}  k _{3} \bigr) \, k _{2\mu} \, \bigl( u \, \si^{\mu} \, \bar{v} \bigr)\ ,  \\
{\cal X}^{3} \ \ &= \ \ \bigl( \xi  _{1}  k _{3} \bigr) \, \bigl( \xi _{2}  k _{1} \bigr) \, \xi _{3\mu} \, \bigl( u \, \si^{\mu} \, \bar{v} \bigr)  \ , \ \ \ \ \  & {\cal X}^{18} \ \ &= \ \ \bigl( \xi  _{1}  k _{4} \bigr) \, \bigl( \xi _{2}  k _{1} \bigr) \, \xi _{3\mu} \, \bigl( u \, \si^{\mu} \, \bar{v} \bigr)\ ,  \\
{\cal X}^{4} \ \ &= \ \ \bigl( \xi  _{2}  k _{1} \bigr) \, \bigl( \xi _{3}  k _{1} \bigr) \, \xi _{1\mu} \, \bigl( u \, \si^{\mu} \, \bar{v} \bigr)  \ , \ \ \ \ \  & {\cal X}^{19} \ \ &= \ \ \bigl( \xi  _{1}  k _{2} \bigr) \, \bigl( \xi _{2}  k _{4} \bigr) \, \xi _{3\mu} \, \bigl( u \, \si^{\mu} \, \bar{v} \bigr)\ ,  \\
{\cal X}^{5} \ \ &= \ \ \bigl( \xi  _{1}  \xi _{3} \bigr) \, \bigl( \xi _{2}  k _{1} \bigr) \, k _{1\mu} \, \bigl( u \, \si^{\mu} \, \bar{v} \bigr)  \ , \ \ \ \ \ & {\cal X}^{20} \ \ &= \ \ \bigl( \xi  _{1}  k _{3} \bigr) \, \bigl( \xi _{2}  k _{4} \bigr) \, \xi _{3\mu} \, \bigl( u \, \si^{\mu} \, \bar{v} \bigr)\ ,  \\
{\cal X}^{6} \ \ &= \ \ \bigl( \xi  _{1}  k _{2} \bigr) \, \bigl( \xi _{2}  k _{3} \bigr) \, \xi _{3\mu} \, \bigl( u \, \si^{\mu} \, \bar{v} \bigr)  \ , \ \ \ \ \  & {\cal X}^{21} \ \ &= \ \ \bigl( \xi  _{2}  k _{4} \bigr) \, \bigl( \xi _{3}  k _{1} \bigr) \, \xi _{1\mu} \, \bigl( u \, \si^{\mu} \, \bar{v} \bigr)\ ,  \\
{\cal X}^{7} \ \ &= \ \ \tfrac{s_{1}}{2\al'} \; \bigl( \xi  _{2}  \xi _{3} \bigr) \, \xi _{1\mu} \, \bigl( u \, \si^{\mu} \, \bar{v} \bigr)  \ , \ \ \ \ \  & {\cal X}^{22} \ \ &= \ \ \bigl( \xi  _{1}  \xi _{3} \bigr) \, \bigl( \xi _{2}  k _{4} \bigr) \, k _{1\mu} \, \bigl( u \, \si^{\mu} \, \bar{v} \bigr) \ , \\
{\cal X}^{8} \ \ &= \ \ \bigl( \xi  _{2}  k _{1} \bigr) \, \bigl( \xi _{3}  k _{2} \bigr) \, \xi _{1\mu} \, \bigl( u \, \si^{\mu} \, \bar{v} \bigr)  \ , \ \ \ \ \  & {\cal X}^{23} \ \ &= \ \ \bigl( \xi  _{1}  k _{4} \bigr) \, \bigl( \xi _{2}  k _{3} \bigr) \, \xi _{3\mu} \, \bigl( u \, \si^{\mu} \, \bar{v} \bigr)\ ,  \\
{\cal X}^{9} \ \ &= \ \ \bigl( \xi  _{2}  \xi _{3} \bigr) \, \bigl( \xi _{1}  k _{2} \bigr) \, k _{1\mu} \, \bigl( u \, \si^{\mu} \, \bar{v} \bigr)  \ , \ \ \ \ \  & {\cal X}^{24} \ \ &= \ \ \bigl( \xi  _{1}  k _{4} \bigr) \, \bigl( \xi _{3}  k _{2} \bigr) \, \xi _{2\mu} \, \bigl( u \, \si^{\mu} \, \bar{v} \bigr)\ ,  \\
{\cal X}^{10} \ \ &= \ \ \bigl( \xi  _{1}  \xi _{2} \bigr) \, \bigl( \xi _{3}  k _{2} \bigr) \, k _{1\mu} \, \bigl( u \, \si^{\mu} \, \bar{v} \bigr)  \ , \ \ \ \ \  & {\cal X}^{25} \ \ &= \ \ \bigl( \xi  _{1} k _{4} \bigr) \, \bigl( \xi _{2}  \xi _{3} \bigr) \, k _{2\mu} \, \bigl( u \, \si^{\mu} \, \bar{v} \bigr)\ ,  \\
{\cal X}^{11} \ \ &= \ \ \bigl( \xi  _{1} k _{2} \bigr) \, \bigl( \xi _{3}  k _{2} \bigr) \, \xi _{2\mu} \, \bigl( u \, \si^{\mu} \, \bar{v} \bigr)  \ , \ \ \ \ \  & {\cal X}^{26} \ \ &= \ \ \bigl( \xi_{1}  k _{4} \bigr) \, \bigl( \xi _{2}  k _{4} \bigr) \, \xi _{3\mu} \, \bigl( u \, \si^{\mu} \, \bar{v} \bigr)\ ,  \\
{\cal X}^{12} \ \ &= \ \ \bigl( \xi  _{2}  \xi _{3} \bigr) \, \bigl( \xi _{1}  k _{2} \bigr) \, k _{2\mu} \, \bigl( u \, \si^{\mu} \, \bar{v} \bigr)  \ , \ \ \ \ \  & {\cal X}^{27} \ \ &= \ \ \tfrac{s_{1}}{2\al'} \; \bigl( \xi  _{1}  \xi _{3} \bigr) \, \xi _{2\mu} \, \bigl( u \, \si^{\mu} \, \bar{v} \bigr)\ ,  \\
{\cal X}^{13} \ \ &= \ \ \bigl( \xi  _{1}  k _{3} \bigr) \, \bigl( \xi _{2}  k _{3} \bigr) \, \xi _{3\mu} \, \bigl( u \, \si^{\mu} \, \bar{v} \bigr)  \ , \ \ \ \ \  & {\cal X}^{28} \ \ &= \ \ \bigl( \xi_{1}  k _{2} \bigr) \, \bigl( \xi _{3}  k _{1} \bigr) \, \xi _{2\mu} \, \bigl( u \, \si^{\mu} \, \bar{v} \bigr) \ , \\
{\cal X}^{14} \ \ &= \ \ \bigl( \xi  _{2}  k _{3} \bigr) \, \bigl( \xi _{3}  k _{1} \bigr) \, \xi _{1\mu} \, \bigl( u \, \si^{\mu} \, \bar{v} \bigr)  \ , \ \ \ \ \  & {\cal X}^{29} \ \ &= \ \ \bigl( \xi_{1}  \xi _{3} \bigr) \, \bigl( \xi _{2}  k _{1} \bigr) \, k _{2\mu} \, \bigl( u \, \si^{\mu} \, \bar{v} \bigr)\ ,  \\
{\cal X}^{15} \ \ &= \ \ \bigl( \xi  _{1}  \xi _{3} \bigr) \, \bigl( \xi _{2}  k _{3} \bigr) \, k _{1\mu} \, \bigl( u \, \si^{\mu} \, \bar{v} \bigr) \ , \ \ \ \ \ & {\cal X}^{30} \ \ &= \ \ \bigl( \xi_{1}  \xi _{2} \bigr) \, \bigl( \xi _{3}  k _{1} \bigr) \, k _{2\mu} \, \bigl( u \, \si^{\mu} \, \bar{v} \bigr)\ .
\end{alignat*}
Further fourteen cases with a $\si^{\mu} \bar{\si}^{\nu} \si^{\la}$ triple product arise:
\begin{alignat*}{3}
{\cal Y}^{1} \ \ &= \ \ \tfrac{1}{2} \; \bigl( \xi _{2}  k _{1} \bigr) \, k_{1\mu}  \, \xi_{1\nu}  \, \xi_{3\la}  \, \bigl( u \, \si^{\mu} \, \bar{\si}^{\nu} \, \si^{\la} \, \bar{v} \bigr) \ , \ \ \ \ \
&{\cal Y}^{2} \ \ &= \ \ \tfrac{1}{2} \; \tfrac{s_{1}}{2\al'} \; \xi_{1\mu} \, \xi_{2\nu}  \, \xi_{3\la}  \, \bigl( u \, \si^{\mu} \, \bar{\si}^{\nu} \, \si^{\la} \, \bar{v} \bigr)  \ ,\\
{\cal Y}^{3} \ \ &= \ \ \tfrac{1}{2} \; \bigl( \xi _{2}  k _{1} \bigr) \, \xi_{1\mu}  \, k_{2\nu}  \, \xi_{3\la} \, \bigl( u \, \si^{\mu} \, \bar{\si}^{\nu} \, \si^{\la} \, \bar{v} \bigr) \ , \ \ \ \ \
&{\cal Y}^{4} \ \ &= \ \ \tfrac{1}{2} \; \bigl( \xi _{1}  k _{2} \bigr) \, k_{1\mu} \, \xi_{2\nu}  \, \xi_{3\la}  \, \bigl( u \, \si^{\mu} \, \bar{\si}^{\nu} \, \si^{\la} \, \bar{v} \bigr) \ , \\
{\cal Y}^{5} \ \ &= \ \ \tfrac{1}{2} \; \bigl( \xi _{1}  \xi _{2} \bigr) \, k_{1\mu}  \, k_{2\nu}  \, \xi_{3\la} \, \bigl( u \, \si^{\mu} \, \bar{\si}^{\nu} \, \si^{\la} \, \bar{v} \bigr) \ , \ \ \ \ \
&{\cal Y}^{6} \ \ &= \ \ \tfrac{1}{2} \; \bigl( \xi _{1}  k _{2} \bigr) \, k_{2\mu}  \, \xi_{2\nu}  \, \xi_{3\la}  \, \bigl( u \, \si^{\mu} \, \bar{\si}^{\nu} \, \si^{\la} \, \bar{v} \bigr) \ , \\
{\cal Y}^{7} \ \ &= \ \ \tfrac{1}{2} \; \bigl( \xi _{1}  k _{3} \bigr) \, k_{2\mu}  \, \xi_{2\nu}  \, \xi_{3\la}  \, \bigl( u \, \si^{\mu} \, \bar{\si}^{\nu} \, \si^{\la} \, \bar{v} \bigr) \ , \ \ \ \ \
&{\cal Y}^{8} \ \ &= \ \ \tfrac{1}{2} \; \bigl( \xi _{3}  k _{1} \bigr) \, \xi_{1\mu}  \, k_{2\nu}  \, \xi_{2\la}  \, \bigl( u \, \si^{\mu} \, \bar{\si}^{\nu} \, \si^{\la} \, \bar{v} \bigr) \ , \\
{\cal Y}^{9} \ \ &= \ \ \tfrac{1}{2} \; \bigl( \xi _{1}  \xi _{3} \bigr) \, k_{1\mu} \, k_{2\nu}  \, \xi_{2\la} \, \bigl( u \, \si^{\mu} \, \bar{\si}^{\nu} \, \si^{\la} \, \bar{v} \bigr) \ , \ \ \ \ \
&{\cal Y}^{10} \ \ &= \ \ \tfrac{1}{2} \; \bigl( \xi _{2}  k _{3} \bigr) \, k_{1\mu} \, \xi_{1\nu}  \, \xi_{3\la}  \, \bigl( u \, \si^{\mu} \, \bar{\si}^{\nu} \, \si^{\la} \, \bar{v} \bigr)  \ ,\\
{\cal Y}^{11} \ \ &= \ \ \tfrac{1}{2} \; \bigl( \xi _{3}  k _{2} \bigr) \, k_{1\mu}  \, \xi_{1\nu}  \, \xi_{2\la} \, \bigl( u \, \si^{\mu} \, \bar{\si}^{\nu} \, \si^{\la} \, \bar{v} \bigr) \ , \ \ \ \ \
&{\cal Y}^{12} \ \ &= \ \ \tfrac{1}{2} \; \bigl( \xi _{2}  \xi _{3} \bigr) \, k_{1\mu}  \, \xi_{1\nu}  \, k_{2\la}  \, \bigl( u \, \si^{\mu} \, \bar{\si}^{\nu} \, \si^{\la} \, \bar{v} \bigr) \ , \\
{\cal Y}^{13} \ \ &= \ \ \tfrac{1}{2} \; \bigl( \xi _{2}  k _{4} \bigr) \, k_{1\mu}  \, \xi_{1\nu}  \, \xi_{3\la} \, \bigl( u \, \si^{\mu} \, \bar{\si}^{\nu} \, \si^{\la} \, \bar{v} \bigr) \ , \ \ \ \ \
&{\cal Y}^{14} \ \ &= \ \ \tfrac{1}{2} \; \bigl( \xi _{1}  k _{4} \bigr) \, k_{2\mu} \, \xi_{2\nu}  \, \xi_{3\la}  \, \bigl( u \, \si^{\mu} \, \bar{\si}^{\nu} \, \si^{\la} \, \bar{v} \bigr)\ .
\end{alignat*}
Finally, the seven point function (\ref{1,6}) introduces a five $\si$--term,
\beq
{\cal Z} \ \ = \ \ \tfrac{1}{4} \; \xi_{1\mu}  \, k_{1\nu}  \, \xi_{2\la}  \, k_{2\rho}  \, \xi_{3\tau}  \, \bigl( u \, \si^{\mu} \, \bar{\si}^{\nu} \, \si^{\la} \, \bar{\si}^{\rho} \, \si^{\tau} \, \bar{v} \bigr) \ . \notag
\eeq

The next Subsection
is devoted to the integrand
${V}_{\te{CKG}}^{-1}\  \langle V_{\Phi_{1}}(z_{1}) \, ... \,
V_{\Phi_{5}}(z_{5}) \rangle$ of \req{0,2}.
Its determination involves SCFT techniques and careful bookkeeping of more
than forty additive contributions. Then the integration over the disk boundary is performed by  picking a particular ordering $\rho$ of the vertex operator insertion points and evaluate the hypergeometric integrals occurring in this subamplitude. The
technology of these integrals has been
already employed in \cite{6GG} for the  five gluon amplitudes, c.f. also
Section \ref{GGGGG}.
Eventually, the partial amplitudes is combined to the full amplitude in Section
5.

\subsection{Computation of the full string amplitude}

The correlator \req{start} factorizes into several independent pieces and
we can concentrate on separate correlation functions.
All necessary space--time correlators are listed in the Appendix \ref{appA}:
for ghosts in \req{1,3a}, for the bosonic space--time coordinate fields
$X^{\mu}$, given in
\req{1,4a}--\req{1,4c}, and finally for the NS-R SCFT with the $\psi^{\mu}$,
$S_{\al}$ and $S^{\dbe}$ fields, given in  \req{1,5b}.
Because of the term proportional to $\psi^{\mu} \psi^{\nu}$ in the
$V^{(0)}_{A}$ vertex operators (\ref{1,1b}) there also occurs a seven--point
correlator involving five NS fermions $\psi^\mu$ and two R spin
fields $S_\al$, $S_{\dbe}$. This correlator, presented in \req{1,6}, is derived
thoroughly in a separate paper \cite{spin}.
Finally the internal Ramond fields combine into  the correlator \req{1,3b}.

In this Subsection, we give the intermediate result of inserting all the subcorrelators and kinematic factors into the full five--parton correlation function. The large number of additive terms shall be thought of being grouped according to the $z_{ij}$ dependences (where 36 terms with distinct functions of the $z_{ij}$ each are omitted along the ... dots at this stage):
\begin{align}
\langle &V_{A^{x}}^{(0)}(z_{1},\xi _{1},k _{1}) \, V_{A^{y}}^{(0)}(z_{2},\xi _{2}, k _{2}) \, V_{A^{z}}^{(-1)}(z_{3},\xi _{3},k _{3}) \, V_{\psi^{\al}_{\be}}^{(-1/2)}(z_{4},u,k_{4}) \, V_{\bar{\psi}^{\be}_{\al}}^{(-1/2)}(z_{5},\bar{v} ,k _{5}) \rangle_{\rho} \notag \\
&= \ \ 2\sqrt{2} \, \al'^{2} \, g_{\te{D}p_{x}} \, g_{\te{D}p_{y}} \, g_{\te{D}p_{z}} \, e^{\phi_{10}} \, t(\rho) \xi_{1\mu}  \, \xi_{2\la}  \, \xi_{3\tau} \, u^{\al} \, \bar{v}^{\dbe} \, \langle e^{-\phi(z_{3})} \, e^{-\phi(z_{4})/2} \, e^{-\phi(z_{5})/2} \rangle\notag \\
&\times \ \langle \Xi^{a \cap b}(z_{4}) \, \bar{\Xi}^{a \cap b}(z_{5}) \rangle \ \Biggl[\ i^{2} \, \Bigl\langle \pa X^{\mu}(z_{1}) \, \pa X^{\la}(z_{2}) \, \prod_{i=1}^{5} e^{ik_i   X(z_{i})} \Bigr\rangle \ \langle \psi^{\tau}(z_{3}) \, S_{\al}(z_{4}) \, S_{\dbe} (z_{5}) \rangle \Biggr. \notag \\
& \ \ \ \ \ \ \ \ \ \ - \ 2i\al' \, \Bigl\langle  \pa X^{\mu}(z_{1}) \, \prod_{i=1}^{5} e^{ik_i   X(z_{i})} \Bigr\rangle \ k_{2\rho} \, \langle \psi^{\la}(z_{2}) \, \psi^{\rho}(z_{2}) \, \psi^{\tau}(z_{3}) \, S_{\al} (z_{4}) \, S_{\dbe}(z_{5}) \rangle \notag \\
& \ \ \ \ \ \ \ \ \ \ - \ 2i\al' \, \Bigl\langle  \pa X^{\la}(z_{2}) \, \prod_{i=1}^{5} e^{ik_i   X(z_{i})} \Bigr\rangle \ k_{1\nu}  \, \langle \psi^{\mu}(z_{1}) \, \psi^{\nu}(z_{1}) \, \psi^{\tau}(z_{3}) \, S_{\al} (z_{4}) \, S_{\dbe}(z_{5}) \rangle \notag \\
& \ \ \ \ \ \ \ \ \ \ + \ 4\al'^{2} \, \Biggl. \Bigl\langle  \prod_{i=1}^{5} e^{ik_i   X(z_{i})} \Bigr\rangle \ k_{1\nu}  \, k_{2\rho}  \, \langle \psi^{\mu}(z_{1}) \, \psi^{\nu}(z_{1}) \, \psi^{\la}(z_{2}) \, \psi^{\rho}(z_{2}) \, \psi^{\tau}(z_{3}) \, S_{\al} (z_{4}) \, S_{\dbe}(z_{5}) \rangle \Biggr] \notag \\
&= \ \ 8 \, \al'^{4} \, g_{\te{D}p_{x}} \, g_{\te{D}p_{y}} \, g_{\te{D}p_{z}} \, e^{\phi_{10}} \, t(\rho)\notag \\
& \times \ \frac{1}{z_{34} \, z_{35} \, z_{45}} \; \prod_{i,j=1 \atop {i<j}}^{5} \bigl| z_{ij} \bigr|^{2\al' k_i   k_j} \ \Biggl[\ \frac{{\cal X}^{1} \, - \, {\cal X}^{2}}{(z_{12})^{2}} \ - \ \frac{{\cal X}^{3}}{z_{12} \, z_{13}}  \ + \ \ \ ... \ \ \ + \ \frac{(z_{45})^{2} \, {\cal Z}}{z_{14} \, z_{15} \, z_{24} \, z_{25}}\ \Biggr]\ ,
\label{1,7}
\end{align}
with the group factor: 
$$t(\rho)=\te{Tr} ( T^{1_\rho} \, T^{2_\rho} \, T^{3_\rho} \, T^{4_\rho} \, T^{5_\rho} )\ .$$

The $PSL(2,\RR)$ symmetry on the disk allows to fix three of the five vertex operator
insertion points. From the ''volume'' factor ${V}_{\te{CKG}}$ of the conformal
Killing group, one has to include a $c$--ghost correlator
$\langle c(z_{i}) \, c(z_{j}) \, c(z_{k}) \rangle \eq z_{ij} \, z_{ik} \, z_{jk} $
evaluated at the fixed  positions $z_{i}$.
It turns out to be most economic to send $z_{k} := z_{5} \mto \infty$ (such that quotients like $\frac{z_{15}}{z_{25}}$ tend to 1), the remaining two ''jokers'' are used to fix $z_{i} := z_{3} = \zeta_{A}$ and $z_{j} := z_{4} = \zeta_{B}$ such that
$\frac{1}{{V}_{\te{CKG}}} \; \frac{1}{z_{34} \, z_{35} \, z_{45}}=
\frac{\langle c(z_{3}) \, c(z_{4}) \, c(z_{5}) \rangle}{z_{34} \, z_{35} \, z_{45}} \;
\de(z_{3} - \zeta_{A}) \, \de(z_{4} - \zeta_{B}) \,
\de(z_{5} - \infty)= \de(z_{3} - \zeta_{A}) \,
\de(z_{4} - \zeta_{B}) \, \de(z_{5} - \infty)$.
The unintegrated, $PSL(2,\RR)$ fixed subamplitude (\ref{1,7}) then simplifies to an expression with eleven distinct $z_{ij}$ dependences only (where it is most convenient to set the finite positions $\zeta_{A}$, $\zeta_{B}$ to $0$ and $1$):
\begin{align}
&V_{\te{CKG}}^{-1}\ \langle V_{A^{x}}^{(0)}(z_{1},\xi _{1},k _{1}) \, V_{A^{y}}^{(0)}(z_{2},\xi _{2}, k _{2}) \, V_{A^{z}}^{(-1)}(z_{3},\xi _{3},k _{3}) \, V_{\psi^{\al}_{\be}}^{(-1/2)}(z_{4},u,k_4) \, V_{\bar{\psi}^{\be}_{\al}}^{(-1/2)}(z_{5},\bar{v} ,k _{5}) \rangle_{\rho} \notag \\
& \ \ = \ \ 8 \, \al'^{4} \ g_{\te{D}p_{x}} \ g_{\te{D}p_{y}} \ g_{\te{D}p_{z}} \ e^{\phi_{10}} \ t(\rho) \ \de(z_{3} \, - \ \zeta_{A}) \, \de(z_{4} \, - \ \zeta_{B}) \, \de(z_{5} \, - \, \infty)   \,\notag \\
&\times\bigl| z_{12} \bigr|^{s_{1}}\ \bigl| z_{13} \bigr|^{s_{4} - s_{1} - s_{2}} \  \bigl| z_{14} \bigr|^{s_{2} - s_{4} - s_{5}} \ \bigl| z_{23} \bigr|^{s_{2}} \, \bigl| z_{24}\ \bigr|^{s_{5}- s_{2}-s_{3}} \, \bigl| z_{34} \bigr|^{s_{3}}  \,\notag \\
&\times\Biggl[  \frac{{\cal K}^{1}}{z_{12} \, z_{13}} \ + \ \frac{{\cal K}^{2}}{z_{12} \, z_{23}} \ + \ \frac{{\cal K}^{3}}{z_{13} \, z_{23}}\ + \ \frac{{\cal K}^{4}}{z_{12} \, z_{14}} \ + \ \frac{{\cal K}^{5}}{z_{12} \, z_{24}} \ + \ \frac{{\cal K}^{6}}{z_{13} \, z_{24}}  \Biggr. \notag \\
& \hskip1cm\Biggl. \ + \ \frac{{\cal K}^{7}}{z_{23} \, z_{14}} \ + \ \frac{{\cal K}^{8}}{z_{14} \, z_{24}} \ + \ \frac{z_{14} \, {\cal K}^{9}}{z_{12} \, z_{13} \, z_{24}} \ + \ \frac{{\cal K}^{10}}{(z_{12})^{2}} \ + \ \frac{z_{14} \, {\cal K}^{11}}{(z_{12})^{2} \, z_{24}} \Biggr]\ .
\label{1,10}
\end{align}
We have introduced the following set of kinematic factors ${\cal K}^{i}$
\begin{align}
{\cal K}^{1} \ \ &= \ \  -{\cal X}^{3} \ + \ {\cal X}^{4} \ - \  {\cal X}^{5}\ , \notag \\
{\cal K}^{2} \ \ &= \ \  {\cal X}^{6} \ - \ {\cal X}^{7} \ + \  {\cal X}^{8} \ + \ {\cal X}^{9} \ - \ {\cal X}^{10} \ - \ {\cal X}^{11} \ + \ {\cal X}^{12}\ , \notag \\
{\cal K}^{3} \ \ &= \ \ {\cal X}^{13} \ - \ {\cal X}^{14} \ + \ {\cal X}^{15} \ - \ {\cal X}^{16} \ + \ {\cal X}^{17}\ , \notag \\
{\cal K}^{4} \ \ &= \ \ -{\cal X}^{18} \ - \ {\cal Y}^{1}\ , \notag \\
{\cal K}^{5} \ \ &= \ \ {\cal X}^{19} \ - \  {\cal Y}^{2} \ + \ {\cal Y}^{3} \ + \ {\cal Y}^{4} \ - \ {\cal Y}^{5} \ + \ {\cal Y}^{6}\ , \notag \\
{\cal K}^{6} \ \ &= \ \ {\cal X}^{20} \ - \ {\cal X}^{21} \ + \ {\cal X}^{22} \ + \ {\cal Y}^{7} \ - \ {\cal Y}^{8} \ + \ {\cal Y}^{9}\ , \notag \\
{\cal K}^{7} \ \ &= \ \ {\cal X}^{23} \ - \ {\cal X}^{24} \ + \ {\cal X}^{25} \ + \ {\cal Y}^{10} \ - \ {\cal Y}^{11} \ + \ {\cal Y}^{12}\ , \notag \\
{\cal K}^{8} \ \ &= \ \ {\cal X}^{26} \ + \ {\cal Y}^{13} \ + \ {\cal Y}^{14} \ + \ {\cal Z}\ , \notag \\
{\cal K}^{9} \ \ &= \ \ {\cal X}^{27} \ - \ {\cal X}^{28} \ - \ {\cal X}^{29} \ + \ {\cal X}^{30}\ , \notag \\
{\cal K}^{10} \ \ &= \ \ {\cal X}^{1} \ - \ {\cal X}^{2}\ , \notag \\
{\cal K}^{11} \ \ &= \ \ -s_{1} \, {\cal X}^{1} \ + \ {\cal X}^{2}\ ,
\label{1,11}
\end{align}
which is related
to the set of kinematic factors ${\cal X}^{j}$, ${\cal Y}^{k}$ and ${\cal Z}$
defined in Subsection \ref{KINEMATICS}.

\subsection{Partial subamplitudes}

After having worked out  the correlator \req{start} in the previous
Subsections, we now turn to the integration within a fixed subamplitude.
The correlator \req{1,10} gives rise to the partial amplitude
\bea
&&\hskip-0.5cm\ds{{\cal M}_{\rho} \bigl[A^{x} (\xi _{1} , k _{1}), A^{y} (\xi _{2}, k _{2}),
A^{z} (\xi _{3} , k _{3}),\psi^{\al}_{\be}( u , k _{4}) ,
\bar{\psi}^{\be}_{\al}(\bar{v}, k _{5}) \bigr]=
t(\rho)\ A(1_\rho,2_\rho,3_\rho,4_\rho,5_\rho)=
\frac{e^{-\phi_{10}}}{2 \, \al'^{2}} \ V_{\te{CKG}}^{-1}}\\[3mm]
&&\hskip-0.5cm\ds{\times\int\limits_{{\cal I}_{\rho}}\ \lf(\prod_{k=1}^{5} \dd z_{k}\ri)\
\langle V_{A^{x}}^{(0)}(z_{1},\xi _{1},k _{1}) \, V_{A^{y}}^{(0)}(z_{2},\xi _{2},
k _{2}) \, V_{A^{z}}^{(-1)}(z_{3},\xi _{3},k _{3}) \,
V_{\psi^{\al}_{\be}}^{(-1/2)}(z_{4},u,k_4) \,
V_{\bar{\psi}^{\be}_{\al}}^{(-1/2)}(z_{5},\bar{v} ,k _{5})
\rangle_{\rho},} \label{2,99}
\eea
which refers to a specific ordering $\rho$ of the five vertex operators along the
boundary of the disk. Independent on the permutation $\rho$, there is a universal prefactor
$\frac{e^{-\phi_{10}}}{2\al'^{2}}$ in each subamplitude whose origin is
discussed in \cite{PO}. We set the Chan Paton trace
aside and focus on the color--stripped subamplitude~$A$.

Generically there are $24$ different $S_{5} / \ZZ_{5}$ permutations
$\rho$ to place the five vertex operators along the boundary of the disk.
However, not all of them yield a non--vanishing group trace.
It follows that all three gauge bosons must be associated either to one of the two stacks $a$ or $b$.

First of all, we have to distinguish two topologically inequivalent cases:
\begin{itemize}
\item [($i$)] no gauge boson vertex operator is located between the two quark
positions $z_4,z_5$ yielding the group trace $\Tr(T^{a_i}T^{a_j}T^{a_k}T^{\alpha_4}_{\beta_4}T^{\beta_5}_{\alpha_5})$ for
 $(i,j,k)$ any permutation of $(1,2,3)$;
 \item [($ii$)] one $A^{b}$ vertex operator is placed between the two
the two quark positions $z_4,z_5$ such that the Chan Paton factor reads
$\Tr(T^{a_i}T^{a_j}T^{\alpha_4}_{\beta_4} T^{b_k} T^{\beta_5}_{\alpha_5})$.
 \end{itemize}
Clearly the case of two gluon vertices between the
two quark vertices is equivalent to case~$(ii)$.

\medskip
\noindent
\underline{Case $(i)$:}\\
\noindent
Case $(i)$ leads to six non--vanishing contributions, displayed in the following
Figure  \ref{gggqqall}.
\begin{figure}[H]
\centering
\includegraphics[width=0.9\textwidth]{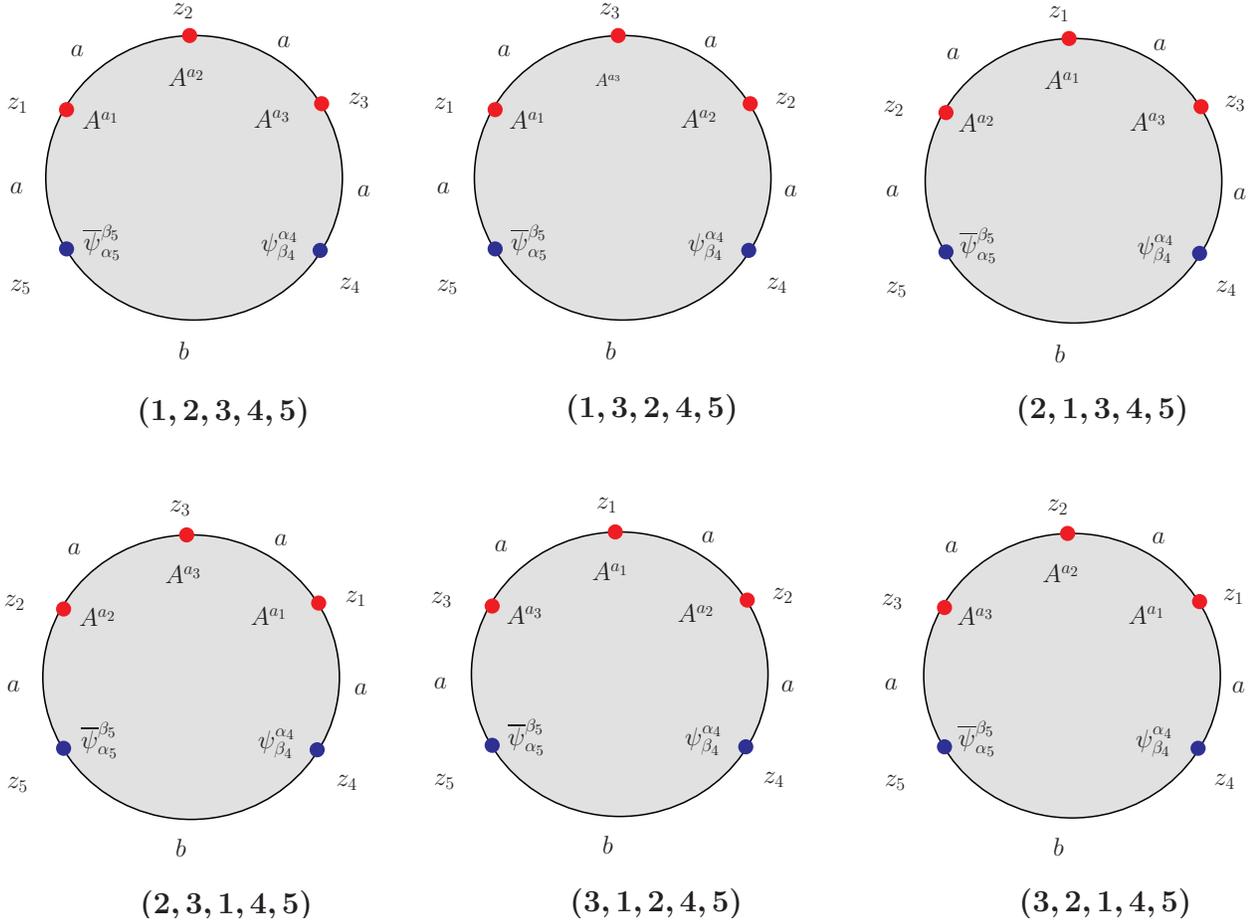}    
\caption{Orderings of gauge boson and fermion vertex operators for case $(i)$.}
\label{gggqqall}
\end{figure}
\noindent
According to the orderings of vertex positions in the case $(i)$
we fix the remaining two
vertex operator positions as
\beq
z_{3} \eq 0 \co z_{4} \eq 1 \ ,
\label{choice1a}
\eeq
while the remaining two variables $z_1$ and $z_2$
are free to choose along the boundary of the disk subject
to the six orderings encountered above.
A convenient parameterization is
\beq
z_1 \eq x \, y \co z_2 \eq x\ ,
\label{choice1b}
\eeq
with $x,y\in \RR$.
Their region of integration -- denoted as ${\cal I}_{\rho}$ in the following -- specifies the six orderings of case $(i)$.

\beq
\begin{array}{|c|c|rr|} \hline \rho & z_{1},z_{2} & {\cal I}_{\rho} & \\\hline
(1,2,3,4,5) & -\infty<z_1<z_2<0  &-\infty<x<0 &1<y<\infty \\
(1,3,2,4,5) & -\infty< z_{1} < 0 < z_{2} < 1 & 0 < x < 1 &-\infty < y < 0 \\
(2,1,3,4,5) & -\infty< z_{2} < z_{1} < 0 &-\infty < x < 0 &0 < y < 1 \\
(2,3,1,4,5) & -\infty< z_{2}< 0 < z_{1} < 1 &-\infty < x < 0 &\frac{1}{x} < y < 0 \\
(3,1,2,4,5) & 0 < z_{1} < z_{2} < 1 &0<x<1 &0<y<1 \\
(3,2,1,4,5) & 0 < z_{2} < z_{1} < 1 &0<x<1 &1 < y < \frac{1}{x} \\ \hline  \end{array}
\label{REG1}
\eeq

\vskip0.5cm
\noindent
Let us now work out the correlator \req{1,10} for the
choices \req{choice1a} and \req{choice1b} of vertex operator positions.
As discussed before, with the two fermions sitting on neighbouring positions $(4,5)$,
all the three gluons need be to attached to the same stack of branes
(which we will label by $a$). For case $(i)$ the bosonic coupling constants are the same $g_{\te{D}p_{a}}$.
Then integrating \req{1,10} yields
\bea
\ds{A(1_\rho,2_\rho,3_\rho,4_\rho,5_\rho)}&=& \ds{4 \, \al'^{2} \
g_{\te{D}p_{a}}^{3} \int \limits_{ {\cal I}_{\rho}} \dd x\  \dd y
\ |x|^{s_{4}} \,  |y|^{s_{4} - s_{1} - s_{2}} \, |1-x|^{s_{5} - s_{2} - s_{3}} \,
|1-y|^{s_{1}} \, |1-xy|^{s_{2}-s_{4}-s_{5}} }\\
&\times& \ds{\Biggl[ \ - \
\frac{{\cal K}^{1}}{x \, y \, (1-y)} \ - \ \frac{{\cal K}^{2}}{x \, (1-y)} \ +
\ \frac{{\cal K}^{3}}{x \, y} \ + \ \frac{{\cal K}^{4}}{(1-y) \, (1-xy)} \ + \ \frac{{\cal K}^{5}}{(1-x) \,
 (1-y)}
\Biggr. }\\
&&\ds{\Biggl.  - \ \frac{{\cal K}^{6}}{y \, (1-x)} \ - \ \frac{{\cal
   K}^{7}}{(1-xy)} \
+ \ \frac{x \, {\cal K}^{8}}{(1-x) \, (1-xy)} \ - \ \frac{(1-xy) \, {\cal K}^{9}}{x \, y \, (1-x) \,
 (1-y)}} \\
&&\ds{\Biggl. + \ \frac{{\cal K}^{10}}{x \, (1-y)^{2}} \ + \ \frac{(1-xy) \, {\cal K}^{11}}{x\, (1-x)
 \, (1-y)^{2}}  \ \Biggr] :=4 \ \al'^{2} \ g_{\te{D}p_{a}}^{3} \
\sum_{i=1}^{11} {H}_{i}^{\rho} \ {\cal K}^{i}\ ,}
\label{2,101}
\eea
where ${\cal I}_{\rho}$ is defined by \req{REG1} and the eleven integrals
${H}_{i}^{\rho}$ refer to a given permutation $\rho$.
Note the Jacobian $\left| \frac{\pa (z_{1},z_{2})}{\pa(x,y)} \right| = x$
which is already included into \req{2,101}.

Each partial amplitude ${\cal M}_{\rho}$ respects gauge invariance.
In fact, checking gauge invariance is a powerful tool to
cast the amplitude \req{2,101} into shorter form.
The basic idea is that pure gauge configurations for
the gluons, $A_{\mu} \sim \pa_{\mu} \chi$, cannot yield any nonzero physical
observables. More precisely, setting $\xi_i = k_i$ for one of the
polarization vectors $i=1,2,3$ makes the amplitude vanish.
One can now go back to the list of kinematic structures in Subsection
\ref{KINEMATICS} and check the behaviour of the ${\cal
 X}^{i}$, ${\cal Y}^{j}$ and ${\cal Z}$ under $\xi_l \to k_l$. Of
course, many previously linearly independent terms then coincide due to
effects like $(\xi_1  k_2) \to (k_1 k_2) =
\frac{s_{1}}{2\al'}$. Each of the ''reduced'' kinematics gives an equation
between its $s_{i}$-dependent coefficients with which they enter the
amplitude. Many of these equations are trivial because they simply tell us
that e.g. ${\cal X}^{3}$ and ${\cal X}^{5}$ have to be included into the same
${\cal K}^{i}$ collection, c.f. Eq.  (\ref{1,11}).
However, many non--trivial identities between the eleven functions
${H}_{i}^{\rho}$ follow this way. They are collected in Appendix \ref{appB}. These relations are almost sufficient to express the eleven integrals in (\ref{2,101}) in terms of a two dimensional basis, we need to read off one additional relation from the polynomials in~(\ref{2,101})
\beq
{H}_{1}^{\rho} = {H}_{2}^{\rho} - {H}_{3}^{\rho}\ .
\label{Addition}
\eeq

\begin{figure}[H]
\centering
\includegraphics[width=0.75\textwidth]{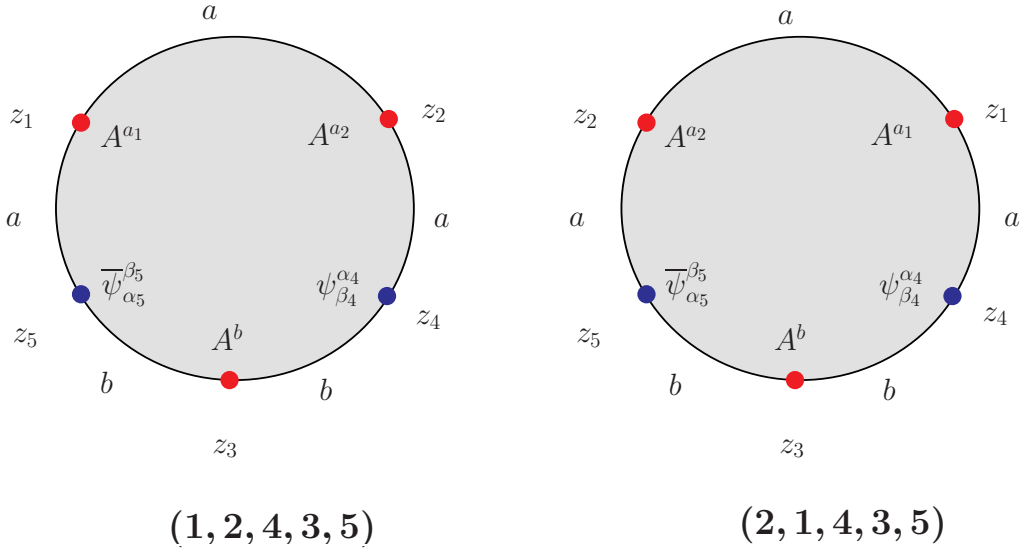}
\caption{Orderings of gauge boson and fermion vertex operators, for case $(ii)$.}
\label{gggqqall1}
\end{figure}

\medskip
\noindent
\underline{Case $(ii)$:}\\
\noindent
For the case $(ii)$ we may choose the third gluon vertex placed
in between the two quarks. Then two orderings are possible, shown in
Figure \ref{gggqqall1}.
\noindent
These orderings enter the subamplitudes with group trace
$\Tr(T^{a_i}T^{a_j}T^{\alpha_4}_{\beta_4}T^{a_3}T^{\beta_5}_{\alpha_5})$ for either
$(i,j) = (1,2)$ or $(i,j) = (2,1)$.
According to the orderings of vertex positions in the case $(ii)$
we fix the remaining two vertex operator positions as
\beq
z_{3} \eq 1 \co z_{4} \eq 0
\label{choice2}
\eeq
while the remaining two variables $z_1$ and $z_2$
are free to be chosen along the boundary of the disk subject
to the two orderings encountered above.
We parametrize the positions $z_{1,2}$ by real numbers $x,y$,
\beq
z_1 \eq x \co z_2 \eq x \, y \ ,
\label{choice2b}
\eeq
similarly to \req{choice1b}. The region ${\cal I}_{\rho}$ of $(x,y)$ integration specifies the two orderings of case $(ii)$:

\beq
\begin{array}{|c|c|rr|} \hline \rho & z_{1},z_{2} & {\cal I}_{\rho} & \\\hline
(1,2,4,3,5) & -\infty<z_1<z_2<0  &-\infty<x<0 &0<y<1 \\
(2,1,4,3,5) & -\infty< z_{2} < z_{1} < 0 & -\infty < x < 0 &1 < y < \infty \\ \hline \end{array}
\label{REG2}
\eeq
\vskip0.5cm
\noindent
The different choices \req{choice2} and \req{choice2b} of vertex operator positions clearly yield a new form for the correlator \req{1,10}. We have two neighbouring gluons $A^{a_{1}},A^{a_{2}}$ originating from the stack $a$ of branes, while the third gluon placed in between  the two fermions is attached to a distinct stack $b$. This is why the coupling of the correlator is modified as $g_{\te{D}p_{a}}^{3} \to g_{\te{D}p_{a}}^{2} g_{\te{D}p_{b}}$. The analogue of \req{2,101} becomes:
\bea
\ds{A(1_\rho,2_\rho,3_\rho,4_\rho,5_\rho)}&=&\ds{4\, \al'^{2} \, g_{\te{D}p_{a}}^{2} \,
g_{\te{D}p_{b}}\ \int \limits_{ {\cal I}_{\rho}} \dd x\  \dd y
\ |x|^{s_{1}-s_{3}-s_{4}} \,  |y|^{s_{5} - s_{2} - s_{3}} \, |1-x|^{s_{4} - s_{1} - s_{2}} \,
|1-y|^{s_{1}} \, |1-xy|^{s_{2}} }\\
&\times&\ds{ \Biggl[ \ - \
\frac{{\cal K}^{1}}{(1-x) \, (1-y)} \ - \ \frac{{\cal K}^{2}}{ (1-y) \, (1-xy) } \ +
\ \frac{x \, {\cal K}^{3}}{(1-x) \, (1-xy)} \ + \ \frac{{\cal K}^{4}}{x \, (1-y)}
\Biggr.}\\
&&\ds{\Biggl. + \ \frac{{\cal K}^{5}}{x\,y \, (1-y)} \  - \ \frac{{\cal K}^{6}}{y \, (1-x)} \ - \ \frac{{\cal
   K}^{7}}{1-xy} \
+ \ \frac{ {\cal K}^{8}}{x \, y}- \ \frac{ {\cal K}^{9}}{ y \, (1-x) \,
 (1-y)} }\\
&&\ds{\Biggl.  + \ \frac{{\cal K}^{10}}{x \, (1-y)^{2}} \ + \ \frac{ {\cal K}^{11}}{x\,y \, (1-y)^{2}}  \
\Biggr] :=4\ \al'^{2} \ g_{\te{D}p_{a}}^{2} \ g_{\te{D}p_{b}} \
\sum_{i=1}^{11} {H}_{i}^{\rho} \ {\cal K}^{i}\ .}
\label{new2,101}
\eea
Again, gauge invariance can be used to find relations between the eleven
functions ${H}_{i}^{\rho}$. In fact, since \req{new2,101} assumes exactly
the same form as \req{2,101} we obtain the same constraints for the functions
${H}_{i}^{\rho}$ as given in Appendix
\ref{appB}. Moreover, in the present case the equation \req{Addition} also
holds.

To summarize in both cases $(i)$ and $(ii)$ we have the same set of 
relations \req{B1}--\req{B3} and Eq. \req{2,101}, which allow to express 
any of the eleven functions ${H}_{i}^{\rho}$ in terms of two linearly independent basis functions.
For any given ordering $\rho$ from the eleven functions ${H}_{i}^{\rho}$ 
we may choose the following pair:
\beq\label{2,8}
h_{1}^{\rho} = {H}^{\rho}_3\ \ \ ,\ \ \ h_{2}^{\rho} = {H}^\rho_7\ .
\eeq
Thus in the  case $(i)$ we work with the following basis of two integrals
\bea
\ds{h^\rho_{1} }&=&\ds{\int \limits_{ {\cal I}_{\rho}} \dd x\  \dd y
\ |x|^{s_{4}-1} \,  |y|^{s_{4} - s_{1} - s_{2}-1} \, |1-x|^{s_{5} - s_{2} - s_{3}} \,
|1-y|^{s_{1}} \, |1-xy|^{s_{2}-s_{4}-s_{5}}\ , }\\
\ds{h^\rho_{2}}&=&\ds{-\int \limits_{ {\cal I}_{\rho}} \dd x\  \dd y
\ \ |x|^{s_{4}} \,  |y|^{s_{4} - s_{1} - s_{2}} \, |1-x|^{s_{5} - s_{2} - s_{3}} \,
|1-y|^{s_{1}} \, |1-xy|^{s_{2}-s_{4}-s_{5}-1}\ ,}
\label{BASIS1}
\eea
while in the  case $(ii)$ we choose the basis:
\bea
\ds{h^\rho_{1} }&=&\ds{\int \limits_{ {\cal I}_{\rho}} \dd x\  \dd y
\ |x|^{s_1-s_3-s_4+1} \,  |y|^{s_5-s_2-s_3} \, |1-x|^{s_4-s_1-s_2-1} \,
|1-y|^{s_{1}} \, |1-xy|^{s_{2}-1}\ ,}\\
\ds{h^\rho_{2}}&=&\ds{-\int \limits_{ {\cal I}_{\rho}} \dd x\  \dd y
\ \ |x|^{s_1-s_3-s_4} \,  |y|^{s_5 - s_2 - s_3} \, |1-x|^{s_4 - s_1 - s_2} \,
|1-y|^{s_{1}} \, |1-xy|^{s_{2}-1}\ .}
\label{BASIS2}
\eea
For a given ordering $\rho$ all the other nine functions ${H}_{i}^{\rho}$ 
can be expressed as linear combination of the basis \req{2,8} as:
\beq
{H}^\rho_{i} \eq C_{1}^{i}  \ h_{1}^{\rho} + \ C_{2}^{i}  \ h_{2}^{\rho}\ \ \ ,\ \ \ 
i=1,\ldots,11\ .
\label{coeff}
\eeq
Eventually, the final result for the generic  $gggq\bar q$ 
subamplitudes \req{2,101} and \req{new2,101} can be cast into
\bea
\ds{A(1_\rho,2_\rho,3_\rho,4_\rho,5_\rho) }&=&\ds{ 4\, 
\al'^{2}\, g_{\te{D}p_{a}}^{3}  \ 
\lf\{\ \left( \sum_{i=1}^{9} C^{i}_{1} \, {\cal K}^{i} \ + \ D_{1}^{1} \, {\cal X}^{1} \ 
+ \ D_{1}^{2} \, {\cal X}^{2} \right) \, h_{1}^{\rho} \ri.}\\[5mm]
&+&\ds{\lf.\left( \sum_{i=1}^{9} C^{i}_{2} \, {\cal K}^{i} \ + \ D_{2}^{1} \, 
{\cal X}^{1} \ + \ D_{2}^{2} \, {\cal X}^{2} \right) \, h_{2}^{\rho}\ \ri\}\ ,}
\label{FINAL}
\eea
for case $(i)$, with the two basis functions \req{BASIS1}.
In the other case $(ii)$ one simply has to perform the replacement 
$g_{\te{D}p_{a}}^{3} \ra g_{\te{D}p_{a}}^{2} g_{\te{D}p_{b}}$ in \req{FINAL}
and use the basis \req{BASIS2}.
In \req{FINAL} the coefficients $C_{1}^{i} $, $C_{2}^{i},\ i=1,\ldots,9$,
which are determined by the equations \req{B1}, \req{B2} and \req{B3}, 
can be read off from \req{2,9}, while
the coefficients $D_i^j,\ i,j=1,2$ are displayed in \req{eff1}.

\subsection[Full $ggg  q\bar q$ amplitude and its relation to $ggggg$]{Full $\bm{ggg  q\bar q}$ amplitude and its relation to $\bm{ggggg}$}
\label{RELbetween}

In order to obtain an explicit form of the full $gggq\bar q$ amplitude and compare it to the well-known $ggggg$ result, we first of all consider the partial amplitude
${\cal M}_{(12345)}(g^+_1,g^-_2, g^-_3, q^-_4, \bar{q}^+_5)$
associated to the canonical ordering $\rho=(1,2,3,4,5)$,
with the Chan-Paton factor
\begin{equation}
t^{\alpha_{4}\beta_{5}a_{1}a_{2}a_{3}}_{\alpha_{5}\beta_{4}} \ \ \equiv \ \ \big( T^{a_{1}}  T^{a_{2}}  T^{a_{3}}\big)^{\alpha_4}_{\alpha_5} \, \delta^{\beta_5}_{\beta_4} \ .
\end{equation}
As with all helicity amplitudes, substantial simplifications can be achieved by appropriate choices of the reference vectors for gluon polarizations. In our case, it is convenient to choose the quark momentum $k_4$ for the positive helicity  gluon $g_1$ and the antiquark momentum $k_5$ for the negative helicity  gluons $g_2$ and $g_3$, respectively.
Thus
\begin{equation}
\xi^{+}_{1 \mu} \eq - \; \frac{\bar u_{\dot\al}(k_{1}) \,
\bar\sigma_{\mu}^{\dot\al\al} \, u_{\al}(k_{4})}{\sqrt{2}\,\langle 14\rangle}
\label{ximu}\end{equation}
and
\begin{equation} \xi^{-}_{2\mu} \eq - \;\frac{u^{\al}(k_{2}) \,
\sigma_{\mu \al\dot\al} \, \bar u^{\dot\al}(k_{5})}{\sqrt{2}\;[25]}~~,
\qquad\quad \xi^{-}_{3\mu} \eq -\; \frac{u^{\al}(k_{3}) \,
\sigma_{\mu \al\dot\al} \, \bar u^{\dot\al}(k_{5})}{\sqrt{2}\;[35]}~~,
\label{ximu1}\end{equation}
where we used the standard notation
\begin{equation}\langle i\, j\rangle \ \ \equiv \ \ u^{\al}(k_{i}) \, u_{\al}(k_{j})\quad ,\quad[i\, j] \ \ \equiv \ \
\bar u_{\dot\al}(k_i) \, \bar u^{\dot\al}(k_j) \ .\end{equation}
{}For this choice of reference vectors, from all ${\cal X}^i$ listed in Subsection 4.1, the only non-vanishing
ones are those with $i=5,10,15,22,29,30$ while all ${\cal Y}^i={\cal Z}=0$.
It follows that the only non-vanishing kinematic factors (\ref{1,11}) are:
\beq 
\big({\cal K}^1, {\cal K}^2, {\cal K}^3, {\cal K}^6, {\cal K}^9\big) =
\frac{[15]^2}{\sqrt{2} \, [25] \, [35]}\ \big(\langle 12\rangle\langle 34\rangle[15]~,\langle 23\rangle\langle 24\rangle[25]~,\langle 23\rangle\langle 34\rangle[35]~,\langle 24\rangle\langle 34\rangle[45]~,\langle 23\rangle\langle 42\rangle[25]\big)~. \label{KKKKK} 
\eeq
Their coefficients $H_{i=1,2,3,6,9}^{(12345)}$ can be expressed in terms
of the functions $f_1$ and $f_2$, introduced in \req{BASIS}, 
which describe five gluon processes (c.f. Appendix \ref{appC}):
\bea
\ds{H_1^{(12345)}}&=&\ds{\lf(-\frac{s_2 s_5}{s_1
   s_3}-\frac{s_2 s_5}{s_1
   s_4}\ri)\, f_1 \ + \ (s_1-s_3-s_4)\,
\lf(\frac{s_2}{s_1 s_4}+\frac{s_5}{s_1
 s_3}-\frac{1}{s_1}\ri)\, f_2 \ ,}\\[5mm]
\ds{H_2^{(12345)}}&=&\ds{-\lf(1+\frac{s_2 s_5}{s_1 s_3}+\frac{s_2 s_5}{s_1
  s_4}+\frac{s_2}{s_3}+\frac{s_5}{s_4}\ri)
\, f_1  }\\
&&\ds{\ \ \ \ \ \ + \ (s_1-s_3-s_4)\, \lf(\frac{s_2}{s_1 s_4}+\frac{s_5}{s_1
  s_3}-\frac{1}{s_1}+\frac{1}{s_3}+\frac{1}{s_4}\ri)\, f_2 \ ,}\\[5mm]
\ds{H_3^{(12345)}}&=&\ds{-\lf(1+\fc{s_2}{s_3}+\fc{s_5}{s_4}\ri)\, f_1 \ + \
(s_1-s_3-s_4)\, \lf(\fc{1}{s_3}+\fc{1}{s_4}\ri)\, f_2\ ,}\\[5mm]
\ds{H_6^{(12345)}}&=&\ds{-\fc{s_2}{s_3}\, f_1 \ + \ \fc{s_1-s_3-s_4}{s_3}\, f_2 \ ,}\\[5mm]
\ds{H_9^{(12345)}}&=&\ds{-\lf(\frac{s_2 s_5}{s_1
   s_3}+\frac{s_2 s_5}{s_1
   s_4}+\frac{s_2}{s_3}\ri)\, f_1+(s_1-s_3-s_4) \, \lf(\frac{s_2}{s_1 s_4}+\frac{s_5}{s_1
   s_3}-\frac{1}{s_1}+\frac{1}{s_3}\ri)\, f_2 \ .}
   \label{H12345}
\eea
After using the ($\rho$ universal) relations
${H}_2^{\rho}={H}_1^{\rho}+{H}_3^{\rho}$ and
${H}_9^{\rho}={H}_1^{\rho}+{H}_6^{\rho}$ for $\rho=(1,2,3,4,5)$ together with Schouten's identity and momentum conservation, Eqs. (\ref{2,99}) and (\ref{2,101}) yield
\bea
\ds{{\cal M}_{(12345)}(g^+_1,g^-_2, g^-_3, q^-_4, \bar{q}^+_5)}&=&\ds{2\sqrt{2} \, \ap^2 \, g^3 \, t^{\alpha_{4}\beta_{5}a_{1}a_{2}a_{3}}_{\alpha_{5}\beta_{4}} \; \frac{[15]^3}{[25] \, [35]}}\\[4mm]
&\times&\ds{  [\langle 12\rangle  \langle 34\rangle ( {H}_1^{(12345)} + {H}_6^{(12345)} ) +  \langle 14\rangle  \langle 23\rangle  ({H}_6^{(12345)}- {H}_3^{(12345)})]\ ,}
\eea
which by using Eqs. (\ref{H12345}) becomes\footnote{Recall that $g\equiv g_{D_{p_a}}$, c.f. Footnote 4.}:
\beq
{\cal M}_{(12345)}(g^+_1,g^-_2, g^-_3, q^-_4, \bar{q}^+_5)=2\sqrt{2} \, g^3 \, t^{\alpha_{4}\beta_{5}a_{1}a_{2}a_{3}}_{\alpha_{5}\beta_{4}} \; \frac{[14] \, [15]^3}{[12][23][34][45][51]}\ \Big(s_2  s_5\  f_1  +  \ap^2\  \langle 12\rangle \, [23] \, \langle 35\rangle [51]\ f_2 \,\Big)\ .
\eeq
Using the definitions (\ref{vfac}), (\ref{pfac}) of the five gluon formfactors $V^{(5)}$ and $P^{(5)}$, the above result can be written in a factorized form as follows\footnote{All the $\ap$ factors are absorbed into the kinematic invariants via
    \[\ap \, \langle i \, j \rangle \, [j \, i] \eq s_{ij}
\co \ap^2 \, \langle i \, j \rangle \, [j \, l] \, \langle l \, m \rangle \, [m \, i] \eq \frac{1}{2} \; \Bigl( s_{ij} \, s_{lm} \ - \ s_{il} \, s_{jm} \ + \ s_{im} \, s_{jl} \ - \ 4i \, \epsilon(i,j,k,l) \Bigr)\ . \]} :
\begin{equation}\label{m5q}
{\cal M}_{(12345)}(g^+_1,g^-_2, g^-_3, q^-_4, \bar{q}^+_5) \eq
[ V^{(5)}(s_j) \ - \ 2i\,
P^{(5)}(s_j)\, \epsilon(1,2,3,4) ]\ {\cal M}_{(12345)}^{\te{QCD}}\!(g^+_1,g^-_2, g^-_3, q^-_4, \bar{q}^+_5)\ . \end{equation}
The QCD amplitude reads
\begin{equation}
{\cal M}_{(12345)}^{\te{QCD}}\!(g^+_1,g^-_2, g^-_3, q^-_4, \bar{q}^+_5) \eq 2\sqrt{2} \, g^3 \, t^{\alpha_{4}\beta_{5}a_{1}a_{2}a_{3}}_{\alpha_{5}\beta_{4}} \; \frac{[14] \, [15]^3}{[12][23][34][45][51]}\ ,
\label{QCDgggqq} \end{equation}
and the prefactor is given by exactly the same function as it appears in the five-gluon amplitude (\ref{m5}). By using the $\cal C$-functions defined in equation (\ref{cfun}),
the amplitude (\ref{m5q}) can be written as:
\begin{equation}\label{ampc}
{\cal M}_{(12345)}(g^+_1,g^-_2, g^-_3, q^-_4, \bar{q}^+_5)\eq 2\sqrt{2} \, g^3 \,
t^{\alpha_{4}\beta_{5}a_{1}a_{2}a_{3}}_{\alpha_{5}\beta_{4}} \, [14] \, [15]^3\, {\cal C}(k_1,k_2,k_3,k_4,k_5)~.\end{equation}
In order to obtain the full amplitude describing a given helicity configuration, we need the amplitudes
associated to all other Chan-Paton factors, with the orderings $\rho=(1_{\rho},2_{\rho},3_{\rho},4_{\rho},5_{\rho})$,
where $\rho$ are the relevant permutations of $(1,2,3,4,5)$ and $i_{\rho}:=\rho(i)$. For all three gluons associated to a single D-brane, these permutations are (1,3,2,4,5), (2,1,3,4,5), (2,3,1,4,5), (3,1,2,4,5) and (3,2,1,4,5).
By explicitly evaluating the amplitudes associated to various orderings of the vertex operators, we find a striking
result that they can be obtained from the original amplitude (\ref{ampc}) by simply permuting the arguments of the function $\cal C$, with the same permutation applied inside the Chan-Paton factor:
\begin{equation}\label{m51q}
{\cal M}_{\rho}(g^+_1,g^-_2, g^-_3, q^-_4, \bar{q}^+_5) \eq 2\sqrt{2} \, g^3 \, [14] \, [15]^3\; {\cal C}(k_{1_{\rho}},k_{2_{\rho}},k_{3_{\rho}},
k_{4_{\rho}},k_{5})\,t^{\alpha_{4}\beta_{5}a_{1_{\rho}}a_{2_{\rho}}
a_{3_{\rho}}}_{\alpha_{5}\beta_{4}}\ . \end{equation}
Hence the full amplitude, obtained by summing all orderings, is given by
\begin{equation}\label{qqfin}
{\cal M}(g^+_1,g^-_2, g^-_3, q^-_4, \bar{q}^+_5) \eq 2\sqrt{2} \ g^3 \  [14] \,[15]^3 \sum_{\rho\in \Pi_q} {\cal C}(k_{1_{\rho}},k_{2_{\rho}},k_{3_{\rho}},
k_{4_{\rho}},k_{5})\,t^{\alpha_{4}\beta_{5}a_{1_{\rho}}a_{2_{\rho}}
a_{3_{\rho}}}_{\alpha_{5}\beta_{4}}\ ,~\end{equation}
where
\begin{equation}\label{pqset}
\Pi_q \ \ \equiv \ \ \Big\{ (1,2,3,4,5),~ (1,3,2,4,5),~(2,1,3,4,5),~ (2,3,1,4,5), ~(3,1,2,4,5),~(3,2,1,4,5)\Big\}\ .
\end{equation}

If one of gauge bosons originates from a different stack, say boson number 3 is from stack $b$, then the Chan-Paton factor becomes
\begin{equation}
t^{\alpha_{4}\beta_{5}a_{1}a_{2}b}_{\alpha_{5}\beta_{4}} \eq \big( T^{a_{1}}  T^{a_{2}} \big)^{\alpha_4}_{\alpha_5} \, (T^b)^{\beta_5}_{\beta_4}
\end{equation}
and the amplitude reads
\begin{equation}\label{qqfinb}
{\cal M}(g^+_1,g^-_2, b^-_3, q^-_4, \bar{q}^+_5) \eq 2\sqrt{2} \, g^2 \, g_{b} \, [14] \, [15]^3\sum_{\rho\in \Pi_q^{\prime}} {\cal C}(k_{1_{\rho}},k_{2_{\rho}},k_{3_{\rho}},
k_{4_{\rho}},k_{5})\,t^{\alpha_{4}\beta_{5}a_{1_{\rho}}a_{2_{\rho}}
b}_{\alpha_{5}\beta_{4}}~,\end{equation}
where
\begin{equation}\label{pqsetb}
\Pi'_q \ \ \equiv \ \ \Big\{ (1,2,4,3,5), ~(2,1,4,3,5)\Big\}\ .
\end{equation}
Here $g_b\equiv g_{D_{p_b}}$. As in the case of five-gluon amplitudes, the dependence of the full amplitude (\ref{qqfin})
on the helicity configuration is limited to the $[14][15]^3$ factor. All other ``mostly minus'' amplitudes, with the single positive helicity gluon labeled by arbitrary $i$ instead of  1, can be obtained from Eq. (\ref{qqfin}) by simply replacing the helicity factor
$[14][15]^3\to [i4][i5]^3$. ``Mostly plus'' amplitudes are obtained by complex conjugation.

The fact that both purely gluonic $ggggg$ and $gggq\bar q$ amplitudes, given in Eqs. (\ref{m5}) and (\ref{m5q}), contain the same string formfactors can
be explained in the following way. At the disk level, the scattering of gluons and gluinos takes place on a single D--brane stack, say $a$. On the other hand, quarks are localized on the intersection of stack $a$ with another one, say $b$, with stack $b$ intersecting stack $a$ at a certain angle $\theta$. One can consider the formal limit
$\theta\to 0$ which puts stack $b$ on top of stack $a$, with quarks appearing now as gauginos of the enhanced group associated to the generators $T^a\oplus\, T^b$. In this limit, the amplitudes involving quarks and gluons fall into the universal class of gluino--gluon amplitudes, see Eq. (\ref{universal}). Taking the limit $\theta\to 0$ requires some care, however it becomes particularly simple if  only one quark-antiquark pair is present because the corresponding amplitudes are $\theta$-independent. This essentially follows from the trivial form of the two-point correlation function of the boundary changing operators, see Eq. (\ref{1,3b}). Thus $gggq\bar q$ amplitudes are related to the universal $ggggg$ amplitudes by a combination of the trivial $\theta\to 0$ limit with spacetime supersymmetry.

It is clear that the above argument extends to  amplitudes involving more gluons or
gauginos, i.e. to the set of amplitudes \req{UNIVERSAL}.
Moreover, also amplitudes involving more than two chiral fermions boil down to
universal gauge amplitudes in the limit $\theta\ra0$. In that case, the amplitudes
do depend on the intersection angle and the limit $\theta\ra0$ has to be taken with
some care. At the more technical level before undertaking this limit
some Poisson resummations have to be performed
on the instanton part such that the amplitudes is rendered finite in the limit
$\theta\ra0$.
E.g. for quarks from the same intersecting stack the four--quark amplitude
becomes the four--gaugino amplitude in the limit $\theta\ra0$
\beq\label{wehave}
{\cal M}(q_1^-,\bar q_2^+,q_3^-,\bar q_4^+)
\ \ \stackrel{\theta\ra0}{\lra}\ \ \
{\cal M}(\chi_1^-,\bar \chi_2^+,\chi_3^-,\bar \chi_4^+)\ ,
\eeq
with the four--gaugino amplitude
${\cal M}(\chi_1^-,\bar \chi_2^+,\chi_3^-,\bar \chi_4^+)=2
g^2\ \vev{13}\ \ket{24}\ B(s,u)$, given in Ref. \cite{MHV}.

To conclude, the kinematics and the non--angle dependent part of a string amplitude
involving an arbitrary number of chiral fermions agrees with the corresponding
amplitude involving the same numbers of gauginos, c.f. also Section \ref{newsec}.

\section{Relations between  $\bm{ggggg}$ and $\bm{gggq\bar q}$ partial amplitudes}
\label{SUPREL}

Although there are {\it a priori} $(N{-}1)!$ partial $N$-gluon tree--level amplitudes
in QCD they satisfy certain linear relations which become particularly useful when
evaluating the squared modulus of the full amplitude. Kleiss--Kuif relations reduce
the number of independent field theory amplitudes to $(N{-}2)!$ \cite{Kleiss,DelDuca},
while
Bern--Carrasco--Johansson relations further reduce their number to $(N{-}3)!$
\cite{Bern}. In string theory, there exists a similar set of relations at the disk level.
It is not surprising that it is hard to further reduce the number of disk amplitudes
because it is known that $(N-3)!$ is the number of independent hypergeometric
functions for $N$--point amplitudes \cite{STi,MHV}.

The string deformation of the QCD relations can be derived by
analytically continuing the real integration of any open string vertex position
to the full complex plane \cite{stieberg,Bohr}.
Analytic continuation of all $N$ vertex positions and successive application of
permutations gives $N!$ relations.
These identities  allow to reduce the number of independent subamplitudes of an
open string $N$--point amplitude to $(N-3)!$. This number is identical to the dimension
of a minimal basis of generalized Gaussian
hypergeometric functions describing the full $N$--point open string amplitude
\cite{STi,MHV}, see also \cite{NMED} for the case $N=5$.

\def\Cc{{\cal C}}
For the concrete five--point case for the partial amplitudes $A$
one has the relation \cite{stieberg}:
\bea\label{DUALv}
\ds{A(1,2,3,4,5)}&+&\ds{e^{2\pi i \ap k_1k_2}\ A(2,1,3,4,5)+
e^{2\pi i\ap(k_4k_5-k_2k_3)}\  A(2,3,1,4,5)}\\
&+&\ds{e^{-2\pi i \ap k_1k_5}\ A(2,3,4,1,5)=0}
\eea
as a result of analytic continuation of the position $z_1$ of the first gluon.
Further relations, in total $120$, follow by applying the same procedure to
the other positions $z_2,\ldots,z_5$ and permutations.

In the case  under consideration, the dependence of $gggq\bar q$ and $ggggg$
amplitudes on the color ordering is limited to the $\cal C$--functions (\ref{cfun}).
Hence with \req{gmhv5}, i.e.
\beq
A(1_\rho,2_\rho,3_\rho,4_\rho,5_\rho)=4\sqrt 2\ g^3\ \ket{15}^4\
{\cal C}(k_{1_\rho},k_{2_\rho},k_{3_\rho},k_{4_\rho},k_{5_\rho})
\eeq
the five gluon relations \req{DUALv}
allow expressing all $\cal C$--factors in terms of a two--element ``basis'' and can be also used to simplify the $gggq\bar q$ amplitudes.
By experimenting with five gluon relations we were able to find a basis which leads to very simple expressions.
It is $\{\Cc(1,3,4,2,5),\ \Cc(2,4,1,3,5)\}$.
In Table 2, we write the components $c_1$ and $c_2$ of the twelve $\Cc$--functions:
\begin {equation}\label{cbasis}
\Cc(1_\rho,2_\rho,3_\rho,4_\rho,5)=
c_1(\rho)\ \Cc(1,3,4,2,5) \ + \ c_2(\rho)\ \Cc(2,4,1,3,5)\ .
\end{equation}
Note that partial amplitudes $\Cc$ are odd under mirror reflections,
$\Cc(1,2,3,4,5)=-\Cc(5,4,3,2,1)$,
therefore there is no need for listing the remaining orderings separately.

\begin{center}
\begin{tabular}{| l | p{6.3cm} | p{7.0cm} |}
    \hline
    $\qquad \rho$ & $\qquad \qquad\qquad ~c_1(\rho)$ & $\qquad\qquad \qquad \qquad c_2(\rho)$ \\ \hline
    $(1,2,3,4)$ & $\begin{matrix}\scriptstyle -\frac{\cot\pi s_1}{\sin\pi s_4}\sin\pi(s_3+s_4)\sin\pi s_5\\
 \scriptstyle - \frac{\cot\pi s_2}{\sin\pi s_4}\sin\pi(s_4+s_5)\sin\pi s_3\\ \scriptstyle
 -\cos\pi(s_3-s_5)\end{matrix}
$ & $\scriptstyle \frac{\sin\pi(s_1+s_2)}{\sin\pi s_1 \sin\pi s_2 \sin\pi s_4}\sin\pi(s_1-s_3-s_4)\sin\pi(s_2-s_4-s_5)$\\ \hline
$(1,3,2,4)$ & $\scriptstyle\frac{1}{\sin\pi s_2\sin\pi s_4}\sin\pi(s_4+s_5)\sin\pi s_3$ & $\scriptstyle -\frac{1}{\sin\pi s_2\sin\pi s_4}\sin\pi(s_1-s_3-s_4)\sin\pi(s_2-s_4-s_5)$\\ \hline
$(2,1,3,4)$ & $\scriptstyle\frac{1}{\sin\pi s_1\sin\pi s_4}\sin\pi(s_3+s_4)\sin\pi s_5$ & $\scriptstyle -\frac{1}{\sin\pi s_1\sin\pi s_4}\sin\pi(s_1-s_3-s_4)\sin\pi(s_2-s_4-s_5)$\\ \hline
$(2,3,1,4)$ & $\scriptstyle\frac{1}{\sin\pi s_2\sin\pi s_4}\sin\pi s_3\sin\pi s_5$ & $\scriptstyle -\frac{1}{\sin\pi s_2\sin\pi s_4}\sin\pi(s_1-s_3-s_4)\sin\pi(s_2-s_5)$\\ \hline
$(3,1,2,4)$ & $\scriptstyle\frac{1}{\sin\pi s_1\sin\pi s_4}\sin\pi s_3\sin\pi s_5$ & $\scriptstyle -\frac{1}{\sin\pi s_1\sin\pi s_4}\sin\pi(s_1-s_3)\sin\pi(s_2-s_4-s_5)$\\ \hline
$(3,2,1,4)$ & $\scriptstyle -\frac{\cot\pi s_2}{\sin\pi s_4}\sin\pi s_3\sin\pi s_5-\frac{\cot\pi s_1}{\sin\pi s_4}\sin\pi s_3\sin\pi s_5$ &$\begin{matrix}\scriptstyle\frac{\cot\pi s_1}{\sin\pi s_4}\sin\pi(s_1-s_3)\sin\pi (s_2-s_4-s_5)\\ \scriptstyle + \frac{\cot\pi s_2}{\sin\pi s_4}\sin\pi(s_2-s_5)\sin\pi (s_1-s_3-s_4)\\ \scriptstyle +\cos\pi(s_1-s_2-s_3+s_5)\end{matrix}
$\\ \hline
$(2,3,4,1)$& $\scriptstyle\frac{1}{\sin\pi s_2}\sin\pi(s_2+s_3)$ & $\scriptstyle\frac{1}{\sin\pi s_2}\sin\pi(s_1-s_3-s_4)$ \\ \hline
$(3,4,1,2)$& $\scriptstyle\frac{1}{\sin\pi s_1}\sin\pi s_5$ & $\scriptstyle\frac{1}{\sin\pi s_1}\sin\pi(s_1+s_2-s_4-s_5)$ \\ \hline
$(1,2,4,3)$& $\scriptstyle\frac{1}{\sin\pi s_1}\sin\pi (s_1+s_5)$ & $\scriptstyle\frac{1}{\sin\pi s_1}\sin\pi(s_2-s_4-s_5)$ \\ \hline
$(1,4,2,3)$& $\scriptstyle\frac{1}{\sin\pi s_2}\sin\pi s_3$ & $\scriptstyle\frac{1}{\sin\pi s_2}\sin\pi(s_1+s_2-s_3-s_4)$ \\ \hline
$(1,3,4,2)$&$1$ & $0$\\ \hline
$(2,4,1,3)$&$0$ & $1$\\ \hline
\end{tabular}
\vskip0.5cm
\begin{flushleft}{\bf Table 2:\ }\it The coefficients of ${\cal C}$--functions expressed in the basis (\ref{cbasis}). Since all twelve permutations leave the fifth position invariant, they are labeled by four numbers.
\end{flushleft}
\end{center}

\vskip0.75cm

\section{One gluon and four fermion amplitudes $\bm{gq\ov qq\ov q}$}
\label{GQQQQ}

The third class of five point--amplitude involves four quarks or leptons
and one gluon
\beq
\vev{V_{A^{a}}^{(0)}(z_{1},\xi_1,k_1) \ V_{\psi^{\al_2}_{\be_2}}^{(-1/2)}(z_{2},u_2,k_2) \
V_{\bar{\psi}^{\be_3}_{\delta_3}}^{(-1/2)}(z_{3},\bar{v}_3 ,k_3) \
V_{\psi^{\delta_4}_{\gamma_4}}^{(-1/2)}(z_{4},u_4,k_4) \
V_{\bar{\psi}^{\gamma_5}_{\al_5}}^{(-1/2)}(z_{5},\bar{v}_5 ,k _{5}) }_{\rho}
\label{4f1}
\eeq
involving the vertex operators introduced in Subsection \ref{VOPs}.
Again, first one has to work out all SCFT correlators and then perform the integration over positions $z_i$ for a given ordering $\rho$ of vertex operators.

\subsection{Computation of the full string amplitude}
\label{sec:TheAmplitudeAtGenericVertexOperatorPositions}

In addition to the contribution of the space--time fields $X,\psi,S$  given in Appendix \ref{appA} there is a four--point function of boundary changing operators $\Xi$ carrying the internal geometric data like intersection angles $\theta^{j}$ of the brane configuration:
\beq
\langle \Xi^{a \cap b}(z_{1}) \, \bar{\Xi}^{b\cap d}(z_{2}) \, \Xi^{d \cap c}(z_{3}) \, \bar{\Xi}^{c\cap a}(z_{4}) \rangle \eq \left( \frac{z_{13} \, z_{24}}{z_{12} \, z_{14} \, z_{23} \, z_{34}} \right)^{3/4} \ I_\rho(\{z_{i}\};\theta^{j}) \ .
\label{4f2}
\eeq
The internal part  of the chiral fermion vertex operators \req{1,2a} and \req{1,2b}
gives rise to non--trivial mapping from the disk world--sheet into the internal target
space. As a result the four--point correlator of these fields \req{4f2} receives
a correction factor $I_\rho(\{z_{i}\};\theta^{j})=I_1(\{z_{i}\};\theta^{j})\ I_{2\rho}(\{z_{i}\};\theta^{j})$ describing disk instantons and the quantum correlator of these fields. The quantum four--point correlator
$I_1(\{z_{i}\};\theta^{j})$ has been already computed in \cite{cvetic1}, while the effect $I_{2\rho}(\{z_{i}\};\theta^{j})$ of world--sheet disk instantons has been  derived in \cite{cvetic2} for intersecting D$6$--branes. Note, that the
additional gluon insertion in \req{4f1} has no effect on the instanton part, since
the gluon vertex operator \req{1,1b}
has vanishing OPE with the internal field $\Xi$.
The factor $I_\rho$ encoding these effects depends on the
specific D--brane setup under consideration and is thoroughly discussed in \cite{LHC}. We shall discuss more details in the next Section, while in this Section it is sufficient
to  give the  value of $I_\rho$ at the relevant
choices of the fermion vertex operator positions $(z_{1},z_{2},z_{3},z_{4})=
(0,x,1,\infty)$:
\beq
I_\rho \bigl( \{0,x,1,\infty\};\theta^{j} \bigr) =I_1(x)\ I_{2\rho}(x):=I_\rho(x)\ .
\label{4f3}
\eeq

Before $PSL(2,\RR)$ gauge fixing, the  correlator \req{4f1} is given by
($\xi_1\equiv\xi,\ k_1\equiv p$)
\begin{align}
&\vev{V_{A^{a}}^{(0)}(z_{1},\xi_1,k_1) \  V_{\psi^{\al_2}_{\be_2}}^{(-1/2)}(z_{2},u_2,k_2) \
V_{\bar{\psi}^{\be_3}_{\delta_3}}^{(-1/2)}(z_{3},\bar{v}_3 ,k_3) \
V_{\psi^{\delta_4}_{\gamma_4}}^{(-1/2)}(z_{4},u_4,k_4) \
V_{\bar{\psi}^{\gamma_5}_{\al_5}}^{(-1/2)}(z_{5},\bar{v}_5 ,k _{5}) }_{\rho}\notag \\
&=g_{D6_a}\ C\ t(\rho) \ \xi_{\mu} \ u_2^{\al} \ \bar{v}^{\dbe}_3\ u^{\ga}_4 \ \bar{v}_5^{\dde}\ \ \langle e^{-\phi(z_{2})/2} \, e^{-\phi(z_{3})/2} \, e^{-\phi(z_{4})/2} \, e^{-\phi(z_{5})/2} \rangle\notag \\
&\times
\vev{\Xi^{a \cap b}(z_{2}) \, \bar{\Xi}^{b\cap c}(z_{3}) \, \Xi^{c \cap d}(z_{4}) \, \bar{\Xi}^{d\cap a}(z_{5})}\  \biggl\{\ i \, \Bigl\langle  \pa X^{\mu}(z_{1}) \, \prod_{i=1}^{5} e^{ik_i   X(z_{i})} \Bigr\rangle \, \langle S_{\al}(z_{2}) \, S_{\dbe}(z_{3}) \, S_{\ga}(z_{4}) \, S_{\dde}(z_{5}) \rangle \ \biggr. \notag \\
&\hskip5cm
\biggl. - \ 2\ap \ p_{\nu} \ \Bigl\langle  \prod_{i=1}^{5} e^{ik_i   X(z_{i})} \Bigr\rangle \, \langle \psi^{\mu}(z_{1}) \, \psi^{\nu}(z_{1}) \, S_{\al}(z_{2}) \, S_{\dbe}(z_{3}) \, S_{\ga}(z_{4}) \, S_{\dde}(z_{5}) \rangle\ \biggr\} \notag \\
&=\tilde C \ t(\rho) \  u_2^{\al} \, \bar{v}_3^{\dbe}  \ u^{\ga}_4 \ \bar{v}_5^{\dde}  \ \frac{(z_{24} \ z_{35})^{1/2}}{z_{23} \ z_{25} \ z_{34} \ z_{45}}\
I_\rho(\{z_{i}\},\theta^{j})\  \prod_{i,j=1 \atop {i<j}}^{5} \bigl| z_{ij} \bigr|^{2\al' k_i   k_j}\ \left\{\ \left(\ \frac{\xi   k_2}{z_{12}}+ \frac{\xi   k_3}{z_{13}}+ \frac{\xi   k_4}{z_{14}} +
\frac{\xi   k_5}{z_{15}} \ \right) \right. \notag \\
&\times\frac{\vep_{\al \ga} \ \vep_{\dbe \dde}}{(z_{24} \, z_{35})^{1/2}} +\left. \ \frac{\xi_{\mu} \, p_{\nu}}{2 \, (z_{24} \, z_{35})^{1/2} } \; \left(\ - \frac{z_{25} \, \si^{\mu}_{\al \dbe} \, \si^{\nu}_{\ga \dde}}{z_{12} \, z_{15}} \ + \ \frac{z_{23} \, \si^{\mu}_{\al \dde} \, \si^{\nu}_{\ga \dbe}}{z_{12} \, z_{13}} \ + \ \frac{z_{34} \, \si^{\mu}_{\ga \dde} \, \si^{\nu}_{\al \dbe}}{z_{13} \, z_{14}} \ + \ \frac{z_{45} \, \si^{\mu}_{\ga \dbe} \, \si^{\nu}_{\al \dde}}{z_{14} \, z_{15}}\  \right)\  \right\} \ ,
\label{4f4}
\end{align}
with the normalizations:
\beq\label{norm}
C=2\pi\ap\ e^{\phi_{10}}\ \ \ ,\ \ \ \tilde C=2\ap\ g_{D6_a}\, C=\pi\ (2\ap)^2\ e^{\phi_{10}}\ g_{D6_a}\ .
\eeq
Note, that the normalization $C$ is fixed by
the four--fermion string amplitude \cite{LHC}, while  the additional
factor of $g_{D6_a}$ stems from the normalization of the gauge vertex \req{1,1b}.
The Chan-Paton group trace factors $t(\rho)$ are
specified in the next Subsection.

The next step is to integrate over the positions $z_i$ of  vertex operators.
The $PSL(2,\RR)$ invariance of the disk world--sheet allows to fix three
of them to arbitrary real values.
This symmetry does not change the cyclic ordering of the vertex positions.
As a consequence we must consider two inequivalent fixings to take into account all possible orderings.
Our choice of variables is the following:
\beq
z_{1}= x\ y\ \ \ ,\ \ \ z_{2} =0\ \ \ ,\ \ \ z_3=x\ \ \ ,\ \ \ z_4=1\ \ \ ,\ \ \
z_{5}=\infty \ .
\label{4f5a}
\eeq
This choice gives rise to the set of orderings $\rho_1$ of the five
vertex positions $z_i$ along the boundary, with $z_3<z_5$.
The other set of orderings $\rho_2$ with  $z_3>z_5$
is obtained by e.g. permuting the labels $3$ and $5$ in the final expressions,
i.e. by choosing:
\beq
z_{1}= x\ y\ \ \ ,\ \ \ z_{2} =0\ \ \ ,\ \ \ z_3=\infty\ \ \ ,\ \ \ z_4=1\ \ \ ,\ \ \
z_{5}=x \ .
\label{4f5b}
\eeq
Hence, in the following we may focus on the choice \req{4f5a} and its orderings
$\rho_1$.

In the following we have to consider three different cases referring
to the helicity configuration of the four fermions.
So far, we have assumed a special helicity configuration, namely the pair of fermions 
 $\psi^{\al}_{\be}$, $\psi_{\ga}^{\de}$  being left handed and the pair
$\bar{\psi}^{\be}_{\delta}$, $\bar{\psi}^{\gamma}_{\al}$  being right handed -- in shorthand $(24)_{L}(35)_{R}$.
Combining Ramond spin fields as $S_{\al} \otimes \bar{\Xi}$ and
$S_{\dbe} \otimes \Xi$ for some of the quarks gives rise to helicities
$(25)_L(34)_R$ and $(23)_L(45)_R$.
These amplitudes are very similar to the $(24)_{L}(35)_{R}$ case apart from a minor modification in the spacetime spin field correlators (\ref{4fA1}) and (\ref{4fA2}).

Although our amplitude results can be used for both quarks and leptons in the following we shall use the notation $q,\bar q$ instead of
$\psi^\al_\bet,\bar \psi^\bet_\al$, respectively.

\ \\
\noindent
\underline{Helicity configuration $(24)_L(35)_R$,\ $(35)_L(24)_R$}

\ \\ \noindent
For the helicity configurations $(24)_L(35)_R$ or $(35)_L(24)_R$ and the choice \req{4f5a} the integration of the amplitude \req{4f4} yields the expression
$$\ba{lcl}
\ds{{\cal M}_{\rho_1}}&:=&\ds{{\cal M}_{\rho_1}(g_1,q_2^-,\bar q_3^+,q_4^-,\bar q_5^+)={V}_{\te{CKG}}^{-1}\
\lf(\ \int\limits_{{\cal I}_{\rho_1}}\prod_{k=1}^{5}\dd z_k\ \ri)}\\[5mm]
&\times&\ds{\vev{V_{A^{a}}^{(0)}(z_{1},\xi,p) \ V_{\psi^{\al_2}_{\be_2}}^{(-1/2)}(z_{2},u_2,k_2) \
V_{\bar{\psi}^{\be_3}_{\delta_3}}^{(-1/2)}(z_{3},\bar{v}_3 ,k_3) \
V_{\psi^{\delta_4}_{\gamma_4}}^{(-1/2)}(z_{4},u_4,k_4) \
V_{\bar{\psi}^{\gamma_5}_{\al_5}}^{(-1/2)}(z_{5},\bar{v}_5 ,k _{5}) }_{\rho}}
\ea$$
\bea
&=&\ds{\tilde C\  t(\rho_1) \, \int \limits _{{\cal I}_{\rho_1}} dx\ dy\
I_{\rho_1}(x)\
\bigl| x \bigr|^{s_{4}-1} \, \bigl| y \bigr|^{s_{1}} \, \bigl| 1- x \bigr|^{s_{3}-1} \, \bigl| 1-y \bigr|^{s_{4}-s_{1}-s_{2}} \, \bigl| 1-xy \bigr|^{s_{2}-s_{4}-s_{5}} }\\
&\times&\ds{\left\{ -\frac{{\cal P}_{1}}{y} \ + \ \frac{{\cal P}_{2}}{1-y} \ + \ \frac{x \, {\cal P}_{3}}{1-xy} \ - \ \frac{{\cal P}_{4}}{y} \ + \ \frac{{\cal P}_{5}}{y \, (1-y)} \ - \ \frac{(1-x) \, {\cal P}_{6}}{(1-y) \, (1-xy)} \ - \ \frac{x \, {\cal P}_{7}}{1-xy} \right\}}\\
&=&\ds{\tilde C\ t(\rho_1)\ \sum_{i=1}^7 G_i^{\rho_1} \ {\cal P}_i\ ,}
\label{4f6}
\eea
where  the seven integrals
$G_i^{\rho_1}$ refer to a given permutation $\rho_1$. The integration
region ${\cal I}_{\rho_1}$ referring to the fixing \req{4f5a} will be specified in the next Subsection.
The Jacobian $x = \left| \det \frac{\pa (z_{1},z_{3})}{\pa(x,y)} \right|$ is already included in \req{4f6}.
We have introduced shorthands for the momentum- and spinor products involved in (\ref{4f4}),
\beq
{\cal P}_{1}=\bigl( u_2 \, u_4 \bigr) \, \bigl( \bar{v}_3 \, \bar{v}_5\bigr) \,
\xi   k_{2}\ \ \ ,\ \ \
{\cal P}_{2}=\bigl( u_2 \, u_4 \bigr) \, \bigl( \bar{v}_3 \, \bar{v}_5\bigr) \,
\xi   k_{3}\ \ \ ,\ \ \
{\cal P}_{3}=\bigl( u_2 \, u_4 \bigr) \, \bigl( \bar{v}_3 \, \bar{v}_5 \bigr) \,
\xi   k_{4}\ ,
\label{calp}
\eeq
as well as further ones arising from the six point correlator (\ref{4fA2}):
\begin{alignat}{3}\notag
{\cal P}_{4} \ \ &= \ \ \tfrac{1}{2} \, \xi_{\mu} \, \bigl( u_2 \, \si^{\mu} \, \bar{v}_3 \bigr) \, p_{\nu} \, \bigl( u_4 \, \si^{\nu} \, \bar{v}_5\bigr) \ , \ \ \ \ \ &{\cal P}_{6} \ \ &= \ \ \tfrac{1}{2} \, \xi_{\mu} \, \bigl( u_4 \, \si^{\mu} \, \bar{v}_5 \bigr) \, p_{\nu} \, \bigl( u_2 \, \si^{\nu} \, \bar{v}_3 \bigr)\ , \\
{\cal P}_{5} \ \ &= \ \ \tfrac{1}{2} \, \xi_{\mu} \, \bigl( u_2 \, \si^{\mu} \, \bar{v}_5 \bigr) \, p_{\nu} \, \bigl( u_4 \, \si^{\nu} \, \bar{v}_3 \bigr) \ , \ \ \ \ \ &{\cal P}_{7} \ \ &= \ \ \tfrac{1}{2} \, \xi_{\mu} \, \bigl( u_4 \, \si^{\mu} \, \bar{v}_3 \bigr) \, p_{\nu} \, \bigl( u_2 \, \si^{\nu} \, \bar{v}_5 \bigr)\ .\label{calp2}
\end{alignat}
\noindent

\ \\ \noindent
\underline{Helicity configuration $(25)_L(34)_R$,\ $(34)_L(25)_R$}

\ \\ \noindent
For the helicity configuration $(25)_L(34)_R$,\ $(34)_L(25)_R$ and the choice \req{4f5a} we get
\bea
\ds{{\cal M}'_{\rho_1}}&:=&
\ds{{\cal M}'_{\rho_1}(g_1,q_2^-,\bar q_3^+,q_4^+,\bar q_5^-)} \\[3mm]
&=&\ds{ \tilde C\ t(\rho_1) \, \int\limits_{{\cal I}_{\rho_1}} dx\ dy \
I_{\rho_1}(x) \bigl| x \bigr|^{s_{4}-1} \, \bigl| y \bigr|^{s_{1}} \, \bigl| 1- x \bigr|^{s_{3}-3/2} \, \bigl| 1-y \bigr|^{s_{4}-s_{1}-s_{2}} \, \bigl| 1-xy \bigr|^{s_{2}-s_{4}-s_{5}}} \\
&\times&\ds{\left\{ -\frac{{\cal Q}_{1}}{y} \ + \ \frac{{\cal Q}_{2}}{1-y} \ + \ \frac{x \, {\cal Q}_{3}}{1-xy} \ - \ \frac{{\cal Q}_{4}}{y \, (1-xy)} \ + \ \frac{{\cal Q}_{5}}{y \, (1-y)} \ - \ \frac{ {\cal Q}_{6}}{1-y} \ +
\ \frac{x \, {\cal Q}_{7}}{1-xy} \right\}}\\
&=&\ds{\tilde C\ t(\rho_1)\ \sum_{i=1}^7 \gp{i} \ {\cal Q}_i\ ,}
\label{4f7a}
\eea
with:
\bea
{\cal Q}_{1}&=&\ds{\bigl( u_2 \, u_5 \bigr) \, \bigl( \bar{v}_3 \, \bar{v}_4 \bigr) \, \xi   k_{2}\ \ \ ,\ \ \
{\cal Q}_{2}=\bigl( u_2 \, u_5 \bigr) \, \bigl( \bar{v}_3 \, \bar{v}_4 \bigr) \,
\xi   k_{3}\ \ \ ,\ \ \
{\cal Q}_{3}=\bigl( u_2 \, u_5 \bigr) \, \bigl( \bar{v}_3 \, \bar{v}_4 \bigr) \,
\xi   k_{4} \ ,}\\[2mm]
{\cal Q}_{4}&=&\ds{\tfrac{1}{2} \, \xi_{\mu} \, \bigl( u_2 \, \si^{\mu} \, \bar{v}_3 \bigr) \, p_{\nu} \, \bigl( u_5 \, \si^{\nu} \, \bar{v}_4 \bigr) \ , \ \ \ \ \
{\cal Q}_{6}=\tfrac{1}{2} \, \xi_{\mu} \, \bigl( u_5 \, \si^{\mu} \, \bar{v}_4 \bigr) \, p_{\nu} \, \bigl( u_2 \, \si^{\nu} \, \bar{v}_3 \bigr)\ ,}\\[2mm]
{\cal Q}_{5}&=&\ds{\tfrac{1}{2} \, \xi_{\mu} \, \bigl( u_2 \, \si^{\mu} \, \bar{v}_4 \bigr) \, p_{\nu} \, \bigl( u_5 \, \si^{\nu} \, \bar{v}_3 \bigr) \ , \ \ \ \ \
{\cal Q}_{7}=\tfrac{1}{2} \, \xi_{\mu} \, \bigl( u_5 \, \si^{\mu} \, \bar{v}_3 \bigr) \, p_{\nu} \, \bigl( u_2 \, \si^{\nu} \, \bar{v}_4 \bigr) \ .}\label{calq}\eea
%%%%%%%%%%%%%%%%%%%%%%%%%%%%%%
 \goodbreak

\ \\ \noindent
\underline{Helicity configuration $(23)_L(45)_R$,\ $(45)_L(23)_R$}

\ \\ \noindent
For the helicity configuration $(23)_L(45)_R$,\ $(45)_L(23)_R$ and the choice
\req{4f5a} we obtain
$$\ba{lcl}
\ds{{\cal M}''_{\rho_1}}&:=&\ds{{\cal M}''_{\rho_1}(g_1,q_2^-,\bar q_3^-,q_4^+,\bar q_5^+)}\\[3mm]
&=&\ds{\tilde C\ t(\rho_1) \, \int\limits_{{\cal I}_{\rho_1}} dx\ dy\ I_{\rho_1}(x) \ \bigl| x \bigr|^{s_{4}-3/2} \, \bigl| y \bigr|^{s_{1}} \,
\bigl| 1- x \bigr|^{s_{3}-1} \, \bigl| 1-y \bigr|^{s_{4}-s_{1}-s_{2}} \, \bigl| 1-xy \bigr|^{s_{2}-s_{4}-s_{5}} }\\
&\times&\ds{\left\{ -\frac{{\cal R}_{1}}{y} \ + \ \frac{{\cal R}_{2}}{1-y} \ + \ \frac{x \, {\cal R}_{3}}{1-xy} \ - \ \frac{{\cal R}_{4}}{y} \ + \ \frac{{\cal R}_{5}}{y \, (1-xy)} \ + \ \frac{(1-x) \, {\cal R}_{6}}{(1-y) \, (1-xy)} \ - \
\frac{ {\cal R}_{7}}{1-y}\right\}}
\ea$$
\bea
&=&\ds{\tilde C\  t(\rho_1)\ \sum_{i=1}^7 \gpp{i} \ {\cal R}_i\ ,}
\label{4f7b}
\eea
with:
\bea
{\cal R}_{1}&=&\ds{\bigl( u_2 \, u_3 \bigr) \, \bigl( \bar{v}_4 \, \bar{v}_5 \bigr) \,
\xi   k_{2}\ \ \ ,\ \ \
{\cal R}_{2}=\bigl( u_2 \, u_3 \bigr) \, \bigl( \bar{v}_4 \, \bar{v}_5 \bigr) \,
\xi   k_{3}\ \ \ ,\ \ \
{\cal R}_{3}= \bigl( u_2 \, u_3 \bigr) \, \bigl( \bar{v}_4 \, \bar{v}_5 \bigr) \,
\xi   k_{4}\ ,}\\[2mm]
\ds{{\cal R}_{4}}&=&\ds{\tfrac{1}{2} \, \xi_{\mu} \, \bigl( u_2 \, \si^{\mu} \, \bar{v}_4 \bigr) \, p_{\nu} \, \bigl( u_3 \, \si^{\nu} \, \bar{v}_5 \bigr)\ , \ \ \ \ \
{\cal R}_{6}=\tfrac{1}{2} \, \xi_{\mu} \, \bigl( u_3 \, \si^{\mu} \, \bar{v}_5 \bigr) \, p_{\nu} \, \bigl( u_2 \, \si^{\nu} \, \bar{v}_4 \bigr) \ ,}\\[2mm]
{\cal R}_{5}&=&\ds{\tfrac{1}{2} \, \xi_{\mu} \, \bigl( u_2 \, \si^{\mu} \, \bar{v}_5 \bigr) \, p_{\nu} \, \bigl( u_3 \, \si^{\nu} \, \bar{v}_4 \bigr) \ , \ \ \ \ \
{\cal R}_{7}= \tfrac{1}{2} \, \xi_{\mu} \, \bigl( u_3 \, \si^{\mu} \, \bar{v}_4 \bigr) \, p_{\nu} \, \bigl( u_2 \, \si^{\nu} \, \bar{v}_5 \bigr)\ .}
\label{calr}
\eea
Note that the polynomials associated with ${\cal Q}_{4,...,7}$, ${\cal R}_{4,...,7}$ slightly differ from the ${\cal P}_{4,...,7}$ counterparts, and -- what we also observed for the gluonless four quark amplitude -- the exponents of $x$ and $(1-x)$ receive further modification by $1/2$.

Next, we cast the expressions \req{4f6}, \req{4f7a} and \req{4f7b} into
shorter form. As it is known from the other two classes of five--point amplitudes
$ggggg$ and $gggq\bar q$
a two--dimensional basis of functions should be sufficient to specify the full amplitude for each of the three helicity configurations $(i), (ii)$ and $(iii)$.
This is achieved by deriving some functional relations, which hold among each set of
the seven functions and allow for a two--dimensional solution.
One of these relations can be derived by studying gauge invariance of the
partial amplitudes \req{4f6}, \req{4f7a} and \req{4f7b}.
The latter have to stay
invariant under a gauge transformation of the gluon polarization $\xi \to \xi + p$, in particular ${\cal M} \bigl. \bigr|_{\xi = p}$ has to vanish. For $\xi= p$, the three sets of kinematics ${\cal P}_{i}$, ${\cal Q}_{i}$ and ${\cal R}_{i}$,
reduce to two linearly independent expressions. 

Let us demonstrate this for the case $(i)$.
Upon the substitution $\xi = p$, the kinematic factors ${\cal P}_{1,2,3}$ simply become $ \sim (u_2 u_4 ) ( \bar{v}_3 \, \bar{v}_5 )   (p   k_{2},p   k_{3}, p   k_4)$ whereas the remaining ${\cal P}_{A,...,D} \bigl. \bigr|_{\xi = p}$ are connected by a $\si$ matrix identity,
\begin{align}
\eta^{\mu \nu} \, &\vep_{\al \ga} \, \vep_{\dbe \dde} \eq \si^{(\mu}_{\al \dde} \, \si^{\nu)}_{\ga \dbe} \ - \ \si^{(\mu}_{\al \dbe} \, \si^{\nu)}_{\ga \dde} \co \eta^{\mu \nu} \, p_{\mu} \, p_{\nu} \eq 0 \notag \\
 &\Longrightarrow \ \ \ p_{\mu} \, \bigl( u_2 \, \si^{\mu} \, \bar{v}_3 \bigr) \, p_{\nu} \, \bigl( u_4 \, \si^{\nu} \, \bar{v}_5 \bigr) \eq p_{\mu} \, \bigl( u_2 \, \si^{\mu} \, \bar{v}_5 \bigr) \, p_{\nu} \, \bigl( u_4 \, \si^{\nu} \, \bar{v}_3 \bigr) \ .
\label{4f8}
\end{align}
We therefore find:
\begin{align}
0 \ \ &\stackrel{!}{=} \ \ {\cal M}_{\rho_1}\Bigl. \Bigr|_{\xi = p}
\sim \ \ \int_{{\cal I}_{\rho_1}} dx\ dy\ I_{\rho_1}(x)\ \bigl| x \bigr|^{s_{4}-1} \, \bigl| y \bigr|^{s_{1}} \, \bigl| 1- x \bigr|^{s_{3}-1} \, \bigl| 1-y \bigr|^{s_{4}-s_{1}-s_{2}} \, \bigl| 1-xy \bigr|^{s_{2}-s_{4}-s_{5}} \notag \\
&\times \ \biggl[\  \bigl( u_2 \, u_4 \bigr) \,
\bigl( \bar{v}_3 \, \bar{v}_5 \bigr) \;
\lf\{\ -\frac{s_1}{y} \ + \ \frac{s_4 - s_1 - s_2}{1-y} \ + \
\frac{x \, (s_2 - s_4 - s_5)}{1-xy}\  \ri\}\notag \\
&+ \ p_{\mu} \, \bigl( u_2 \, \si^{\mu} \, \bar{v}_3 \bigr) \, p_{\nu} \,
\bigl( u_4 \, \si^{\nu} \, \bar{v}_5 \bigr) \; \biggl( \underbrace{-\frac{1}{y} \ +
\ \frac{1}{y \, (1-y)} \ - \ \frac{1-x}{(1-y) \, (1-xy)} \ - \
\frac{x}{1-xy}}_{= \ 0} \biggr)\ \biggr] \notag \\
&= -( u_2 \, u_4 ) \, ( \bar{v}_3 \, \bar{v}_5 ) \
\int_{{\cal I}_{\rho_1}} dx\ dy\  I_{\rho_1}(x)\ \bigl| x \bigr|^{s_{4}-1} \,
\bigl| 1- x \bigr|^{s_{3}-1} \ \frac{d}{d y} \; \lf\{\ \bigl| y \bigr|^{s_{1}} \,
\bigl| 1-y \bigr|^{s_{4}-s_{1}-s_{2}} \, \bigl| 1-xy \bigr|^{s_{2}-s_{4}-s_{5}} \ \ri\}.
\label{4f9}
\end{align}
Since the boundary term
$| y |^{s_{1}} | 1-y|^{s_{4}-s_{1}-s_{2}} | 1-xy |^{s_{2}-s_{4}-s_{5}} \bigl. \bigr|^{0}_{-\infty}$ vanishes, we have confirmed\footnote{Exactly the same mechanism applies to the remaining helicities -- the coinciding ${\cal Q}_{A,...,D} \bigl. \bigr|_{\xi = p}$ and ${\cal R}_{A,...,D} \bigl. \bigr|_{\xi = p}$ are multiplied by identically vanishing terms $0=-\frac{1}{y  (1-xy)} + \frac{1}{y  (1-y)}- \frac{1}{1-y} + \frac{x}{1-xy}$ or $0=-\frac{1}{y} + \frac{1}{y  (1-xy)}+ \frac{1-x}{(1-y)(1-xy)} - \frac{1}{1-y}$, and the bispinors $(u_2  u_5 )( \bar{v}_2  \bar{v}_3 )$, $( u_2 u_3 )( \bar{v}_4 \bar{v}_5)$ come along with a total derivative $\frac{\dd}{\dd y} \bigl( | y |^{s_{1}} | 1-y|^{s_{4}-s_{1}-s_{2}} | 1-xy |^{s_{2}-s_{4}-s_{5}} \bigr)$.} gauge invariance from (\ref{4f9}) and obtained   the identity
\beq\label{wefound}
s_1\ G^{\rho_1}_1+(s_4-s_1-s_2)\ G_2^{\rho_1}+(s_2-s_4-s_5)\ G^{\rho_1}_3\ ,
\eeq
which holds in all three cases $(i), (ii)$ and $(iii)$. However for
each configuration $(i), (ii)$ and $(iii)$ there hold various additional
relations between their set of  functions $G_i^\rho, \gp{i}$ and $\gpp{i}$.
For the set $G_i^\rho$ relevant to case $(i)$ we find
\beq\label{relations1}
G^{\rho_1}_4=G^{\rho_1}_1\ \ \ ,\ \ \ G^{\rho_1}_5=-G^{\rho_1}_1+G^{\rho_1}_2\ \ \ ,\ \ \ G^{\rho_1}_6=G^{\rho_1}_3-G^{\rho_1}_2\ \ \ ,\ \ \ G^{\rho_1}_7=-G^{\rho_1}_3\ ,
\eeq
while for the set $\gp{i}$ relevant to case $(ii)$
\beq\label{relations2}
\gp{4}=\gp{1}-\gp{3}\ \ \ ,\ \ \ \gp{5}=-\gp{1}+\gp{2}\ \ \ ,\ \ \
\gp{6}=-\gp{2}\ \ \ ,\ \ \ \gp{7}=\gp{3}\ ,
\eeq
and for the set $\gpp{i}$ relevant to case $(iii)$:
\beq\label{relations3}
\gpp{4}=\gpp{1} \ \ ,\ \ \ \gpp{5}=-\gpp{1}+\gpp{3}
\ \ \ ,\ \ \ \gpp{6}=\gpp{2}-\gpp{3}\ \ \ ,\ \ \ \gpp{7}=-\gpp{2}\ .
\eeq
These relations allow to reduce for each configuration the set of  functions
$G_i^{\rho_1}, \gp{i}$ and $\gpp{i}$ to a minimal basis of two functions.
Note, that these relations hold for any given ordering $\rho_1$.

In the following we choose the two functions
\bea\label{newqq4}
&&\ds{G^{\rho_1}_1\lf[{n_1\atop n_2}\ri]=-\int\limits_{{\cal I}_{\rho_1}} dx\ dy\ I_{\rho_1}(x)\ y^{-1}\
\bigl| x \bigr|^{s_{45}-1-\fc{n_1}{2}} \, \bigl| y \bigr|^{s_{12}} \,
\bigl| 1- x \bigr|^{s_{34}-1-\fc{n_2}{2}} \, \bigl| 1-y \bigr|^{s_{13}} \, \bigl| 1-xy \bigr|^{s_{14}}\ ,}\\
&&\ds{G^{\rho_1}_2\lf[{n_1\atop n_2}\ri]=\int\limits_{{\cal I}_{\rho_1}} dx\ dy\ I_{\rho_1}(x)\  (1-y)^{-1}\
\bigl| x \bigr|^{s_{45}-1-\fc{n_1}{2}} \, \bigl| y \bigr|^{s_{12}} \,
\bigl| 1- x \bigr|^{s_{34}-1-\fc{n_2}{2}} \, \bigl| 1-y \bigr|^{s_{13}} \, \bigl| 1-xy \bigr|^{s_{14}}\ ,}
\eea
with $G^{\rho_1}_i=G^{\rho_1}_i\lf[{0\atop 0}\ri],\ \gp{i}=G^{\rho_1}_i\lf[{0\atop
  1}\ri],\ \gpp{i}=G^{\rho_1}_i\lf[{1\atop 0}\ri],\ i=1,2$ as basis
for the cases $(i)$, $(ii)$, and  $(iii)$, respectively.
With this set of  basis the amplitudes \req{4f6}, \req{4f7a}
and \req{4f7b} can be cast into the shorter form
$$\ba{lcl}
\ds{{\cal M}_{\rho_1}(g_1,q_2^-,\bar q_3^+,q_4^-,\bar q_5^+)}&=&\ds{\tilde C\ t(\rho_1)\
\lf\{\ \lf[\ \Pc_1+\Pc_4-\Pc_5-\fc{s_{12}}{s_{14}}\ (\Pc_3+\Pc_6-\Pc_7)\
 \ri]\ G_1^{\rho_1}\ri.}\\[4mm]
&+&\ds{\lf.\lf[\ \Pc_2+\Pc_3+\Pc_5-\Pc_7+\fc{s_{12}+s_{51}}{s_{14}}\
(\Pc_3+\Pc_6-\Pc_7) \ \ri]\ G_2^{\rho_1}\ \ri\}\ ,}\\[7mm]
\ds{{\cal M}'_{\rho_1}(g_1,q_2^-,\bar q_3^+,q_4^+,\bar q_5^-)}&=&\ds{\tilde C\ t(\rho_1)\
\lf\{\ \lf[\ \Qc_1+\Qc_4-\Qc_5-\fc{s_{12}}{s_{14}}\ (\Qc_3-\Qc_4+\Qc_7)\
\ri]\ \gp{1}\ri.}\\[4mm]
&+&\ds{\lf.\lf[\ \Qc_2+\Qc_5-\Qc_6-\fc{s_{13}}{s_{14}}\ (\Qc_3-\Qc_4+\Qc_7)\ \ri]\ \gp{2}\ \ri\}\ ,}
\ea$$
\bea\label{FINALgqqqq}
\ds{{\cal M}''_{\rho_1}(g_1,q_2^-,\bar q_3^-,q_4^+,\bar q_5^+)}&=&\ds{\tilde C\ t(\rho_1)\
\lf\{\ \lf[\ \Rc_1+\Rc_4-\Rc_5-\fc{s_{12}}{s_{14}}\ (\Rc_3+\Rc_5-\Rc_6)\ \ri]\ \gpp{1}\ri.}\\[4mm]
&+&\ds{\lf.\lf[\ \Rc_2+\Rc_6-\Rc_7-\fc{s_{13}}{s_{14}}\ (\Rc_3+\Rc_5-\Rc_6)\ \ri]\ \gpp{2}\ \ri\}\ ,}
\eea
with the normalization $\tilde C$ given in \req{norm}.

To derive the amplitudes for the orderings $\rho_2$ related to the choice
\req{4f5b} we could repeat all the previous steps,
which lead to the result \req{FINALgqqqq}, i.e. begin at \req{4f4}
supplemented by the insertions \req{4f5b}.
However, as already pointed out after Eq. \req{4f5b}, we may obtain these amplitudes
simply by permuting the labels $3$ and $5$.
More precisely to each ordering $\rho_2$ with $z_5<z_3$ we may associate after permuting $3$ and $5$ one ordering $\rho_1$ with $z_3<z_5$.
In addition, we have to multiply all the expressions by an overall minus sign
to take into account the odd signature of the permutation of the fermions $3$ and $5$. The net effect on the kinematics is to perform the following replacement in $\Mc_{\rho_1}$: $\Pc_1\ra\Pc_1,\ \Pc_2\ra-\Pc_1-\Pc_2-\Pc_3,\ \Pc_3\ra\Pc_3,\ \Pc_4\ra-\Pc_5,\ \Pc_5\ra-\Pc_4,\ \Pc_6\ra-\Pc_7,\ \Pc_7\ra-\Pc_6$. On the other hand, for the amplitude $\Mc'_{\rho_2}$ we start at $\Mc''_{\rho_1}$ and perform the replacement: $\Rc_1\ra\Qc_1,\ \Rc_2\ra-\Qc_1-\Qc_2-\Qc_3,\ \Rc_3\ra\Qc_3,\ \Rc_4\ra-\Qc_5,\ \Rc_5\ra-\Qc_4,\ \Rc_6\ra-\Qc_7,\ \Rc_7\ra-\Qc_6$, while for the amplitude $\Mc''_{\rho_2}$ we start at $\Mc'_{\rho_1}$ and perform the following replacement:
$\Qc_1\ra\Rc_1,\ \Qc_2\ra-\Rc_1-\Rc_2-\Rc_3,\ \Qc_3\ra\Rc_3,\ \Qc_4\ra-\Rc_5,\ \Qc_5\ra-\Rc_4,\ \Qc_6\ra-\Rc_7,\ \Qc_7\ra-\Rc_6$. In addition, all kinematic invariants undergo the permutation $3\leftrightarrow5$.
As a result we obtain
$$\ba{lcl}
\ds{{\cal M}_{\rho_2}(g_1,q_2^-,\bar q_3^+,q_4^-,\bar q_5^+)}&=&\ds{\tilde C\ t(\rho_2)\
\lf\{\ \lf[\ \Pc_1+\Pc_4-\Pc_5-\fc{s_{12}}{s_{14}}\ (\Pc_3+\Pc_6-\Pc_7)\
 \ri]\ G_1^{\rho_2}\ri.}\\[4mm]
&-&\ds{\lf.\lf[\ \Pc_1+\Pc_2+\Pc_4-\Pc_6-\fc{s_{12}+s_{13}}{s_{14}}\
(\Pc_3+\Pc_6-\Pc_7) \ \ri]\ G_2^{\rho_2}\ \ri\}\ ,}\\[7mm]
\ds{{\cal M}'_{\rho_2}(g_1,q_2^-,\bar q_3^+,q_4^+,\bar q_5^-)}&=&\ds{\tilde C\ t(\rho_2)\
\lf\{\ \lf[\ \Qc_1+\Qc_4-\Qc_5-\fc{s_{12}}{s_{14}}\ (\Qc_3-\Qc_4+\Qc_7)\ \ri]\ \gpps{1}\ri.}\\[4mm]
&-&\ds{\lf.\lf[\ \Qc_1+\Qc_2+\Qc_3-\Qc_6+\Qc_7+\fc{s_{15}}{s_{14}}\ (\Qc_3-\Qc_4+\Qc_7)\ \ri]\ \gpps{2}\ \ri\}\ ,}
\ea$$
\bea
\ds{{\cal M}''_{\rho_2}(g_1,q_2^-,\bar q_3^-,q_4^+,\bar q_5^+)}&=&\ds{\tilde C\ t(\rho_2)\
\lf\{\ \lf[\ \Rc_1+\Rc_4-\Rc_5-\fc{s_{12}}{s_{14}}\ (\Rc_3+\Rc_5-\Rc_6)\
\ri]\ \gps{1}\ri.}\\[4mm]
&-&\ds{\lf.\lf[\ \Rc_1+\Rc_2+\Rc_3+\Rc_4-\Rc_7+\fc{s_{15}}{s_{14}}\ (\Rc_3+\Rc_5-\Rc_6)\
\ri]\ \gps{2}\ \ri\}\ ,}\label{FINALgqqqqs}
\eea
with the two integrals
\bea\label{newqq4ss}
&&\ds{G^{\rho_2}_1\lf[{n_1\atop n_2}\ri]=-\int\limits_{{\cal I}_{\rho_1}} dx\ dy\ I_{\rho_2}(x) \ y^{-1}\
\bigl| x \bigr|^{s_{34}-1-\fc{n_1}{2}} \, \bigl| y \bigr|^{s_{12}} \,
\bigl| 1- x \bigr|^{s_{45}-1-\fc{n_2}{2}} \, \bigl| 1-y \bigr|^{s_{15}} \, \bigl| 1-xy \bigr|^{s_{14}}\ ,}\\
&&\ds{G^{\rho_2}_2\lf[{n_1\atop n_2}\ri]=\int\limits_{{\cal I}_{\rho_1}} dx\ dy\ I_{\rho_2}(x)\  (1-y)^{-1}\
\bigl| x \bigr|^{s_{34}-1-\fc{n_1}{2}} \, \bigl| y \bigr|^{s_{12}} \,
\bigl| 1- x \bigr|^{s_{45}-1-\fc{n_2}{2}} \, \bigl| 1-y \bigr|^{s_{15}} \, \bigl| 1-xy \bigr|^{s_{14}}\ ,}
\eea
with $G^{\rho_2}_i=G^{\rho_2}_i\lf[{0\atop 0}\ri],\ \gps{i}=G^{\rho_2}_i\lf[{0\atop
  1}\ri],\ \gpps{i}=G^{\rho_2}_i\lf[{1\atop 0}\ri],\ i=1,2$ as basis
for the three helicity configurations $(i)$, $(ii)$, and  $(iii)$, respectively.

At this point, we can substitute the gluon polarization vectors in order to obtain more explicit expressions for the kinematic functions $\cal P$, $\cal Q$ and $\cal R$, defined in Eqs. (\ref{calp}), (\ref{calp2}), (\ref{calq}) and (\ref{calr}). To that end, we choose the most convenient reference vectors, in both positive and negative gluon polarization cases. For instance with the choice of $k_4$
as the reference vector for $\xi_1^+$, see Eq. (\ref{ximu}),  ${\cal P}_3={\cal P}_6={\cal P}_7=0$
and the only non-vanishing ${\cal P}$'s are
\beq\big({\cal P}_1, {\cal P}_2, {\cal P}_4= {\cal P}_5\big)=
\frac{\vev{24}}{\sqrt{2} \vev{14}}\times\Big(
 \vev{24}\ket{21}\ket{35},\ \vev{34}\ket{31}\ket{35},\ \vev{14}\ket{13}\ket{51}
  \,   \Big)~. \label{calpp1} \eeq
  Similarly, ${\cal Q}_3={\cal R}_3=0$, and:
\begin{eqnarray}\big({\cal Q}_1,&&\hskip -9mm {\cal Q}_2, {\cal Q}_4= {\cal Q}_5, {\cal Q}_6={\cal Q}_7\big) \\ &=&
\frac{1}{\sqrt{2} \vev{14}}\times\Big(
 \vev{25}\vev{24}\ket{21}\ket{34},\ \vev{25}\vev{34}\ket{31}\ket{34},\ \vev{24}\vev{51}\ket{13}\ket{14},\ \vev{12}\vev{45}\ket{13}\ket{14}
  \,   \Big)~, \label{calqq1} \nonumber\\[2mm]
\big({\cal R}_1,&&\hskip -9mm {\cal R}_2, {\cal R}_4= {\cal R}_5, {\cal R}_6={\cal R}_7\big) \\ &=&
\frac{1}{\sqrt{2} \vev{14}}\times\Big(
 \vev{23}\vev{24}\ket{21}\ket{45},\ \vev{23}\vev{34}\ket{31}\ket{45},\ \vev{24}\vev{31}\ket{14}\ket{15},\ \vev{12}\vev{43}\ket{14}\ket{15}
  \,   \Big)~. \label{calrr1} \nonumber
  \end{eqnarray}
In this way, Eqs. (\ref{FINALgqqqq}) yield
\bea\label{FINALgqqqq1}
\ds{{\cal M}_{\rho_1}\lf(\lf. {g_1^+\atop  g_1^-}\ri\},q_2^-,\bar q_3^+,q_4^-,\bar q_5^+\ri)}
&=&\ds{\tilde C\ \fc{t(\rho_1)}{\sqrt 2}\ \times\lf\{
{\fc{\vev{24}^2}{\vev{14}}\, \lf(\, \ket{12}\ket{53}\
G_1^{\rho_1}+\ket{13}\ket{25}\ G_2^{\rho_1}\, \ri)\atop
 \ \ \ \fc{\ket{35}^2}{\ket{14}}\, \lf(\, \vev{12}\vev{53}\
G_1^{\rho_1}+\vev{13}\vev{25}\ G_2^{\rho_1}\, \ri)\ ,}\ri.}\\[8mm]
\ds{{\cal M}'_{\rho_1}\lf(\lf. {g_1^+\atop  g_1^-}\ri\},q_2^-,\bar q_3^+,q_4^+,\bar q_5^-\ri)}
&=&\ds{-\tilde C\ \fc{t(\rho_1)}{\sqrt 2}\ \times\lf\{
{\fc{\vev{25}^2}{\vev{14}}\ \lf(\, \ket{12}\ket{53}\
\gp{1}+\ket{13}\ket{25}\ \gp{2} \ri)\atop
 \ \ \ \fc{\ket{34}^2}{\ket{14}}\ \lf(\, \vev{12}\vev{53}\
\gp{1}+\vev{13}\vev{25}\ \gp{2} \ri)\ ,}\ri.}\\[8mm]
\ds{{\cal M}''_{\rho_1}\lf(\lf. {g_1^+\atop  g_1^-}\ri\},q_2^-,\bar q_3^-,q_4^+,\bar q_5^+\ri)}
&=&\ds{-\tilde C\ \fc{t(\rho_1)}{\sqrt 2}\ \times\lf\{
{\fc{\vev{23}^2}{\vev{14}}\ \lf(\,\ket{12}\ket{53}\
\gpp{1}+\ket{13}\ket{25}\ \gpp{2} \ri)\atop
 \ \ \ \fc{\ket{45}^2}{\ket{14}} \lf(\,\vev{12}\vev{53}\
\gpp{1}+\vev{13}\vev{25}\ \gpp{2} \ri)\ ,}\ri.}
\eea
with the normalization $\tilde C$ given in Eq. \req{norm}.
{}Finally, from  Eqs. (\ref{FINALgqqqqs}), we obtain:
\bea\label{FINALgqqqq2}
\ds{{\cal M}_{\rho_2}\lf(\lf. {g_1^+\atop  g_1^-}\ri\},q_2^-,\bar q_3^+,q_4^-,\bar q_5^+\ri)}
&=&\ds{\tilde C\ \fc{t(\rho_2)}{\sqrt 2}\ \times\lf\{
{\fc{\vev{24}^2}{\vev{14}}\, \lf(\, \ket{12}\ket{53}\
G_1^{\rho_2}-\ket{15}\ket{23}\ G_2^{\rho_2}\, \ri)\atop
 \ \ \ \fc{\ket{35}^2}{\ket{14}}\, \lf(\, \vev{12}\vev{53}\
G_1^{\rho_2}-\vev{15}\vev{23}\ G_2^{\rho_2}\, \ri)\ ,}\ri.}\\[8mm]
\ds{{\cal M}'_{\rho_2}\lf(\lf. {g_1^+\atop  g_1^-}\ri\},q_2^-,\bar q_3^+,q_4^+,\bar q_5^-\ri)}
&=&\ds{-\tilde C\ \fc{t(\rho_2)}{\sqrt 2}\ \times\lf\{
{\fc{\vev{25}^2}{\vev{14}}\ \lf(\,\ket{12}\ket{53}\
\gpps{1}-\ket{15}\ket{23}\ \gpp{2} \ri)\atop
 \ \ \ \fc{\ket{34}^2}{\ket{14}} \lf(\,\vev{12}\vev{53}\
\gpps{1}-\vev{15}\vev{23}\ \gpp{2} \ri)\ ,}\ri.}\\[8mm]
\ds{{\cal M}''_{\rho_2}\lf(\lf. {g_1^+\atop  g_1^-}\ri\},q_2^-,\bar q_3^-,q_4^+,\bar q_5^+\ri)}
&=&\ds{-\tilde C\ \fc{t(\rho_2)}{\sqrt 2}\ \times\lf\{
{\fc{\vev{23}^2}{\vev{14}}\ \lf(\, \ket{12}\ket{53}\
\gps{1}-\ket{15}\ket{23}\ \gp{2} \ri)\atop
 \ \ \ \fc{\ket{45}^2}{\ket{14}}\ \lf(\, \vev{12}\vev{53}\
\gps{1}-\vev{15}\vev{23}\ \gp{2} \ri)\ .}\ri.}
\eea

\subsection{Partial subamplitudes and D--brane setup}
\label{Shh}

In this Subsection we discuss the possible orderings $\rho$ of the four chiral fermions along the boundary of the disk world--sheet.
The number of possible orderings depends on the type of D--brane intersections
we are considering.
For the given orderings $\rho_1$ of the four quarks $z_2<z_3<z_4<z_5$ referring to the choice \req{4f5a} or for the orderings $\rho_2$ with $z_2<z_5<z_4<z_3$ referring to \req{4f5b} there are at most
the following  four possibilities to place the gluon position $z_1$ along the boundary of the disk:
\beq\label{allfour}
\ba{lcr}
(1,2,3,4,5)&\ \ \ ,\ \ \ &(1,2,5,4,3)\ ,\\
(2,3,1,4,5)&\ \ \ ,\ \ \ &(2,5,1,4,3)\ .
\ea
\eeq

\noindent
The four different orderings \req{allfour} are specified by their region of integration ${\cal I}_\rho$.
With the choice \req{4f5a} the first column of \req{allfour} can be realized by the
parameterization displayed below.

\beq
\begin{array}{|c|c|c|rr|} \hline \rho_1 &t(\rho_1)& z_{1},z_{3} & {\cal I}_{\rho_1}& \\\hline
(1,2,3,4,5) & (T^{a_1})^{\al_2}_{\al_5}
\delta^{\al_4}_{\al_3}\delta_{\bet_2}^{\bet_3}\delta_{\bet_4}^{\beta_5}& -\infty< z_{1} < 0 < z_{3} < 1 & 0 < x < 1\ , &-\infty < y < 0 \\[2mm]
(2,3,1,4,5) &(T^{a_1})^{\al_4}_{\al_3}
\delta^{\al_2}_{\al_5}\delta_{\bet_2}^{\bet_3}\delta_{\bet_4}^{\beta_5}& 0 < z_{3} < z_{1} < 1 & 0<x<1\ , &1 < y < \frac{1}{x} \\ \hline
\end{array}
\label{REGG1}
\eeq

\vskip0.25cm
\noindent
To parametrize the other two orderings from the second column of \req{allfour} we need the fixing \req{4f5b}. The partial amplitudes corresponding to these
two orderings may be simply obtained from the two orderings in \req{REGG1}
by performing the permutation $3\leftrightarrow 5$ on all labels, c.f. the previous Subsection for more details.
These two orderings yield the group theoretical combination shown below.

\beq
\begin{array}{|c|c|c|} \hline \rho_2 &  t(\rho_2)\\ \hline
(1,2,5,4,3) & (T^{a_1})^{\al_2}_{\al_3}
\delta^{\al_4}_{\al_5}\delta_{\bet_2}^{\bet_5}\delta_{\bet_4}^{\beta_3} \\[2mm]
(2,5,1,4,3) &(T^{a_1})^{\al_4}_{\al_5}
\delta^{\al_2}_{\al_3}\delta_{\bet_2}^{\bet_5}\delta_{\bet_4}^{\beta_3}  \\ \hline
\end{array}
\label{REGG2}
\eeq

\vskip0.25cm
\noindent

\vskip0.5cm
\noindent
\underline{\sl (i) Four--fermion amplitudes involving two twist--antitwist pairs $(\th,1-\th)$}

\begin{figure}[H]
\centering
\includegraphics[width=0.65\textwidth]{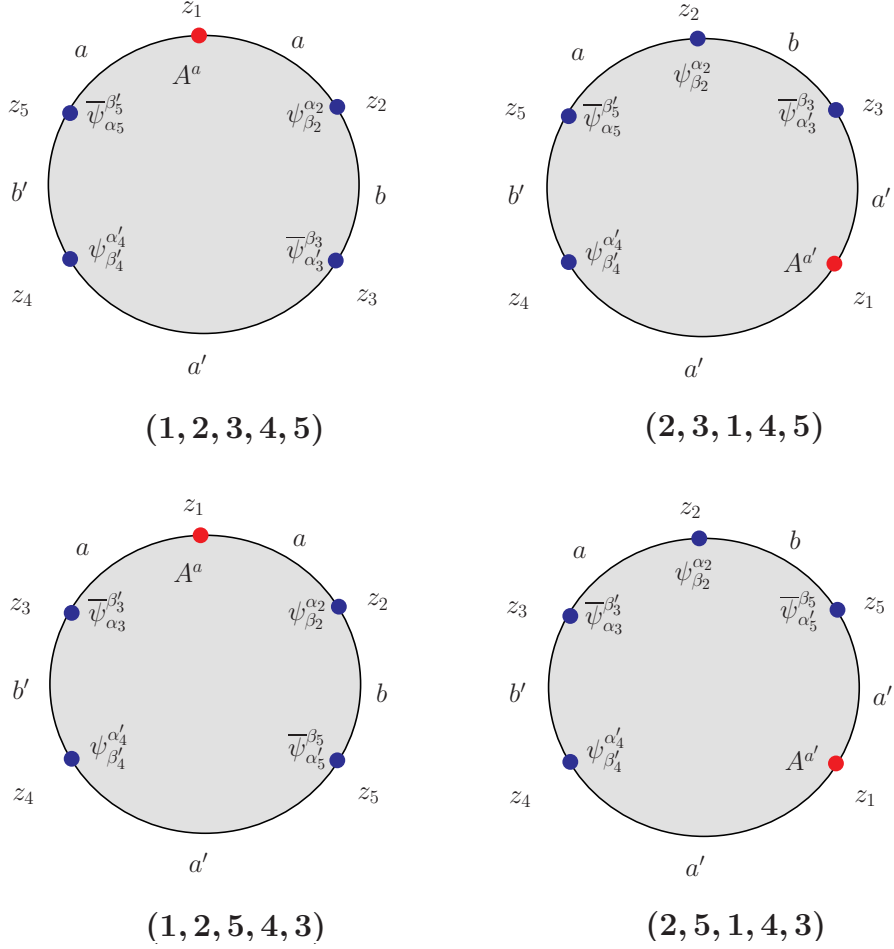}
\caption{Orderings $\rho_1,\rho_2$ of the gauge boson and four fermion vertex operators.}
\label{gqqqq}
\end{figure}

\vskip0.5cm
\noindent
First we consider the case of two pairs $(a,b)$ and $(c,d)\simeq (b',a')$ of D--branes intersecting at the angle $\th$.
In \req{4f2} we have two pairs of twist--antitwist fields $(\th,1-\th)$
from intersections $f_i$ and $f_j$, respectively.
One pair of fermions is related to a twist--antitwist
pair $(\th,1-\th)$
and a second fermion pair to an other  twist--antitwist
pair $(\th,1-\th)$, c.f. \cite{LHC} for more details.
For two pairs of conjugate fermions from the intersecting brane stacks
$a$ and $b$ there are four possibilities to place a gauge boson with gauge group
$G_a$ within the four chiral fermions, c.f. the Figure \ref{gqqqq}.
\noindent
The orderings corresponding to the first column are shown in the first row of Figure \ref{gqqqq},
while the second column is displayed in the second row. Similarly, for a gauge boson of gauge group $G_b$ we obtain four possible orderings.

The functions $I_\rho=I_1I_{2\rho}$, encoding the quantum and instanton part of the correlator \req{4f2}, 
given by \bea\label{givenby1}
\ds{I_1(x)}&=&\ds{\prod_{j=1}^3\lf(\fc{F_j(x)\ F_j(1-x)}{\sin(\pi\th^j)}\ri)^{-1/2}\ ,}\\[5mm]
\ds{I_{2\rho_1}(x)}&=&\ds{\sum_{p_a,p_b\in\ZZ^6}
e^{-\fc{\pi}{\ap}\ \sum\limits_{j=1}^3\sin(\pi\theta^j)\
\lf[\ \lf|p^j_bL_b^j+\delta^j_b\ri|^2\ \tau_j(x)+
\lf|p^j_aL^j_a+\delta^j_a\ri|^2\ \tau_j(1-x)\ \ri]}\ ,}\\
\ds{I_{2\rho_2}(x)}&=&\ds{I_{2\rho_1}(1-x)\ ,}
\eea
with:
$$\tau_j(x)=\fc{F_j(1-x)}{F_j(x)}\ \ \ ,\ \ \ F_j(x)=\FF{2}{1}\lf[\th^j,1-\th^j,1;x\ri]\ ,$$
are  discussed in \cite{LHC}. In this reference in Eqs. (4.17) and (5.49)
the quantities $L$ and $\delta$ may also be looked up, respectively.
In the following for the four possible orderings \req{allfour}  the factorization and the pole structure of the set of amplitudes \req{FINALgqqqq1} and \req{FINALgqqqq2} is discussed.

\ \\
\noindent
\underline{$(1,2,3,4,5)$}

\noindent
For the case $\rho_1=(1,2,3,4,5)$  the pole structure of the three amplitudes \req{4f6}, \req{4f7a} and \req{4f7b} describing the three helicity  limits  are more conveniently obtained after performing the coordinate transformation
\beq\label{trans1}
x\ra\fc{x\ (1-y)}{1-xy}\ \ \ ,\ \ \ y\ra 1-\fc{1}{xy}
\eeq
on the expressions \req{newqq4}, giving:
\bea\label{newq4}
\ds{G^{\rho_1}_1\lf[{n_1\atop n_2}\ri]}&=&
\ds{\int_0^1 dx\ x^{s_2-1-\fc{n_1}{2}}\ (1-x)^{s_3-1-\fc{n_2}{2}}\
\int_0^1\ dy\ y^{s_5-1}\ (1-y)^{s_4-1-\fc{n_1}{2}}}\\[3mm]
&\times&\ds{(1-xy)^{s_1-s_3-s_4+\fc{n_1}{2}+\fc{n_2}{2}}\ I_{\rho_1}\lf(\fc{x(1-y)}{1-xy}\ri)\ ,}\\[5mm]
\ds{G^{\rho_1}_2\lf[{n_1\atop n_2}\ri]}&=&\ds{\int_0^1 dx\ x^{s_2-1-\fc{n_1}{2}}\ (1-x)^{s_3-1-\fc{n_2}{2}}\ \int_0^1\ dy\ y^{s_5-1}\ (1-y)^{s_4-1-\fc{n_1}{2}}}\\[3mm]
&\times&\ds{(1-xy)^{s_1-s_3-s_4+1+\fc{n_1}{2}+\fc{n_2}{2}}\ I_{\rho_1}\lf(\fc{x(1-y)}{1-xy}\ri)\ ,}
\eea
with $G^{\rho_1}_i=G^{\rho_1}_i\lf[{0\atop 0}\ri],\ \gp{i}=G^{\rho_1}_i\lf[{0\atop
  1}\ri],\ \gpp{i}=G^{\rho_1}_i\lf[{1\atop 0}\ri],\ i=1,2$, and $\rho=(1,2,3,4,5)$.
The functions \req{newq4} have  four sets of poles, which occur
in the limits $x\ra0,1$ and $y\ra0,1$. E.g. for $(n_1,n_2)=(0,0)$, which has the most
pole terms, we will find in the following:
\bea\label{Newq4}
\ds{G^{\rho_1}_1\lf[{0\atop 0}\ri]}&=&\ds{\fc{g_{D6_b}^2}{s_1 s_3}+\fc{g_{D6_a}^2}{s_1 s_4}+\fc{g_{D6_a}^2}{s_2 s_4}+\fc{g_{D6_a}^2}{s_2 s_5}+\fc{g_{D6_b}^2}{s_3 s_5}+\ldots\ ,}\\[4mm]
\ds{G^{\rho_1}_2\lf[{0\atop 0}\ri]}&=&\ds{\fc{g_{D6_a}^2}{s_2 s_4}+\fc{g_{D6_a}^2}{s_2 s_5}+\fc{g_{D6_b}^2}{s_3 s_5}+\dots\ .}
\eea

\begin{figure}[H]
\centering
\includegraphics[width=0.85\textwidth]{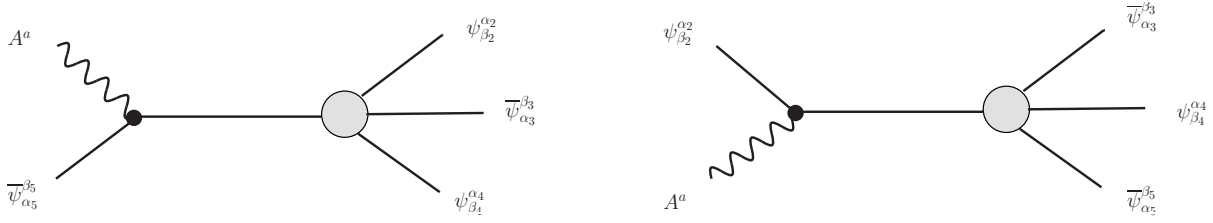}
\caption{Exchange of a quark and $q\bar q q\bar q$ string vertex (for $y\ra 0$ and $\tilde x\ra0$).}\label{new1}
\end{figure}

For the limit $y\ra 0$ we obtain from \req{FINALgqqqq1}
\bea
\ds{\lf.\ba{c}
{\cal M}\lf(\lf. {g_1^+\atop  g_1^-}\ri.\ri.\\
{\cal M}'\lf(\lf. {g_1^+\atop  g_1^-}\ri.\ri.\\
{\cal M}''\lf(\lf. {g_1^+\atop  g_1^-}\ri.\ri.
\ea\ri\}}&\stackrel{y\ra 0}{\lra}&\ds{\sqrt 2\ t(\rho)\ g_{D6_a}\
V_{abab}^{(n_1,n_2)}(s_2,s_3)\ \times\lf\{\hskip-0.25cm\ba{rl}
&\lf(\ds{\fc{\vev{24}^2\ket{32}}{\vev{14}\vev{15}}
  \atop\fc{\ket{35}^2\vev{32}}{\ket{14}\ket{15}} }\ \ \ ,\ \ \
(n_1,n_2)=(0,0)\ ,\ri.\\[5mm]
&\lf(\ds{-\fc{\vev{25}^2\ket{32}}{\vev{14}\vev{15}}
  \atop-\fc{\ket{34}^2\vev{32}}{\ket{14}\ket{15}} }\ \ \ ,\ \ \
(n_1,n_2)=(0,1)\ ,\ri.\\[5mm]
&\lf(\ds{-\fc{\vev{23}^2\ket{32}}{\vev{14}\vev{15}}
  \atop-\fc{\ket{45}^2\vev{32}}{\ket{14}\ket{15}} }\ \ \ ,\ \ \
(n_1,n_2)=(1,0)\ .\ri.\ea\ri.}\label{limit1}
\eea
This limit describes the massless exchange of a quark, coupled to the
four fermion string interaction, c.f. left diagram of Figure \ref{new1}.
\noindent
In \req{limit1} the function
\beq\label{tomi}
V^{(n_1,n_2)}_{abab}(s_2,s_3)=C\ \int_0^1dx\
x^{s_2-1-\fc{n_1}{2}}\ (1-x)^{s_3-1-\fc{n_2}{2}}\ I_{\rho_1}(x)\ ,
\eeq
agrees with the expression (5.70) of \cite{LHC} describing the $q\bar q q
\bar q$ amplitude.
Hence, to further reduce this integral we may apply the results from there and obtain:
\beq
V^{(n_1,n_2)}_{abab}(s_2,s_3)\stackrel{x\ra 0,1}{\lra}
\lf(\fc{g_{D6_a}^2}{s_2-\fc{n_1}{2}}+\fc{g_{D6_b}^2}{s_3-\fc{n_2}{2}}\ri)\ .
\label{Limit1}
\eeq
Note, that \req{limit1} with the limit \req{Limit1}
represents the lowest order of the amplitudes \req{4f6}, \req{4f7a}
and \req{4f7b} focusing on the $s_5$--channel.
Higher orders include exchanges of both universal SR excitations, KK and winding
modes.

\begin{figure}[H]
\centering
\includegraphics[width=0.4\textwidth]{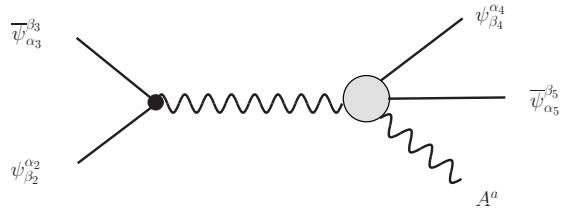}
\caption{Exchange of a gluon and $ggq\bar q$ string vertex (for $x\ra 0$).}
\label{new2}
\end{figure}

On the other hand, for the limit $x\ra0$ we obtain:
\bea
\ds{\lf.\ba{c}
{\cal M}\lf(\lf. {g_1^+\atop  g_1^-}\ri.\ri.\\
{\cal M}'\lf(\lf. {g_1^+\atop  g_1^-}\ri.\ri.\\
{\cal M}''\lf(\lf. {g_1^+\atop  g_1^-}\ri.\ri.
\ea\ri\}}&\stackrel{x\ra 0}{\lra}&\ds{\sqrt 2\ t(\rho)\
\fc{g_{D6_a}^3}{s_2-\fc{n_1}{2}}\
B\lf(s_5,s_4-\fc{n_1}{2}\ri)\ \times
\lf\{\hskip-0.25cm\ba{rl}
&\lf(\ds{\fc{\vev{24}^2\ket{32}\ket{51}}{\vev{14}}
  \atop\fc{\ket{35}^2\vev{32}\vev{51}}{\ket{14}} }\ \ \ ,\ \ \
(n_1,n_2)=(0,0)\ ,\ri.\\[5mm]
&\lf(\ds{-\fc{\vev{25}^2\ket{32}\ket{51}}{\vev{14}}
  \atop-\fc{\ket{34}^2\vev{32}\vev{51}}{\ket{14}} }\ \ \ ,\ \ \
(n_1,n_2)=(0,1)\ ,\ri.\\[5mm]
&\lf(\ds{-\fc{\vev{23}^2\ket{32}\ket{51}}{\vev{14}}
  \atop-\fc{\ket{45}^2\vev{32}\vev{51}}{\ket{14}} }\ \ \ ,\ \ \
(n_1,n_2)=(1,0)\ .\ri.\ea\ri.}\label{limit2}
\eea
The limit \req{limit2} describes at leading order an exchange of a
gluon  coupled to the string amplitude
involving two gluons and two quarks, c.f. the  diagram in Figure \ref{new2}.
\noindent
The string amplitude involving two gluons and two quarks is universal and
encoded in \req{limit2} by the Euler Beta function, defined in
\req{side1}.
Note, that at the NLO also KK modes and windings are exchanged.
In the limit $y \ra 1$ the Euler Beta function in \req{limit2} reduces to:
\beq
B\lf(s_5,s_4-\fc{n_1}{2}\ri)\stackrel{y\ra 1}{\lra}
\fc{1}{s_4-\fc{n_1}{2}}\ .
\label{Limit2}
\eeq
On the other hand, the limit $y\ra 0$ of \req{limit2} is already
contained in \req{Limit1}.

The limit $x\ra 1, y\ra 1$ is more conveniently investigated after performing
 a further transformation
\beq\label{TRANSF}
x\ra\fc{1-\tilde x}{1-\tilde x\tilde y}\ \ \ ,\ \ \ y\ra 1-\tilde x\tilde y
\eeq
on \req{newq4}.
This transformation, which corresponds to a cyclic transformation, gives
\bea
\ds{G_1^{\rho_1}\lf[{n_1\atop n_2}\ri]}&=&
\ds{\int_0^1d\tilde x\ \tilde x^{s_1-1}\
(1-\tilde x)^{s_2-1-\fc{n_1}{2}}\ \int_0^1 d\tilde y\ \tilde
y^{s_4-1-\fc{n_1}{2}}\
(1-\tilde y)^{s_3-1-\fc{n_2}{2}}}\\[5mm]
&\times&\ds{(1-\tilde x\tilde y)^{-s_2-s_3+s_5+\fc{n_1}{2}+\fc{n_2}{2}}\ I_{\rho_1}\lf(\fc{\tilde y(1-\tilde x)}{1-\tilde x\tilde y}\ri)}\\[5mm]
\ds{G_2^{\rho_1}\lf[{n_1\atop n_2}\ri]}&=&\ds{\int_0^1d\tilde x\ \tilde
  x^{s_1}\ (1-\tilde x)^{s_2-1-\fc{n_1}{2}}\ \int_0^1 d\tilde y\
  \tilde y^{s_4-1-\fc{n_1}{2}}\
(1-\tilde y)^{s_3-1-\fc{n_2}{2}}}\\[5mm]
&\times&\ds{(1-\tilde x\tilde y)^{-s_2-s_3+s_5+\fc{n_1}{2}+\fc{n_2}{2}}\ I_{\rho_1}\lf(\fc{\tilde y(1-\tilde x)}{1-\tilde x\tilde y}\ri)\ .}\\[5mm]
\eea
In these new coordinates the limit $x,y\ra1$   is recovered at $\tilde x\ra0$ and yields:
\bea
\ds{\lf.\ba{c}
{\cal M}\lf(\lf. {g_1^+\atop  g_1^-}\ri.\ri.\\
{\cal M}'\lf(\lf. {g_1^+\atop  g_1^-}\ri.\ri.\\
{\cal M}''\lf(\lf. {g_1^+\atop  g_1^-}\ri.\ri.
\ea\ri\}}&\stackrel{\tilde x\ra 0}{\lra}&\ds{\sqrt 2\ t(\rho)\ g_{D6_a}\
V_{abab}^{(n_1,n_2)}(s_4,s_3)\ \times
\ \lf\{\hskip-0.25cm\ba{rl}
&\lf(\ds{\fc{\vev{24}^2\ket{35}}{\vev{12}\vev{14}}
  \atop\fc{\ket{35}^2\vev{35}}{\ket{12}\ket{14}} }\ \ \ ,\ \ \
(n_1,n_2)=(0,0)\ ,\ri.\\[5mm]
&\lf(\ds{-\fc{\vev{25}^2\ket{35}}{\vev{12}\vev{14}}
  \atop-\fc{\ket{34}^2\vev{35}}{\ket{12}\ket{14}} }\ \ \ ,\ \ \
(n_1,n_2)=(0,1)\ ,\ri.\\[5mm]
&\lf(\ds{-\fc{\vev{23}^2\ket{35}}{\vev{12}\vev{14}}
  \atop-\fc{\ket{45}^2\vev{35}}{\ket{12}\ket{14}} }\ \ \ ,\ \ \
(n_1,n_2)=(1,0)\ .\ri.\ea\ri.}
\label{limit11}
\eea
This limit describes the massless exchange of a quark coupled to the four
fermion string interaction, c.f. right diagram of Figure \ref{new1}.

To summarize, for the ordering $\rho_1=(1,2,3,4,5)$ the three amplitudes \req{4f6}, \req{4f7a} and \req{4f7b}
have the following low-energy limits:
$$\ba{lcr}
\ds{{\cal M}_{\rho_1}\lf(\lf. {g_1^+\atop  g_1^-}\ri\},q_2^-,\bar q_3^+,q_4^-,\bar q_5^+\ri)}
&=&\ds{\sqrt 2\ g_{D6_a}\ (T^{a_1})^{\al_2}_{\al_5}
\delta^{\al_4}_{\al_3}\delta_{\bet_2}^{\bet_3}\delta_{\bet_4}^{\beta_5}\ \times\lf\{
\ds{ \fc{\vev{24}^2}{\vev{12}\vev{15}}\ \times\lf(\ g_{D6_a}^2\
  \fc{\vev{25}}{\vev{23}\vev{45}}+g_{D6_b}^2\ \fc{1}{\vev{34}}\ \ri)\atop
  \fc{\ket{35}^2}{\ket{12}\ket{15}}\ \lf(\ g_{D6_a}^2\ \fc{\ket{25}}{\ket{23}\ket{45}}+g_{D6_b}^2\
\fc{1}{\ket{34}}\ \ri)\ ,}\ri.}
\ea$$
\bea\label{FINAL1}
\ds{{\cal M}'_{\rho_1}\lf(\lf. {g_1^+\atop  g_1^-}\ri\},q_2^-,\bar q_3^+,q_4^+,\bar q_5^-\ri)}
&=&\ds{-\sqrt 2\ g_{D6_a}\ (T^{a_1})^{\al_2}_{\al_5}
\delta^{\al_4}_{\al_3}\delta_{\bet_2}^{\bet_3}\delta_{\bet_4}^{\beta_5}\ \times\lf\{
\ds{ g_{D6_a}^2\
  \fc{\vev{25}^3}{\vev{12}\vev{15}\vev{23}\vev{45}}\atop
  g_{D6_a}^2\
  \fc{\ket{34}^2\ket{25}}{\ket{12}\ket{15}\ket{23}\ket{45}}\ ,}\ri.}\\[8mm]
\ds{{\cal M}''_{\rho_1}\lf(\lf. {g_1^+\atop  g_1^-}\ri\},q_2^-,\bar q_3^-,q_4^+,\bar q_5^+\ri)}
&=&\ds{-\sqrt 2\ g_{D6_a}\ (T^{a_1})^{\al_2}_{\al_5}
\delta^{\al_4}_{\al_3}\delta_{\bet_2}^{\bet_3}\delta_{\bet_4}^{\beta_5}\ \times\lf\{
\ds{ g_{D6_b}^2\
  \fc{\vev{23}^2}{\vev{12}\vev{15}\vev{34}}\atop
  g_{D6_b}^2\
  \fc{\ket{45}^2}{\ket{12}\ket{15}\ket{34}}\ .}\ri.}
\eea

\ \\
\noindent
\underline{$(2,3,1,4,5)$}

\noindent
To analyze the pole structure for the case $\rho_1=(2,3,1,4,5)$
we first perform the following transformation on the integrand:
\beq\label{trans1}
x\ra\fc{x\ (1-y)}{1-xy}\ \ \ ,\ \ \ y\ra \fc{1-xy}{1-y}\ .
\eeq
This converts the  expressions \req{newqq4} to
\bea\label{newq5}
\ds{G^{\rho_1}_1\lf[{n_1\atop n_2}\ri]}&=&\ds{-\int_0^1 dx\ x^{s_4-1-\fc{n_1}{2}}\ (1-x)^{-s_1+s_3-s_5-\fc{n_2}{2}}\ \int_0^1\ dy\ y^{-s_1-s_2+s_4}\ (1-y)^{s_2-1-\fc{n_1}{2}}}  \\[4mm]
&\times&\ds{(1-xy)^{s_1-s_3-s_4+\fc{n_1}{2}+\fc{n_2}{2}}\
I_{\rho_1}\lf(\fc{x\ (1-y)}{1-xy}\ri)\ ,}\\[5mm]
\ds{G^{\rho_1}_2\lf[{n_1\atop n_2}\ri]}&=&\ds{-\int_0^1 dx\ x^{s_4-1-\fc{n_1}{2}}\ (1-x)^{-s_1+s_3-s_5-\fc{n_2}{2}}\ \int_0^1\ dy\ y^{-s_1-s_2+s_4}\ (1-y)^{s_2-1-\fc{n_1}{2}}}  \\[4mm]
&\times&\ds{(1-xy)^{s_1-s_3-s_4+\fc{n_1}{2}+\fc{n_2}{2}}\
I_{\rho_1}\lf(\fc{x\ (1-y)}{1-xy}\ri)\ \fc{(1-xy)}{(1-x)\ y}\ ,}
\eea
with $G^{\rho_1}_i=G^{\rho_1}_i\lf[{0\atop 0}\ri],\ \gp{i}=G^{\rho_1}_i\lf[{0\atop
  1}\ri],\ \gpp{i}=G^{\rho_1}_i\lf[{1\atop 0}\ri],\ i=1,2$, and $\rho=(2,3,1,4,5)$.
The integrals \req{newq5} have  three sets of poles in the limits $y\ra0,\ x\ra0,1$ and $x\ra0$. E.g. for $(n_1,n_2)=(0,0)$, which has the most
pole terms, we will find in the following:
\bea\label{Newq5}
\ds{G^{\rho_1}_1\lf[{0\atop 0}\ri]}&=&\ds{-\fc{g_{D6_a}^2}{s_2s_4}-\ldots\ ,}\\[4mm]
\ds{G^{\rho_1}_2\lf[{0\atop 0}\ri]}&=&\ds{-
\fc{g_{D6_a}^2}{s_2s_4}-\fc{g_{D6_a}^2}{s_4\ (-s_1-s_2+s_4)}-
\fc{g_{D6_b}^2}{(-s_1-s_2+s_4)\ (-s_1+s_3-s_5)}-\ldots\ .}
\eea

\begin{figure}[H]
\centering
\includegraphics[width=0.4\textwidth]{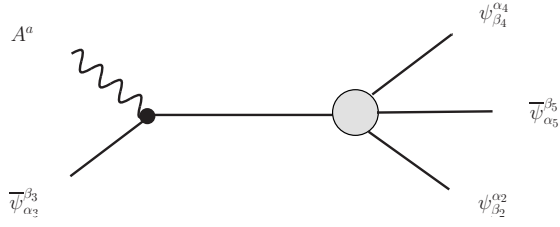}
\caption{Exchange of a quark and $q\bar q q \bar q$ string vertex (for $y\ra0$).}
\label{new3}
\end{figure}

For the limit $y\ra0$ we obtain from \req{FINALgqqqq1}:
\bea
\ds{\lf.\ba{c}
{\cal M}\lf(\lf. {g_1^+\atop  g_1^-}\ri.\ri.\\
{\cal M}'\lf(\lf. {g_1^+\atop  g_1^-}\ri.\ri.\\
{\cal M}''\lf(\lf. {g_1^+\atop  g_1^-}\ri.\ri.
\ea\ri\}}&\stackrel{y\ra 0}{\lra}&\ds{\sqrt 2\ t(\rho)\ g_{D6_a}\
V_{abab}^{(n_1,n_2)}(s_{45},s_{25})\ \times\lf\{\hskip-0.25cm\ba{rl}
&\lf(\ds{-\fc{\vev{24}^2\ket{25}}{\vev{13}\vev{14}}
  \atop-\fc{\ket{35}^2\vev{25}}{\ket{13}\ket{14}} }\ \ \ ,\ \ \
(n_1,n_2)=(0,0)\ ,\ri.\\[5mm]
&\lf(\ds{\fc{\vev{25}^2\ket{25}}{\vev{13}\vev{14}}
  \atop\fc{\ket{34}^2\vev{25}}{\ket{13}\ket{14}} }\ \ \ ,\ \ \
(n_1,n_2)=(0,1)\ ,\ri.\\[5mm]
&\lf(\ds{\fc{\vev{23}^2\ket{25}}{\vev{13}\vev{14}}
  \atop\fc{\ket{45}^2\vev{25}}{\ket{13}\ket{14}} }\ \ \ ,\ \ \
(n_1,n_2)=(1,0)\ .\ri.\ea\ri.}\label{limit3}
\eea
This limit describes the massless exchange of a quark, coupled to the
four fermion string interaction, c.f. Figure \ref{new3}.
\noindent
The function $V_{abab}^{(n_1,n_2)}$ is given in \req{tomi}.
Hence, after applying \req{Limit1} one obtains the lowest order (for $y\ra0,\ x\ra0,1$)
of the amplitudes
\req{4f6}, \req{4f7a} and \req{4f7b} focusing on the $s_{13}$--channel.
Higher orders include exchanges of both universal SR excitations, KK and winding
modes.

\begin{figure}[H]
\centering
\includegraphics[width=0.85\textwidth]{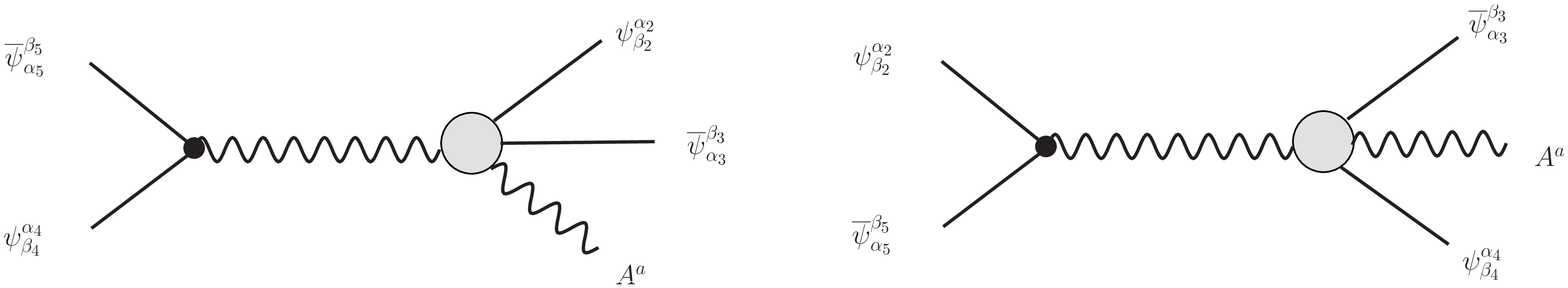}
\caption{Exchange of a gluon and $ggq\bar q$ string vertex (for $x\ra 0$ and $x\ra 1$).}
\label{new4}
\end{figure}

On the other hand, the limit $x\ra 0$ gives
\bea
\ds{{\cal M}\lf( {g_1^+\atop  g_1^-}\ri.}&\stackrel{x\ra 0}{\lra}&
\ds{-\sqrt 2\
t(\rho)\ \fc{g_{D6_a}^3}{s_4}\ \times\lf(\hskip-0.3cm\ba{ll}
&\fc{\vev{24}^2}{\vev{14}}\ \lf[\ \ket{12}\ket{53}\
B\lf(s_{13}+1,s_{23}\ri)+\ket{13}\ket{25}\ B\lf(s_{13},s_{23}\ri)\ \ri]\\[3mm]
&\fc{\ket{35}^2}{\ket{14}}\ \lf[\ \vev{12}\vev{53}\
B\lf(s_{13}+1,s_{23}\ri)+\vev{13}\vev{25}\ B\lf(s_{13},s_{23}\ri)\ \ri]\ ,
\ea\ri.}\\[11mm]
\ds{{\cal M}'\lf({g_1^+\atop  g_1^-}\ri.}&\stackrel{x\ra 0}{\lra}&\ds{\sqrt 2\
t(\rho)\  \fc{g_{D6_a}^3}{s_4}\ \times\lf(\hskip-0.3cm\ba{ll}
&\fc{\vev{25}^2}{\vev{14}}\ \lf[\ \ket{12}\ket{53}\
B\lf(s_{13}+1,s_{23}\ri)+\ket{13}\ket{25}\ B\lf(s_{13},s_{23}\ri)\
\ri]\\[3mm]
&\fc{\ket{34}^2}{\ket{14}}\ \lf[\ \vev{12}\vev{53}\
B\lf(s_{13}+1,s_{23}\ri)+\vev{13}\vev{25}\ B\lf(s_{13},s_{23}\ri)\ \ri]\ ,\ea
\ri.}\\[11mm]
\ds{{\cal M}''\lf( {g_1^+\atop  g_1^-}\ri.}&\stackrel{x\ra 0}{\lra}&\ds{\sqrt 2\
t(\rho)\ \fc{g_{D6_a}^3}{s_4-\h}\ \times\lf(\hskip-0.3cm\ba{ll}
&\fc{\vev{23}^2}{\vev{14}}\ \lf[\ \ket{12}\ket{53}\
B\lf(s_{13}+1,s_{23}-\h\ri)+\ket{13}\ket{25}\
B\lf(s_{13},s_{23}-\h\ri)\ \ri]\\[3mm]
&\fc{\ket{45}^2}{\ket{14}}\ \lf[\ \vev{12}\vev{53}\
B\lf(s_{13}+1,s_{23}-\h\ri)+\vev{13}\vev{25}\
B\lf(s_{13},s_{23}-\h\ri)\ \ri],\ea\ri.}
\label{limit6}
\eea
while for the limit $x\ra 1$ we find:
\bea
\ds{\lf.\ba{c}
{\cal M}\lf(\lf. {g_1^+\atop  g_1^-}\ri.\ri.\\
{\cal M}'\lf(\lf. {g_1^+\atop  g_1^-}\ri.\ri.\\
{\cal M}''\lf(\lf. {g_1^+\atop  g_1^-}\ri.\ri.
\ea\ri\}}&\stackrel{x\ra 1}{\lra}&\ds{\sqrt 2\ t(\rho)\
\fc{g_{D6_a}\ g_{D6_b}^2}{s_{25}-\fc{n_2}{2}}\
B\lf(s_{13},s_{23}+1-\fc{n_1}{2}\ri)\ \times
\lf\{\hskip-0.45cm\ba{rl}
&\lf(\ds{-\fc{\vev{24}^2\ket{25}\ket{13}}{\vev{14}}
  \atop-\fc{\ket{35}^2\vev{25}\vev{13}}{\ket{14}} }\ ,\
(n_1,n_2)=(0,0)\ ,\ri.\\[5mm]
&\lf(\ds{\fc{\vev{25}^2\ket{25}\ket{13}}{\vev{14}}
  \atop\fc{\ket{34}^2\vev{25}\vev{13}}{\ket{14}} }\ ,\
(n_1,n_2)=(0,1)\ ,\ri.\\[5mm]
&\lf(\ds{\fc{\vev{23}^2\ket{25}\ket{13}}{\vev{14}}
  \atop\fc{\ket{45}^2\vev{25}\vev{13}}{\ket{14}} }\ ,\
(n_1,n_2)=(1,0)\ .\ri.\ea\ri.}\label{limit9}
\eea
The limits \req{limit6} and \req{limit9} describe at leading order an
exchange of a gluon coupled to the string amplitude
involving two gluons and two quarks, c.f. left diagram in Figure \ref{new4}.
\noindent
In the limit $y\ra 1$ Eq. \req{limit6} reduces to:
\bea
\ds{\lf.\ba{c}
{\cal M}\lf(\lf. {g_1^+\atop  g_1^-}\ri.\ri.\\
{\cal M}'\lf(\lf. {g_1^+\atop  g_1^-}\ri.\ri.\\
{\cal M}''\lf(\lf. {g_1^+\atop  g_1^-}\ri.\ri.
\ea\ri\}}&\stackrel{x\ra 0,\ y\ra 1}{\lra}&\ds{\sqrt 2\ t(\rho)\
\fc{g_{D6_a}^3}{s_{4}-\fc{n_1}{2}}\  \fc{1}{s_{2}-\fc{n_1}{2}}\ \times
\lf\{\hskip-0.25cm\ba{rl}
&\lf(\ds{-\fc{\vev{24}^2\ket{15}\ket{23}}{\vev{14}}
  \atop-\fc{\ket{35}^2\vev{15}\vev{23}}{\ket{14}} }\ \ \ ,\ \ \
(n_1,n_2)=(0,0)\ ,\ri.\\[5mm]
&\lf(\ds{\fc{\vev{25}^2\ket{15}\ket{23}}{\vev{14}}
  \atop\fc{\ket{34}^2\vev{15}\vev{23}}{\ket{14}} }\ \ \ ,\ \ \
(n_1,n_2)=(0,1)\ ,\ri.\\[5mm]
&\lf(\ds{\fc{\vev{23}^2\ket{15}\ket{23}}{\vev{14}}
  \atop\fc{\ket{45}^2\vev{15}\vev{23}}{\ket{14}} }\ \ \ ,\ \ \
(n_1,n_2)=(1,0)\ .\ri.\ea\ri.}\label{Limit6}
\eea
On the other hand, the limit $y\ra 0$ of \req{limit6} has been already discussed after
Eq. \req{limit3}.

Finally, the limit $x\ra1,\ y\ra 1$ does not yield any massless exchange.
Again, this limit is most conveniently anticipated by first performing the transformation \req{TRANSF}, which yields for the functions:
\bea
\ds{G_1^{\rho_1}\lf[{n_1\atop n_2}\ri]}&=&\ds{\int_0^1 d\tilde x\ \tilde x^{s_2-s_4-s_5-\fc{n_2}{2}}\ (1-\tilde x)^{s_4-1-\fc{n_1}{2}}\ \int_0^1\ d\tilde y\ \tilde y^{s_2-1}\ (1-\tilde y)^{-s_1+s_3-s_5-\fc{n_1}{2}-\fc{n_2}{2}} }\\[3mm]
&\times&\ds{ (1-\tilde x\tilde y)^{-s_2-s_3+s_5+n_1+n_2}\
 I_{\rho_1}\lf(\fc{\tilde y\ (1-\tilde x)}{1-\tilde x\tilde y}\ri)\ ,}\\[5mm]
\ds{G_2^{\rho_1}\lf[{n_1\atop n_2}\ri]}&=&\ds{\int_0^1 d\tilde x\ \tilde x^{s_2-s_4-s_5-\fc{n_2}{2}}\ (1-\tilde x)^{s_4-1-\fc{n_1}{2}}\ \int_0^1\ d\tilde y\ \tilde y^{s_2-1}\ (1-\tilde y)^{-s_1+s_3-s_5-1-\fc{n_1}{2}-\fc{n_2}{2}} }\\[3mm]
&\times&\ds{ (1-\tilde x\tilde y)^{-s_2-s_3+s_5+n_1+n_2}\
 I_{\rho_1}\lf(\fc{\tilde y\ (1-\tilde x)}{1-\tilde x\tilde y}\ri)\ .}
\label{newqd5}
\eea
In this form the limit arises for $\tilde x\ra0$, which does not furnish
massless exchanges in \req{FINALgqqqq}.

To summarize, for the ordering $\rho_1=(2,3,1,4,5)$ the three amplitudes \req{FINALgqqqq1}
have the following low-energy limits:
$$\ba{lcr}
\ds{{\cal M}_{\rho_1}\lf(\lf. {g_1^+\atop  g_1^-}\ri\},q_2^-,\bar q_3^+,q_4^-,\bar q_5^+\ri)}
&=&\ds{-\sqrt 2\ g_{D6_a}\ (T^{a_1})^{\al_4}_{\al_3}
\delta^{\al_2}_{\al_5}\delta_{\bet_2}^{\bet_3}\delta_{\bet_4}^{\beta_5}\ \times\lf\{
\ds{ \fc{\vev{24}^2}{\vev{13}\vev{14}}\ \lf(\ g_{D6_a}^2\
  \fc{\vev{34}}{\vev{23}\vev{45}}+g_{D6_b}^2\ \fc{1}{\vev{25}}\ \ri)\atop
  \fc{\ket{35}^2}{\ket{13}\ket{14}}\ \lf(\ g_{D6_a}^2\
\fc{\ket{34}}{\ket{23}\ket{45}}+g_{D6_b}^2\ \fc{1}{\ket{25}}\ \ri)\ ,}\ri.}
\ea$$
\bea\label{FINAL2}
\ds{{\cal M}'_{\rho_1}\lf(\lf. {g_1^+\atop  g_1^-}\ri\},q_2^-,\bar q_3^+,q_4^+,\bar q_5^-\ri)}
&=&\ds{\sqrt 2\ g_{D6_a}\ (T^{a_1})^{\al_4}_{\al_3}
\delta^{\al_2}_{\al_5}\delta_{\bet_2}^{\bet_3}\delta_{\bet_4}^{\beta_5}\ \times \lf\{
\ds{ g_{D6_a}^2\
  \fc{\vev{25}^2\vev{34}}{\vev{13}\vev{14}\vev{23}\vev{45}}\atop
  g_{D6_a}^2\
  \fc{\ket{34}^3}{\ket{13}\ket{14}\ket{23}\ket{45}}\ ,}\ri.}\\[8mm]
\ds{{\cal M}''_{\rho_1}\lf(\lf. {g_1^+\atop  g_1^-}\ri\},q_2^-,\bar q_3^-,q_4^+,\bar q_5^+\ri)}
&=&\ds{\sqrt 2\ g_{D6_a}\ (T^{a_1})^{\al_4}_{\al_3}
\delta^{\al_2}_{\al_5}\delta_{\bet_2}^{\bet_3}\delta_{\bet_4}^{\beta_5}\ \times\lf\{
\ds{ g_{D6_b}^2\
  \fc{\vev{23}^2}{\vev{13}\vev{14}\vev{25}}\atop
  g_{D6_b}^2\
  \fc{\ket{45}^2}{\ket{13}\ket{14}\ket{25}}\ .}\ri.}
\eea

\ \\
\noindent
\underline{$(1,2,5,4,3)$}

\noindent
As explained after Eqs. \req{FINALgqqqq} and  \req{REGG1} the amplitude for the ordering $\rho_2=(1,2,5,4,3)$ can by be obtained
from the ordering $\rho_1=(1,2,3,4,5)$ by successive exchanging the labels $3$ and $5$ and multiplying by an overall minus sign.
Furthermore, the helicity configurations in the amplitudes $\Mc'_{\rho_2}$ and $\Mc''_{\rho_2}$ are obtained by considering $\Mc''_{\rho_1}$ and $\Mc'_{\rho_1}$, respectively.
This way we obtain the low--energy expansion of the amplitudes \req{FINALgqqqq2} from the expressions \req{FINAL1}:
\bea\label{FINAL3}
\ds{{\cal M}_{\rho_2}\lf(\lf.
{g_1^+\atop  g_1^-}\ri\},q_2^-,\bar q_3^+,q_4^-,\bar q_5^+\ri)}
&=&\ds{\sqrt 2\ g_{D6_a}\ (T^{a_1})^{\al_2}_{\al_3}
\delta^{\al_4}_{\al_5}\delta_{\bet_2}^{\bet_5}\delta_{\bet_4}^{\beta_3}\ \times \lf\{
\ds{ \fc{\vev{24}^2}{\vev{12}\vev{13}}\ \lf(\ g_{D6_a}^2\
  \fc{\vev{23}}{\vev{25}\vev{34}}+g_{D6_b}^2\ \fc{1}{\vev{45}}\ \ri)\atop
  \fc{\ket{35}^2}{\ket{12}\ket{13}}\ \lf(\ g_{D6_a}^2\
\fc{\ket{23}}{\ket{25}\ket{34}}+g_{D6_b}^2\
\fc{1}{\ket{45}}\ \ri)\ ,}\ri.}\\[8mm]
\ds{{\cal M}'_{\rho_2}\lf(\lf. {g_1^+\atop  g_1^-}\ri\},q_2^-,\bar q_3^+,q_4^+,\bar q_5^-\ri)}
&=&\ds{-\sqrt 2\ g_{D6_a}\ (T^{a_1})^{\al_2}_{\al_3}
\delta^{\al_4}_{\al_5}\delta_{\bet_2}^{\bet_5}\delta_{\bet_4}^{\beta_3}\ \times \lf\{
\ds{ g_{D6_b}^2\
  \fc{\vev{25}^2}{\vev{12}\vev{13}\vev{45}}\atop
  g_{D6_b}^2\
  \fc{\ket{34}^2}{\ket{12}\ket{13}\ket{45}}\ ,}\ri.}\\[8mm]
\ds{{\cal M}''_{\rho_2}\lf(\lf. {g_1^+\atop  g_1^-}\ri\},q_2^-,\bar q_3^-,q_4^+,\bar q_5^+\ri)}
&=&\ds{-\sqrt 2\ g_{D6_a}\ (T^{a_1})^{\al_2}_{\al_3}
\delta^{\al_4}_{\al_5}\delta_{\bet_2}^{\bet_5}\delta_{\bet_4}^{\beta_3}\ \times \lf\{
\ds{ g_{D6_a}^2\
  \fc{\vev{23}^3}{\vev{12}\vev{13}\vev{25}\vev{34}}\atop
  g_{D6_a}^2\
  \fc{\ket{45}^2\ket{23}}{\ket{12}\ket{13}\ket{25}\ket{34}}\ .}\ri.}\\[8mm]
\eea

\ \\
\noindent
\underline{$(2,5,1,4,3)$}

\noindent
Similarly, the amplitude for the ordering $\rho_2=(2,5,1,4,3)$ can by be obtained
from the ordering $\rho_1=(2,3,1,4,5)$ by successive exchanging the labels $3$ and $5$ and multiplying by an overall minus sign. Furthermore, the helicity configurations in the amplitudes $\Mc'_{\rho_2}$ and $\Mc''_{\rho_2}$ are obtained by considering $\Mc''_{\rho_1}$ and $\Mc'_{\rho_1}$, respectively.
This way we obtain the low--energy expansion of the amplitudes \req{FINALgqqqq2} from the expressions \req{FINAL2}:
\bea\label{FINAL4}
\ds{{\cal M}_{\rho_2}\lf(\lf. {g_1^+\atop  g_1^-}\ri\},q_2^-,\bar q_3^+,q_4^-,\bar q_5^+\ri)}
&=&\ds{\sqrt 2\ g_{D6_a}\ (T^{a_1})^{\al_4}_{\al_5}
\delta^{\al_2}_{\al_3}\delta_{\bet_2}^{\bet_5}\delta_{\bet_4}^{\beta_3}\ \times\lf\{
\ds{ \fc{\vev{24}^2}{\vev{15}\vev{14}}\ \lf(\ g_{D6_a}^2\
  \fc{\vev{45}}{\vev{25}\vev{34}}+g_{D6_b}^2\ \fc{1}{\vev{23}}\ \ri)\atop
  \fc{\ket{35}^2}{\ket{15}\ket{14}}\ \lf(\ g_{D6_a}^2\
\fc{\ket{45}}{\ket{25}\ket{34}}+g_{D6_b}^2\ \fc{1}{\ket{23}}\ \ri)\ ,}\ri.}\\[8mm]
\ds{{\cal M}'_{\rho_2}\lf(\lf. {g_1^+\atop  g_1^-}\ri\},q_2^-,\bar q_3^+,q_4^+,\bar q_5^-\ri)}
&=&\ds{-\sqrt 2\ g_{D6_a}\ (T^{a_1})^{\al_4}_{\al_5}
\delta^{\al_2}_{\al_3}\delta_{\bet_2}^{\bet_5}\delta_{\bet_4}^{\beta_3}\ \times\lf\{
\ds{ g_{D6_b}^2\
  \fc{\vev{25}^2}{\vev{15}\vev{14}\vev{23}}\atop
  g_{D6_b}^2\
  \fc{\ket{34}^2}{\ket{15}\ket{14}\ket{23}}\ ,}\ri.}\\[8mm]
\ds{{\cal M}''_{\rho_2}\lf(\lf. {g_1^+\atop  g_1^-}\ri\},q_2^-,\bar q_3^-,q_4^+,\bar q_5^+\ri)}
&=&\ds{-\sqrt 2\ g_{D6_a}\ (T^{a_1})^{\al_4}_{\al_5}
\delta^{\al_2}_{\al_3}\delta_{\bet_2}^{\bet_5}\delta_{\bet_4}^{\beta_3}\ \times \lf\{
\ds{ g_{D6_a}^2\
  \fc{\vev{23}^2\vev{45}}{\vev{15}\vev{14}\vev{25}\vev{34}}\atop
  g_{D6_a}^2\
  \fc{\ket{45}^3}{\ket{15}\ket{14}\ket{25}\ket{34}}\ .}\ri.}\\[8mm]
  \eea

\break
%\vskip0.5cm
\noindent
\underline{\sl (ii) Four-fermion amplitudes involving two twist--antitwist pairs
$(\th,1-\th)$ and $(\nu,1-\nu)$}

\begin{figure}[H]
\centering
\includegraphics[width=0.69\textwidth]{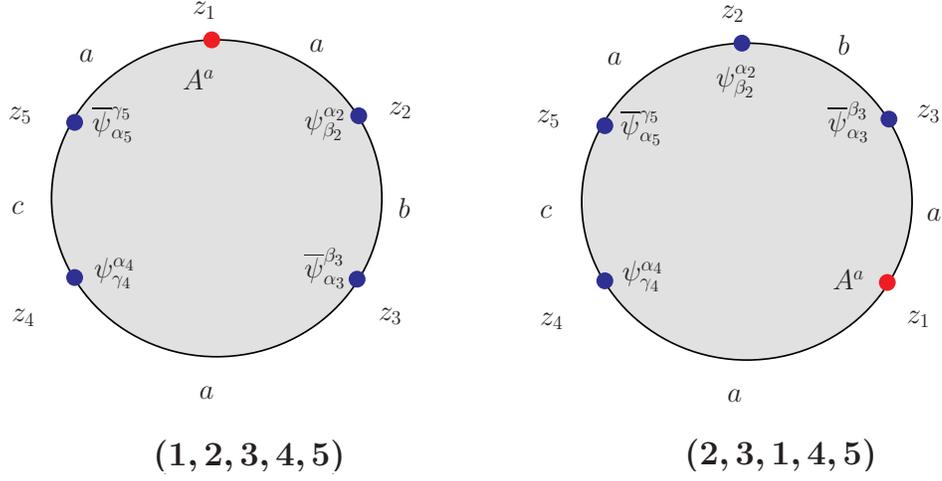}
\caption{Orderings $\rho_1$ of the gauge boson and four fermion vertex operators.}
\label{gqqqq1}
\end{figure}

\vskip0.5cm
\noindent
Here we consider the generic case of three different stacks of D$6$--branes
$a,b,c$ and $d$ with two pairs of twist--antitwist fields,
$(\th,1-\th)$ and $(\nu,1-\nu)$. We have two
intersection angles $\th$ and $\nu$ referring to the intersection pairs $(a,b)$ and
$(c,a)$, respectively.
For this setup there are two possibilities to place a gauge boson with gauge group
$G_a$ within the four chiral fermions, c.f. the Figure \ref{gqqqq1}.
\noindent
These two orderings $\rho_1$ correspond to the first column of \req{allfour}.

The functions $I_\rho=I_1I_{2\rho}$, encoding the quantum and instanton part of the correlator \req{4f2}, 
given by
\bea\label{givenby2}
\ds{I_1(x)}&=&\ds{(2\pi)^{3/2}\ \prod_{j=1}^3\lf[\ B(1-\th^j,\nu^j)\ F_{1;j}(1-x)\ K_{2;j}(x)+B(\th^j,1-\nu^j)\
F_{2;j}(1-x)\ K_{1;j}(x)\ \ri]^{-1/2},}\\[5mm]
\ds{I_{2\rho_1}(x)}&=&\ds{\sum_{p_a,p_b\in\ZZ^6}
e^{-\fc{\pi}{\ap}\ \sum\limits_{j=1}^3\sin(\pi\theta^j)\
\lf\{\ \lf|v^j_b\ri|^2\ \lf(\tau_j(x)+\fc{\beta^j}{2}\ri)+
\lf|v_a^j+\fc{\beta^j}{2}\ v_b^j\ri|^2\
\fc{1}{\tau_j(x)+\fc{\beta^j}{2}}\ \ri\}}\ \ \ ,\ \ \
I_{2\rho_2}(x)=0\ .}
\eea
with $v_r=p_rL_r+\delta_r,\ \delta_c^j,\delta_b^j=0$ and
$\beta^j=-\fc{\sin[\pi(\th^j-\nu^j)]}{\sin(\pi\nu^j)}$ are  discussed in \cite{LHC}. Above we have introduced
the hypergeometric functions $K_{1;j}(x)=\FF{2}{1}[1-\th^j,\nu^j,1;x]$,\
$K_{2;j}(x)=\FF{2}{1}[\th^j,1-\nu^j,1;x]$,\
$F_{1;j}(x)=\FF{2}{1}[1-\th^j,\nu^j;1-\th^j+\nu^j;x]$, and
$F_{2;j}(x)=\FF{2}{1}[1-\nu^j,\th^j,1+\th-\nu^j;x],$
and the Euler Beta function \req{Euler}.
Finally, we have:
\beq\label{TAU}
\tau_j(x)=\fc{1}{\pi}\ B(1-\th^j,\nu^j)\ \sin(\pi\th^j)\
\fc{F_{1;j}(1-x)}{K_{1;j}(x)}\ .
\eeq

\break 
%  \ \\

\noindent
\underline{$(1,2,3,4,5)$}

\noindent
As discussed in \cite{LHC} the underlying four--fermion amplitudes
share only  massless gauge boson exchange in their $s$--channel.
As a consequence the three amplitudes
\req{4f6}, \req{4f7a} and \req{4f7b} give only rise to
a massless gauge boson exchange in their $s_2$-- and $s_4$--channels.
Hence from the two diagrams in Figure \ref{new1} only the $s_2$-- and $s_4$--channels are realized, while the diagram in Figure \ref{new2} contributes in full.
Therefore, to extract the  factorization and pole structure of the set of amplitudes
\req{FINALgqqqq1} for the  ordering $\rho_1=(1,2,3,4,5)$ we may simply borrow the result \req{FINAL1} and erase the terms proportional to $g_{D6_b}^2$.
As a result for the ordering $\rho_1$ the three amplitudes \req{4f6}, \req{4f7a} and \req{4f7b}
have the following low--energy limits:
\bea\label{FINAL1a}
\ds{{\cal M}_{\rho_1}\lf(\lf. {g_1^+\atop  g_1^-}\ri\},q_2^-,\bar q_3^+,q_4^-,\bar q_5^+\ri)}
&=&\ds{\sqrt 2\ g_{D6_a}^3\ (T^{a_1})^{\al_2}_{\al_5}
\delta^{\al_4}_{\al_3}\delta_{\bet_2}^{\bet_3}\delta_{\gamma_4}^{\gamma_5}\ \times
\lf\{\ds{ \fc{\vev{24}^2}{\vev{12}\vev{15}}\
  \fc{\vev{25}}{\vev{23}\vev{45}}\atop
  \fc{\ket{35}^2}{\ket{12}\ket{15}} \fc{\ket{25}}{\ket{23}\ket{45}}\ ,}\ri.}\\[8mm]
\ds{{\cal M}'_{\rho_1}\lf(\lf. {g_1^+\atop  g_1^-}\ri\},q_2^-,\bar q_3^+,q_4^+,\bar q_5^-\ri)}
&=&\ds{-\sqrt 2\ g_{D6_a}^3\ (T^{a_1})^{\al_2}_{\al_5}
\delta^{\al_4}_{\al_3}\delta_{\bet_2}^{\bet_3}\delta_{\gamma_4}^{\gamma_5}\ \times
\lf\{\ds{\fc{\vev{25}^3}{\vev{12}\vev{15}\vev{23}\vev{45}}\atop
   \fc{\ket{34}^2\ket{25}}{\ket{12}\ket{15}\ket{23}\ket{45}}\ ,}\ri.}\\[8mm]
\ds{{\cal M}''_{\rho_1}\lf(\lf. {g_1^+\atop  g_1^-}\ri\},q_2^-,\bar q_3^-,q_4^+,\bar q_5^+\ri)}&=&\ds{0\ .}
\eea

\ \\
\noindent
\underline{$(2,3,1,4,5)$}

\noindent
Again, for this ordering the three amplitudes
\req{4f6}, \req{4f7a} and \req{4f7b} give rise to
a massless gauge boson exchange only in their $s_2$-- and $s_4$--channels.
Hence from the  diagram in Figure \ref{new3} only the  $s_4$--channel is realized.
On the other hand, the left diagram of Figure \ref{new4} contributes in full, while
its right diagram does not give rise to a massless gauge boson exchange.
Therefore, to extract the  factorization and pole structure of the set of amplitudes
\req{FINALgqqqq2} for the  ordering $\rho_1=(2,3,1,4,5)$ we may simply borrow the result \req{FINAL2} and erase the terms proportional to $g_{D6_b}^2$.
As a result for the ordering $\rho_1$ the three amplitudes \req{4f6}, \req{4f7a} and \req{4f7b}
have the following low-energy limits:
\bea\label{FINAL2a}
\ds{{\cal M}_{\rho_1}\lf(\lf. {g_1^+\atop  g_1^-}\ri\},q_2^-,\bar q_3^+,q_4^-,\bar q_5^+\ri)}
&=&\ds{-\sqrt 2\ g_{D6_a}^3\ (T^{a_1})^{\al_4}_{\al_3}
\delta^{\al_2}_{\al_5}\delta_{\bet_2}^{\bet_3}\delta_{\gamma_4}^{\gamma_5}\ \times
\lf\{\ds{ \fc{\vev{24}^2}{\vev{13}\vev{14}}\ \fc{\vev{34}}{\vev{23}\vev{45}}\atop
  \fc{\ket{35}^2}{\ket{13}\ket{14}}\
\fc{\ket{34}}{\ket{23}\ket{45}}\ ,}\ri.}\\[8mm]
\ds{{\cal M}'_{\rho_1}\lf(\lf. {g_1^+\atop  g_1^-}\ri\},q_2^-,\bar q_3^+,q_4^+,\bar q_5^-\ri)}
&=&\ds{\sqrt 2\ g_{D6_a}^3\ (T^{a_1})^{\al_4}_{\al_3}
\delta^{\al_2}_{\al_5}\delta_{\bet_2}^{\bet_3}\delta_{\gamma_4}^{\gamma_5}\  \times
\lf\{\ds{  \fc{\vev{25}^2\vev{34}}{\vev{13}\vev{14}\vev{23}\vev{45}}\atop
    \fc{\ket{34}^3}{\ket{13}\ket{14}\ket{23}\ket{45}}\ ,}\ri.}\\[8mm]
\ds{{\cal M}''_{\rho_1}\lf(\lf. {g_1^+\atop  g_1^-}\ri\},q_2^-,\bar q_3^-,q_4^+,\bar q_5^+\ri)}&=&\ds{0\ .}
\eea

%%%%%%%%%%%%%%%%%%%%%%%%%%%%%%%%%%%%%%%%%%%%%%%%%%%%%%%%%%%%%%%%

\break
% \vskip0.5cm
\noindent
\underline{\sl (iii) Four--fermion amplitudes involving fermions from four different
intersections}

\begin{figure}[H]
\centering
\includegraphics[width=0.35\textwidth]{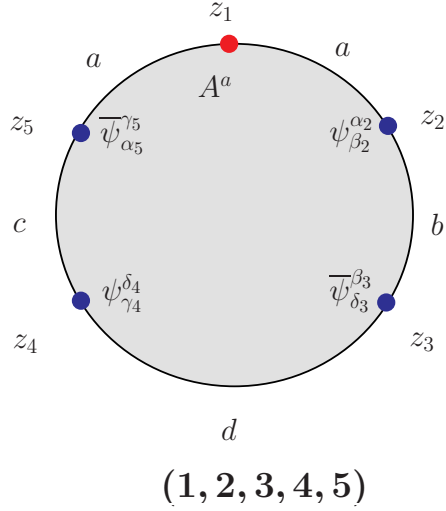}
\caption{Ordering of the gauge boson and four fermion vertex operators.}
\label{gqqqq2}
\end{figure}

\vskip0.5cm
\noindent
The most general case involves four fermions from four
different intersections $f_i$ with the four angles
$\th_i$ and $\th_4=2-\th_1-\th_2-\th_3$.
This setup allows for only one possibility to place the gauge boson with gauge group
$G_a$ within the four chiral fermions, c.f. the Figure \ref{gqqqq2}.
\noindent
Hence, only the first entry $\rho_1=(1,2,3,4,5)$ of the list of orderings \req{allfour} is realized giving rise to the group trace
$$t(\rho_1)=(T^{a_1})^{\al_2}_{\al_5}\ \delta_{\bet_2}^{\bet_3}\ \delta_{\gamma_4}^{\gamma_5}\ \delta^{\delta_4}_{\delta_3}$$
in the amplitudes \req{FINALgqqqq}.

The functions $I_\rho=I_1I_{2\rho}$, encoding the quantum and instanton part of the correlator \req{4f2},
given by
\bea\label{givenby3}
\ds{I_1(x)}&=&\ds{(2\pi)^{3/2}\prod_{j=1}^3\lf[\fc{\Gamma(1-\th^j_1)\ \Gamma(\th_3^j)\ F_{1;j}(1-x)\ K_{2;j}(x)}{\Gamma(\th_2^j+\th_3^j)\
\Gamma(\th_3^j+\th_4^j)}+\fc{\Gamma(\th^j_1)\
\Gamma(1-\th_3^j)\ F_{2;j}(1-x)\ K_{1;j}(x)}{\Gamma(\th_1^j+\th_2^j)\ \Gamma(\th_1^j+\th_4^j)}\ri]^{-1/2},}\\[5mm]
\ds{I_{2\rho_1}(x)}&=&\ds{\sum_{p_a,p_b\in\ZZ^6}
e^{-\fc{\pi}{\ap}\ \sum\limits_{j=1}^3\sin(\pi\theta_2^j)\
\fc{\lf|v^j_b\ \tau_j-v^j_d\ri|^2+\gamma^j\tilde\gamma^j\
\lf|v^j_b\ (\beta^j+\tau_j)+v_d^j\ (1+\alpha^j\ \tau_j)\ri|^2}
{\beta^j+2\ \tau_j+\alpha^j\ \tau_j^2}}\ \ \ ,\ \ \ I_{2\rho_2}(x)=0\ ,}
\eea
with $v_r=p_rL_r+\delta_r,\ \delta_c^j,\delta_b^j=0$  are  discussed in \cite{LHC}. Above we have introduced
the hypergeometric functions
$K_{1;j}(x)=\FF{2}{1}[\th_2^j,1-\th_4^j,\th_1^j+\th_2^j;x]$,\
$K_{2;j}(x)=\FF{2}{1}[\th_4^j,1-\th_2^j,\th_3^j+\th_4^j;x]$,\
$F_{1;j}(x)=\FF{2}{1}[\th_2^j,1-\th_4^j;\th^j_2+\th^j_3;x]$, and
$F_{2;j}(x)=\FF{2}{1}[\th_4^j,1-\th_2^j,\th^j_1+\th^j_4;x],$\
$\beta^j=-\fc{\sin[\pi(\th_2^j+\th_3^j)]}{\sin(\pi\th^j_3)},\ \alpha^j=
-\fc{\sin[\pi(\th_1^j+\th_2^j)]}{\sin(\pi\th^j_1)}$ and
$\gamma^j=\fc{\Gamma(1-\th^j_2)\ \Gamma(1-\th^j_4)}{\Gamma(\th^j_1)\
\Gamma(\th^j_3)},\ \tilde\gamma^j=\fc{\Gamma(\th^j_2)\ \Gamma(\th^j_4)}
{\Gamma(1-\th^j_1)\ \Gamma(1-\th^j_3)}$.
Finally, we have:
\beq\label{TTAU}
\tau_j(x)=\fc{B(\th_2^j,\th_3^j)}{B(\th_1^j,\th_2^j)}\
\fc{F_{1;j}(1-x)}{K_{1;j}(x)}\ .
\eeq

As discussed in \cite{LHC} the underlying four--fermion amplitude
does not allow for massless gauge boson exchange in any channel.
As a consequence the three amplitudes
\req{4f6}, \req{4f7a} and \req{4f7b} do not have any massless
exchanges in their field--theory limits.

\section{Relations between gaugino $\bm{g\chi\bar\chi\chi\bar\chi}$ and quark $\bm{gq\bar q q\bar q}$ partial amplitudes}
\label{newsec}

We continue the discussion of Section \ref{SUPREL} to  relations between subamplitudes  involving one vector and four fermions.
It has been proven in \cite{stieberg}, that the subamplitude relation \req{DUALv}
and permutations thereof hold for open strings from both
Neveu--Schwarz and Ramond sectors, i.e. bosons and fermions.

{}From amplitudes involving chiral matter fermions (or scalars) we may extract those
amplitudes with the chiral fermions $\psi$ replaced by gauginos $\chi$,
c.f. Section \ref{RELbetween}.
More precisely, for amplitudes with chiral fermions represented by open strings stretched between
the two intersecting stacks $a$ and $b$ we simply obtain the corresponding
amplitudes with gauginos associated to stack $a$ by performing first a
Poisson resummation w.r.t. the momenta $p_a$ on the instanton
factor $I_{2\rho}$ and then taking the limit $\th^j\ra0$, i.e. altogether
substituting
\beq\label{substi}
C\, I_{\rho}\ \lra\  g_{D6_a}^2
\eeq
in the $gq\bar q q\bar q$ amplitudes. This simply translates to the identification $g_{D6_a}^2=g_{D6_b}^2$ in the low--energy expansions of the $gq\bar q q\bar q$ amplitudes.
The full expressions $\Mc_\rho, \Mc'_\rho$ and $\Mc''_\rho$  of the latter are computed in the previous Section \ref{GQQQQ} and  given in  Eqs. \req{FINALgqqqq1} and \req{FINALgqqqq2}  subject to the four orderings \req{allfour}.
From them by performing the substitution \req{substi} we may extract the following set of
$g\chi\bar\chi\chi\bar\chi$ amplitudes
\bea\label{SETSi}
{\cal M}_\rho(g_1^+,\chi_2^-,\bar \chi_3^+,\chi_4^-,\bar \chi_5^+)&=&t(\rho)\
A(1_\rho,2_\rho,3_\rho,4_\rho,5_\rho)\ ,\\
{\cal M}'_\rho(g_1^+,\chi_2^-,\bar \chi_3^+,\chi_4^+,\bar \chi_5^-)&=&t(\rho)\
A'(1_\rho,2_\rho,3_\rho,4_\rho,5_\rho)\ ,\\
{\cal M}''_\rho(g_1^+,\chi_2^-,\bar \chi_3^-,\chi_4^+,\bar \chi_5^+)&=&t(\rho)\
A''(1_\rho,2_\rho,3_\rho,4_\rho,5_\rho)
\eea
for the four orderings  \req{allfour}. E.g. from \req{FINALgqqqq1} we obtain:
\bea
\ds{A(1,2,3,4,5)}&=&\ds{
\sqrt{2}\ g_{D6_a}^3\  \fc{\vev{24}^2}{\vev{14}}\, \lf(\, \ket{12}\ket{53}\
G_1\lf[{0\atop 0}\ri]+\ket{13}\ket{25}\ G_2\lf[{0\atop 0}\ri]\ \ri)\ ,}\\[3mm]
\ds{A'(1,2,3,4,5)}&=&\ds{
-\sqrt{2}\ g_{D6_a}^3\ \fc{\vev{25}^2}{\vev{14}}\ \lf(\, \ket{12}\ket{53}\
G_1\lf[{0\atop 1}\ri]+\ket{13}\ket{25}\ G_2\lf[{0\atop 1}\ri]\ \ri) \ ,}\\[3mm]
\ds{A''(1,2,3,4,5)}&=&\ds{
-\sqrt{2}\ g_{D6_a}^3\ \fc{\vev{23}^2}{\vev{14}}\ \lf(\,\ket{12}\ket{53}\
G_1\lf[{1\atop 0}\ri]+\ket{13}\ket{25}\ G_2\lf[{1\atop0}\ri] \ri) \ ,}
\eea
with:
\beq\label{newq4}
\lf.{G_1\lf[{n_1\atop n_2}\ri]\atop G_2\lf[{n_1\atop n_2}\ri]}\ri\}=
\int_0^1 \hskip-0.2cm dx\ x^{s_{23}-1-\fc{n_1}{2}}\ (1-x)^{s_{34}-1-\fc{n_2}{2}}
\int_0^1 \hskip-0.2cm dy\ y^{s_{51}-1}\ (1-y)^{s_{45}-1-\fc{n_1}{2}}\ (1-xy)^{s_{35}+\fc{n_1}{2}+\fc{n_2}{2}}\ \times\lf\{{1\atop (1-xy)\ .}\ri.
\eeq
Similar expressions are obtained for the other three orderings $\rho$ of \req{allfour} and the
other gluon polarization~$g_1^-$.

For each helicity configuration the identity Eq. \req{DUALv} gives rise to one relation between  these four orderings.
For the first set $A(1_\rho,2_\rho,3_\rho,4_\rho,5_\rho)$ of four $g\chi\bar\chi\chi\bar\chi$ amplitudes \req{SETSi}  this identity translates into the relation
\beq\label{translate}
A(1,2,3,4,5)-e^{\pi i s_{12}}\ A(1,2,5,4,3)+e^{\pi i(s_{45}-s_{23})}\ A(2,3,1,4,5)-
e^{-\pi i s_{15}}\ A(2,5,1,4,3)=0
\eeq
among the four partial subamplitudes introduced in \req{SETSi}, respectively.
Identical relations follow for the set of partial amplitudes $A'(1_\rho,2_\rho,3_\rho,4_\rho,5_\rho)$ and $A''(1_\rho,2_\rho,3_\rho,4_\rho,5_\rho)$ describing the other two helicity configurations.
Note, that these relations hold for all three sets of
helicity amplitudes, independently.
Furthermore, these equations hold for each kinematic factor
$\Pc_i,\Qc_i$ and $\Rc_i$, separately.
It is straightforward to prove these identities for the lowest order expressions
\req{FINAL1}, \req{FINAL2}, \req{FINAL3} and \req{FINAL4} subject to the identification $g_{D6_a}^2=g_{D6_b}^2$.
On the other hand, to prove \req{translate} for
the full fledged amplitudes \req{FINALgqqqq} and \req{FINALgqqqqs} (adapted to
$g\chi\bar\chi\chi\bar\chi$) we use
the explict integral formula \req{2,7} and use some hypergeometric function
identities.
For the set of amplitudes $A(1_\rho,2_\rho,3_\rho,4_\rho,5_\rho)$ the identity \req{translate} gives rise to two real equations, which
allow to express all four subamplitudes in terms of two
\bea\label{ALLOW}
\ds{A(1,2,5,4,3)}&=&\ds{\sin\pi(s_1+s_5)^{-1}\ \lf[\ \sin\pi s_5\ A(1,2,3,4,5)+
\sin\pi (-s_2+s_4+s_5)\ A(2,3,1,4,5)\ \ri]\ ,}\\
\ds{A(2,5,1,4,3)}&=&\ds{\sin\pi(s_1+s_5)^{-1}\ \lf[\ \sin\pi s_1\ A(1,2,3,4,5)+
\sin\pi (s_1+s_2-s_4)\ A(2,3,1,4,5)\ \ri]\ ,}
\eea
and equivalent expressions for the set of partial amplitudes $A'$ and $A''$.

Let us now discuss the set of partial $gq\bar qq\bar q$ amplitudes
\bea\label{SETS}
{\cal M}_\rho(g_1,q_2^-,\bar q_3^+,q_4^-,\bar q_5^+)&=&t(\rho)\
A(1_\rho,2_\rho,3_\rho,4_\rho,5_\rho)\ ,\\
{\cal M}'_\rho(g_1,q_2^-,\bar q_3^+,q_4^+,\bar q_5^-)&=&t(\rho)\
A'(1_\rho,2_\rho,3_\rho,4_\rho,5_\rho)\ ,\\
{\cal M}''_\rho(g_1,q_2^-,\bar q_3^-,q_4^+,\bar q_5^+)&=&t(\rho)\
A''(1_\rho,2_\rho,3_\rho,4_\rho,5_\rho)\ ,
\eea
given in Eqs. \req{FINALgqqqq1} and \req{FINALgqqqq2} for the four orderings  \req{allfour} and study  subamplitude relations for them.
In the case of amplitudes with
more than two chiral fermions (or scalars)  the relation \req{DUALv} has to be handled with some care. We first clarify its meaning.
In deriving the relation \req{DUALv}, the monodromy properties of contours in the complex plane of the vertex position $z_1$ are studied \cite{stieberg}. If this position $z_1$
refers to the gluon position the instanton part of the $gq\bar qq\bar q$ amplitudes has no influence on the monodromy and Eq. \req{DUALv} stays valid for these amplitudes.  However, for  $gq\bar qq\bar q$ amplitudes in Eq. \req{DUALv} the two amplitudes $A(2,1,3,4,5)$ and $A(2,3,4,1,5)$ refer to the case, where the color quantum numbers $\al_i,\bet_i$,
i.e. the open string ends of the four fermions are permuted w.r.t. to the other two subamplitudes $A(1,2,3,4,5)$ and $A(2,3,1,4,5)$. Note, that this manipulation has no effect on the contour integral in the complex $z_1$--plane and only affects the instanton factor.
On the other hand, this makes sure, that in all four subamplitudes the same
instanton factor $I_{2\rho_1}(\{z_i\};\th^j)$ referring to the ordering
$z_3<z_5$ appears.
A fact, which is also necessary to prove \req{DUALv} for this type of amplitudes.
To summarize, the relation~\req{DUALv} takes over to $gq\bar qq\bar q$ subamplitudes \req{SETS}
\beq\label{translate1}
A(1,2,3,4,5)+e^{\pi i s_{12}}\ \tilde A(2,1,3,4,5)+e^{\pi i(s_{45}-s_{23})}\  A(2,3,1,4,5)+
e^{-\pi i s_{15}}\ \tilde A(2,3,4,1,5)=0\ ,
\eeq
with the two amplitudes $\tilde A(2,1,3,4,5)$ and $\tilde A(2,3,4,1,5)$ subject to an exchange of the color indices of the open string ends. 
Identical relations follow for the partial amplitudes $A'$ and $A''$ describing the other two helicity configurations.
It is straightforward to prove these identities  for the lowest order expressions \req{FINAL1}, \req{FINAL2}, \req{FINAL3} and \req{FINAL4} after performing the replacement $g_{D6_a}^2\leftrightarrow g_{D6_b}^2$ in  the amplitudes
\req{FINAL3} and \req{FINAL4}, which corresponds to exchanging the color quantum numbers of the open string ends.

It is interesting to note, that the same relation \req{DUALv}, which holds for
the $ggggg$ subamplitudes, gives rise to the same identity for
the $g\chi\bar\chi\chi\bar\chi$ and $gq\bar qq\bar q$ subamplitudes. Moreover, it is shown in
Section~\ref{SUPREL} that these relations also reveal the structure of $gggq\bar q$ subamplitudes.

\section{From amplitudes to parton cross sections}

The purpose of this Section is to present the squared moduli of five-parton disk amplitudes, summed over helicities and colors of all partons, with all external momenta treated as those of outgoing particles. The squared amplitudes describing specific collision channels, averaged over helicities and colors of incident partons and summed over helicities and colors of the outgoing particles can be then simply obtained by dividing by the number of
all possible initial helicity/color configurations.
Also, if some of the partons shall be thought of as incident, their momenta need to be reversed, $k\to -k$, in order to obtain $2\to 3$ cross sections. In this Section, we use the standard mass dimension two kinematic invariants $s_{ij}$ and $s_i$, see Eqs. (\ref{sidim}) and (\ref{sinodim}), although the dimensionless variables  $\hat s_{ij}$ and $\hat s_i$ [normalized in string units, see Eq. (\ref{sijnodim})] enter implicitly as the arguments of dimensionless functions $V^{(5)}(\hat s_i)$ and  $P^{(5)}(\hat s_i)$ defined in Eqs. (\ref{vfac}) and (\ref{pfac}). Furthermore, in the preceding sections we implicitly used the metrics with $(---\,+)$ signature, so if one wants to revert to the standard $(+--\, -)$  (with e.g.\ Mandelstam's $s>0$), a change of sign is required for all scalar products. Finally, the QCD coupling constant $g\equiv g_{Dp_{a}}=g_{D6_{a}}$
while $g_b\equiv g_{Dp_{b}}=g_{D6_{b}}$ is the ``electroweak'' coupling constant associated to stack $b$.
\subsection{Five gluons}

The starting expression is (\ref{gmhv5}) describing one specific helicity configuration. We take its modulus squared, sum over all helicity configurations and color (adjoint) indices of five gluons:
\begin{equation}\label{square5}
\bigl| {\cal M}(g_1,g_2, g_3, g_4, g_5) \bigr|^2 \eq 64\, g^6 \, \left( \sum_{i<j} s_{ij}^4 \right) \, \sum_{\lambda,\lambda'\in \Pi_5}\,
{\cal C}_{\lambda} \, {\cal S}_{\lambda\lambda'} \, {\cal C}^*_{\lambda'}
\end{equation}
We have introduced shorthands for the twelve-vector of ${\cal C}$ permutations
\begin{equation}
{\cal C}_{\lambda} \ \ \equiv \ \  {\cal C}(k_{1_{\lambda}},k_{2_{\lambda}},k_{3_{\lambda}},k_{4_{\lambda}},k_{5_{\lambda}})
\end{equation}
and the color matrix
\begin{equation}\label{matr5}
{\cal S}_{\lambda\lambda'} \eq \sum_{a_1,\dots,a_5}t^{a_{1_{\lambda}}a_{2_{\lambda}}
a_{3_{\lambda}}a_{4_{\lambda}}
a_{5_{\lambda}}} \, \bigl(t^{a_{1_{\lambda'}}a_{2_{\lambda'}}a_{3_{\lambda'}}
a_{4_{\lambda'}}a_{5_{\lambda'}}} \bigr)^* \ .
\end{equation}
The matrix elements can be evaluated by using elementary methods. We will consider two cases separately:
with all five bosons associated to non-abelian gauge group generators and with
one or more bosons associated to a $U(1)$ subgroup of $U(N)$.
In the ``all-non-abelian'' case which, in particular, covers the case of five-gluon scattering in QCD,
the color matrix has the form written in Tables 3~and~4.
\goodbreak

\vskip1cm
\begin{tabular}{|c|c|c|c|c|c|c|c|c|c|c|c|c|}\hline
 & $\scriptstyle(1234)$ & $\scriptstyle(1243)$ & $\scriptstyle(1342)$
& $\scriptstyle(1324)$ & $\scriptstyle(1423)$ & $\scriptstyle(1432)$
& $\scriptstyle(2134)$ & $\scriptstyle(2143)$ & $\scriptstyle(2314)$
& $\scriptstyle(2413)$ & $\scriptstyle(3124)$ & $\scriptstyle(3214)$\\ \hline
$\scriptstyle(1234)$ & $D$ & ${+}X$ & ${+}Y$ & ${+}X$ & ${+}Y$ & ${-}X$ & ${+}X$ & ${-}Y$ & ${+}Y$ &
$0$ & ${+}Y$ & ${-}X$\\ \hline
$\scriptstyle(1243)$ &  ${+}X$  & $D$ & ${-}X$ & ${+}Y$ & ${+}X$ & ${+}Y$ & ${-}Y$ & ${+}X$ & $0$ &
${+}Y$ & ${+}X$ & ${-}Y$\\ \hline
$\scriptstyle(1342)$ &  ${+}Y$  & ${-}X$ & $D$&  ${+}X$ & ${+}Y$ & ${+}X$ & ${+}X$ & ${-}Y$ & ${-}Y$ &
${-}X$ & ${-}Y$ & $0$\\ \hline
$\scriptstyle(1324)$ &  ${+}X$  & ${+}Y$ & ${+}X$ & $D$ & ${-}X$ & ${+}Y$ & ${+}Y$ & $0$ & ${-}X$ &
${+}Y$ & ${+}X$ & ${+}Y$\\ \hline
$\scriptstyle(1423)$ &  ${+}Y$  & ${+}X$ & ${+}Y$ & ${-}X$ & $D$ & ${+}X$ & $0$ & ${+}Y$ & ${+}Y$ &
${-}X$ & ${-}Y$ & ${-}X$\\ \hline
$\scriptstyle(1432)$ &  ${-}X$  & ${+}Y$ & ${+}X$ & ${+}Y$ & ${+}X$ & $D$ & ${-}Y$ & ${+}X$ & ${-}X$ &
${-}Y$ & $0$ & ${+}Y$\\ \hline
$\scriptstyle(2134)$ &  ${+}X$  & ${-}Y$ & ${+}X$ & ${+}Y$ & $0$ & ${-}Y$ & $D$ & ${+}X$ & ${+}X$ &
${+}Y$ & ${-}X$ & ${+}Y$\\ \hline
$\scriptstyle(2143)$ &  ${-}Y$  & ${+}X$ & ${-}Y$ & $0$ & ${+}Y$ & ${+}X$ & ${+}X$ & $D$ & ${+}Y$ &
${+}X$ & ${-}Y$ & ${+}X$\\ \hline
$\scriptstyle(2314)$ &  ${+}Y$  & $0$ & ${-}Y$ & ${-}X$ & ${+}Y$ & ${-}X$ & ${+}X$ & ${+}Y$ & $D$ &
${-}X$ & ${+}Y$ & ${+}X$\\ \hline
$\scriptstyle(2413)$ &  $0$  & ${+}Y$ & ${-}X$ & ${+}Y$ & ${-}X$ & ${-}Y$ & ${+}Y$ & ${+}X$ &
${-}X$ & $D$ & ${-}X$ & ${-}Y$\\ \hline
$\scriptstyle(3124)$ &  ${+}Y$  & ${+}X$ & ${-}Y$ & ${+}X$ & ${-}Y$ & $0$ & ${-}X$ & ${-}Y$ &
${+}Y$ & ${-}X$ & $D$ & ${+}X$\\ \hline
$\scriptstyle(3214)$ &  ${-}X$  & ${-}Y$ & $0$ & ${+}Y$ & ${-}X$ & ${+}Y$ & ${+}Y$ & ${+}X$ &
${+}X$ & ${-}Y$ & ${+}X$ & $D$ \\ \hline
\end{tabular}
\begin{flushleft}{\bf Table 3}. \it Matrix elements ${\cal S}_{\lambda\lambda'}$. Since the elements  of the permutation set $\Pi_5$, see Eq. (\ref{p5set}), have the common last number equal 5, the rows and columns are labeled by first four numbers. The entries $D$, $X$ and $Y$ depend on the gauge group and are listed in Table 4. Note that the matrix is symmetric.\\[3mm]
\end{flushleft}

\vskip1cm
\begin{center}
\begin{tabular}{|c|c|c|c|c|c|c|c|}\hline Group & $C_A$ & $C_F$ & $N_A$  &$D/N_A$ &$X/N_A$ & $Y/N_A$ \\ \hline\hline $\displaystyle
SU(N)$ & $N$ & $\frac{N^{2}-1}{2N}$ &$N^{2}-1$  & $\frac{N^{4}-4N^{2}+10}{16N}$ & $\frac{2-N^{2}}{8N}$ & $\frac{1}{8N}$ \\ \hline
$SO(N)$ & $\frac{N-2}{2}$ & $\frac{N-1}{4}$ & $\frac{N(N-1)}{2}$  & $\frac{(N-2)(N^{2}-2N+2)}{128}$ & $\frac{(N-2)^{2}}{128}$ & $\frac{N-2}{64}$ \\ \hline
$Sp(N)$ & $\frac{N+2}{2}$ & $\frac{N+1}{4}$ & $\frac{N(N+1)}{2}$  & $\frac{(N+2)(N^{2}+2N+2)}{128}$ & $-\frac{(N+2)^{2}}{128}$ & $\frac{N+2}{64}$\\ \hline
\end{tabular}
\begin{flushleft}{\bf Table 4}. {\it The elements $D$, $X$, $Y$ (or 0) of ${\cal S}_{\lambda\lambda'}$ evaluated for semi-simple gauge groups generated by stacks of D-branes. For completeness, we also list the Casimir operators $C_A$ and $C_F$ for the adjoint and fundamental representations, respectively, and the dimension $N_A$ of the adjoint representation.}\\[3mm]
\end{flushleft}
\end{center}

\noindent
If one of particles, say number 1, is a $U(1)$ gauge boson, the corresponding generator is replaced by
\begin{equation}
T^{a_1} \eq Q \mathds{1}\ .
\end{equation}
Although for a $U(1)$ subgroup of $U(N)$, the canonical normalization  is $Q=1/\sqrt{2N}$,
we treat $Q$ as an arbitrary parameter. Note that four non-abelian gauge bosons are then necessarily $SU(N)$. The color factors (\ref{chanp5}) become
\begin{equation}
t^{a_1a_2a_3a_4a_5} \eq iQ \, \Big(f^{a_2 a_3 n} \, d^{a_4 a_5 n} \  + \  f^{a_4 a_5 n} \, d^{a_2 a_3 n}\Big) \ \ \equiv \ \ t^{a_2a_3a_4a_5}_Q
\end{equation}
and the amplitude (\ref{gmhv5}) simplifies considerably in the color sector:
\begin{align}
{\cal M}(\gamma^+_1,g^-_2,g^-_3, g^-_4, g^+_5) \ \ &= \ \ 4\sqrt{2} \, g^3 \, [15]^4 \, \Bigl[\,
\big( {\cal C}_{(1234)}-{\cal C}_{(1432)}+{\cal C}_{(2134)}+{\cal C}_{(2314)}\big) \, t_Q^{a_2a_3a_4a_5} \Bigr. \notag \\
& \ \ \ \ \Bigl. + \ \big( {\cal C}_{(1243)}-{\cal C}_{(1342)}+{\cal C}_{(2143)}+{\cal C}_{(2413)}\big) \, t_Q^{a_2a_4a_3a_5} \Bigr. \notag \\
& \ \ \ \ \Bigl. + \ \big( {\cal C}_{(1324)}-{\cal C}_{(1423)}+{\cal C}_{(3124)}+{\cal C}_{(3214)}\big) \, t_Q^{a_3a_2a_4a_5}\Bigr]
\end{align}
After adding squared amplitudes describing all possible helicity configurations, and summing over color indices of four gluons, we obtain
\begin{align}
\bigl|{\cal M}(\gamma_1,g_2, &g_3, g_4,g_5) \bigr|^2 \eq 8 \, g^6 \, Q^2 \, (N^2{-}1) \, (N^2{-}4) \, \left( \sum_{i<j} s_{ij}^4 \right) \notag \\
&\times \Bigl[
\bigl| {\cal C}_{(1234)}-{\cal C}_{(1432)}+{\cal C}_{(2134)}+{\cal C}_{(2314)} \bigr|^2 \ + \ \bigl|{\cal C}_{(1243)}-{\cal C}_{(1342)}+{\cal C}_{(2143)}+{\cal C}_{(2413)} \bigr|^2 \Bigr.  \notag \\
& \ \ \ \ \ \ \ \ \ \ \  \Bigl. + \ \bigl|{\cal C}_{(1324)}-{\cal C}_{(1423)}+{\cal C}_{(3124)}+{\cal C}_{(3214)}\bigr|^2\Bigr] \ . \label{squareab1}
\end{align}
In the case of two abelian gauge bosons, $T^{a_1}=T^{a_2}=Q \mathds{1}$, the color factor becomes
\begin{equation}
t^{a_1a_2a_3a_4a_5} \eq \frac{iQ^2}{2} \; f^{a_3a_4a_5} \ \ \equiv \ \ t^{a_3a_4a_5}_{QQ}\ ,
\end{equation}
and the amplitude (\ref{gmhv5}) simplifies to
\begin{align}
{\cal M}&(\gamma^+_1,\gamma^-_2,g^-_3, g^-_4, g^+_5) \eq 4\sqrt{2} \, g^3 \, t^{a_3a_4a_5}_{QQ} \, [15]^4 \, \Bigl[\,
{\cal C}_{(1234)}-{\cal C}_{(1432)}+{\cal C}_{(2134)}+{\cal C}_{(2314)}\Bigr. \notag \\
& \ \Bigl. - \ {\cal C}_{(1243)}+{\cal C}_{(1342)}-{\cal C}_{(2143)}-{\cal C}_{(2413)}+
 {\cal C}_{(1324)}-{\cal C}_{(1423)}+{\cal C}_{(3124)}+{\cal C}_{(3214)} \Bigr] \ .
\end{align}
After adding squared amplitudes describing all possible helicity configurations, and summing over color indices of three gluons, we obtain
\begin{align}
\bigl|{\cal M}&(\gamma_1,\gamma_2, g_3, g_4,g_5) \bigr|^2 \eq 16 \,g^6 \,Q^4 \, N \, (N^2{-}1) \, \left( \sum_{i<j} s_{ij}^4 \right)
\bigl|{\cal C}_{(1234)}-{\cal C}_{(1432)} +{\cal C}_{(2134)} + {\cal C}_{(2314)} \bigr. \nonumber\\
& \ \bigl.  - \, {\cal C}_{(1243)}+{\cal C}_{(1342)}-{\cal C}_{(2143)}-{\cal C}_{(2413)}+
{\cal C}_{(1324)}-{\cal C}_{(1423)}+{\cal C}_{(3124)}+{\cal C}_{(3214)}\big|^2 \ . \label{squareab2}
\end{align}
The amplitudes with three or more abelian gauge bosons vanish identically.

The squared amplitudes (\ref{square5}), (\ref{squareab1}) and (\ref{squareab2}) can be further simplified by expressing all $\cal C$-functions in terms of the two-element basis
${\cal C}_{(1342)}$, ${\cal C}_{(2413)}$, using the relations listed in Table 2.

\subsection{Three gluons, two quarks}

We begin with the case of all three non-abelian gluons originating from stack $a$, with the amplitude given in (\ref{qqfin}). After taking its modulus squared, summing over all helicity configurations and over color indices of quarks and gluons, we obtain
\begin{equation}\label{squareq1}
\bigl|{\cal M}(g_1,g_2, g_3, q_4, {\bar q}_5) \bigr|^2 \eq  16 \, g^6\sum_{i=1,2,3} \bigl( s_{i4}^3 \, s_{i5} \, + \, s_{i4} \, s_{i5}^3 \bigr)\sum_{\lambda,\lambda'\in \Pi_q}\,
{\cal C}_{\lambda} \, {\cal P}_{\lambda\lambda'} \, {\cal C}^*_{\lambda'}\ ,
\end{equation}
where the entries of the matrix $\cal P$ can be extracted from Table 5. 

\vskip1cm
\begin{center}
\begin{tabular}{|c|c|c|c|c|c|c|}\hline
 & $\scriptstyle(123)$ & $\scriptstyle(132)$ & $\scriptstyle(213)$
& $\scriptstyle(231)$ & $\scriptstyle(312)$ & $\scriptstyle(321)$\\ \hline
$\scriptstyle(123)$ & $D_q$ & $X_q$ & $X_q$ & $Y_q$ & $Y_q$ & $Z_q$\\ \hline
$\scriptstyle(132)$ &  $X_q$  & $D_q$ & $Y_q$ & $Z_q$ & $X_q$ & $Y_q$\\ \hline
$\scriptstyle(213)$ &  $X_q$  & $Y_q$ & $D_q$&  $X_q$ & $Z_q$ & $Y_q$\\ \hline
$\scriptstyle(231)$ &  $Y_q$  & $Z_q$ & $X_q$ & $D_q$ & $Y_q$ & $X_q$\\ \hline
$\scriptstyle(312)$ &  $Y_q$  & $X_q$ & $Z_q$ & $Y_q$ & $D_q$ & $X_q$\\ \hline
$\scriptstyle(321)$ &  $Z_q$  & $Y_q$ & $Y_q$ & $X_q$ & $X_q$ & $D_q$\\ \hline
\end{tabular}
\begin{flushleft}{\bf Table 5}. \it Matrix elements ${\cal P}_{\lambda\lambda'}$. Since the elements  of the permutation set $\Pi_q$, see Eq. (\ref{pqset}), have common last two numbers (4,5), the rows and columns are labeled by first three numbers. The entries $D_q$, $X_q$, $Y_q$ and $Z_q$ depend on the gauge group and are given in Eqs. (\ref{dq11})-(\ref{dq12}). Note that the matrix is symmetric.\\[3mm]
\end{flushleft}
\end{center}

\noindent
Its elements can be written as
\begin{align}
D_q \ \ &= \ \ \bigl[N \, C_F^3\bigr]_a[N]_b\label{dq11}\\
X_q  \ \ &= \ \ \left[N \, C_F^2 \, \left(C_F\, - \, \frac{C_A}{2}\right)\right]_a [N]_b\\
Y_q \ \ &= \ \ \left[N \, C_F \,\left(C_F \, - \, \frac{C_A}{2}\right)^2 \right]_a[N]_b\\
Z_q \ \ &= \ \ \left[N \, C_F \, \left(C_F \, - \, \frac{C_A}{2}\right) \, \bigl(C_F \, - \, C_A \bigr)\right]_a[N]_b\label{dq12}
\end{align}
where the group constants enclosed by square brackets with subscripts $a$ and $b$ refer to the respective stacks and can be read out from Table 4.

If one of the vector bosons is associated to the $U(1)$ subgroup of $U(N)$ and the other two are $SU(N)$ gauge bosons, the amplitude (\ref{qqfin}) simplifies to
\begin{align}
{\cal M}(\gamma^+_1,g^-_2,g^-_3, q^-_4, {\bar q}^+_5) \ \ = \ \ 2\sqrt{2} \, &g^3 \, Q \, [14] \, [15]^3
\Bigl[ \bigl( {\cal C}_{(1234)}+{\cal C}_{(2134)}+{\cal C}_{(2314)} \bigr) \, (T^{a_2}T^{a_3})^{\alpha_4}_{\alpha_5} \, \delta^{\beta_5}_{\beta_4} \Bigr. \notag \\
& \ \ \ \ \Bigl. + \ \bigl( {\cal C}_{(1324)}+{\cal C}_{(3124)}+{\cal C}_{(3214)} \bigr) \, (T^{a_3}T^{a_2})^{\alpha_4}_{\alpha_5} \, \delta^{\beta_5}_{\beta_4} \Bigr] \ .
\end{align}
After adding squared amplitudes describing all possible helicity configurations, and summing over all color indices, we obtain
\begin{align}
\bigl|{\cal M}&(\gamma_1,g_2, g_3, q_4, {\bar q}_5) \bigr|^2 \eq 4 \, g^6 \, Q^2 \, [N]_b \, \left[\frac{N^2-1}{N}\right]_a\sum_{i=1,2,3} \bigl( s_{i4}^3 \, s_{i5} \, + \, s_{i4} \, s_{i5}^3 \bigr)
\notag \\
&\times \ \biggl[ \bigl[N^2-1 \bigr]_a\, \Bigl( \bigl|{\cal C}_{(1234)}+{\cal C}_{(2134)}+{\cal C}_{(2314)} \bigr|^2 \ + \ \bigl| {\cal C}_{(1324)}+{\cal C}_{(3124)}+{\cal C}_{(3214)} \bigr|^2 \Bigr) \biggr. \notag  \\
& \ \ \ \ \ \ \ \ \ \biggl. - \ 2 \, \Re \Bigl\{ ({\cal C}_{(1234)}+{\cal C}_{(2134)}+{\cal C}_{(2314)}) \, ({\cal C}_{(1324)}^*+{\cal C}_{(3124)}^*+{\cal C}_{(3214)}^*) \Bigr\} \biggr]
\end{align}
With two and three abelian gauge bosons, the amplitudes acquire even simpler form:
\begin{eqnarray}
{\cal M}(\gamma^+_1,\gamma^-_2,g^-_3, q^-_4, {\bar q}^+_5)&=&2\sqrt{2} \, g^3 \, Q^2 \, (T^{a_3})^{\alpha_4}_{\alpha_5} \, \delta^{\beta_5}_{\beta_4} \, [14] \, [15]^3
\sum_{\lambda\in\Pi_q}{\cal C}_{\lambda} \\
{\cal M}(\gamma^+_1,\gamma^-_2,\gamma^-_3, q^-_4, {\bar q}^+_5)&=&2\sqrt{2} \, g^3 \, Q^3 \, \delta^{\alpha_4}_{\alpha_5} \, \delta^{\beta_5}_{\beta_4} \,[14] \, [15]^3
\sum_{\lambda\in\Pi_q}{\cal C}_{\lambda}
\end{eqnarray}
and the respective squares become
\begin{eqnarray}
\bigl|{\cal M}(\gamma_1,\gamma_2, g_3, q_4, {\bar q}_5) \bigr|^2&=& 8 \, g^6 \, Q^4 \, [N]_b \, \bigl[N^2-1 \bigr]_a\sum_{i=1,2,3} \bigl(s_{i4}^3 \, s_{i5} \, + \, s_{i4} \, s_{i5}^3)
\,\biggl| \sum_{\lambda\in\Pi_q}{\cal C}_{\lambda}\biggr|^2\ , \\
\bigl|{\cal M}(\gamma_1,\gamma_2, \gamma_3, q_4, {\bar q}_5) \bigr|^2&=& 16 \, g^6 \, Q^6 \, [N]_b \, [N]_a\sum_{i=1,2,3} \bigl(s_{i4}^3 \, s_{i5} \, + \, s_{i4} \,s_{i5}^3 \bigr)\,\biggl|
\sum_{\lambda\in\Pi_q}{\cal C}_{\lambda}\biggr|^2\ .
\end{eqnarray}

Next, we consider two non-abelian gluons from stack $a$ and one from stack $b$, with the amplitude given by equation (\ref{qqfinb}). After taking its modulus squared, summing over all helicity configurations and over color indices of quarks and gluons, we obtain
\begin{align}
&\bigl|{\cal M}(g_1,g_2, b_3, q_4, {\bar q}_5)|^2\eq  16 \, g^4 \, g^2_{b} \, [N \, C_F]_b \, [N \, C_F]_a\sum_{i=1,2,3} \bigl(s_{i4}^3 \, s_{i5} \, + \, s_{i4} \, s_{i5}^3 \bigr) \notag \\
& \, \times \, \biggl\{[C_F]_a \, \Bigl(\bigl|{\cal C}_{(1243)} \bigr|^2 \, + \, \bigl|{\cal C}_{(2143)} \bigr|^2 \Bigr) \ + \ \left[C_F \, - \, \frac{C_A}{2} \right]_a \bigl({\cal C}_{(1243)}{\cal C}_{(2143)}^* \, + \, {\cal C}_{(1243)}^*{\cal C}_{(2143)} \bigr) \biggr\} \ .
\end{align}
A simple replacement of group factors in the above formula
covers also the case of $U(1)$ gauge boson from stack $b$:
\begin{equation}\label{brepl}
\bigl|{\cal M}(g_1,g_2, \gamma_{b3}, q_4, {\bar q}_5) \bigr|^2 \eq \bigl|{\cal M}(g_1,g_2, b_3, q_4, {\bar q}_5) \bigr|^2 \Bigl. \Bigr|_{  [NC_F]_b\to [NQ^2]_b }
\end{equation}
Finally, we consider the case when one of the two vector bosons from stack $a$ is associated to $U(1)$:
\begin{equation}
\bigl| {\cal M}(\gamma_1,g_2, b_3, q_4, {\bar q}_5) \bigr|^2 \eq 16 \, g^4 \, g^2_{b} \, [N\, C_F]_b \, [Q^2 \, N \, C_F]_a\sum_{i=1,2,3} \bigl(s_{i4}^3 \, s_{i5} \, + \, s_{i4} \, s_{i5}^3 \bigr)  \bigl| {\cal C}_{(1243)} + {\cal C}_{(2143)} \bigr|^2
\end{equation}
If the second gauge boson from stack $a$ and/or the gauge boson from stack $b$ are associated to $U(1)s$, the corresponding formulas can be obtained from the above equation by replacements $NC_F\to NQ^2$, as in (\ref{brepl}).

\subsection{One gluon and two quark-antiquark pairs}

The amplitudes involving one gluon and two quark-antiquark pairs are more complicated and in some way more interesting than other five-point amplitudes because they are non-universal and depend on the geometry of compact dimensions. As explained in Ref. \cite{Anchordoqui:2009mm}, such amplitudes can be used as ``precision tests''
of low mass string theory that relate geometry of compact dimensions to the collider data. The results presented in Section 6 are derived in the specific context of orientifold compactifications with intersecting stacks of D$6$--branes and, similarly to four--fermion amplitudes discussed in Section 6.3 of \cite{LHC}, need some explanations before employing them in collider phenomenology.

The main problem with non--universal amplitudes discussed in Section 6 as well as in Section 6.3 of \cite{LHC} is that already in the leading low-energy approximation ($\alpha'\to 0$ limit) they include the effects of  particles that are expected to acquire masses of order $M_{\rm string}$ while on the other hand, they are missing some contributions of intermediate electro-weak gauge bosons. As an example, consider  the process $u_L(1)d_L(2)\to u_L(3)d_L(4)+g(5)$ which is described by the amplitudes ${\cal M}_{\rho_1}$ of Eq. (\ref{FINALgqqqq1}),
receiving contributions from the orderings $\rho_1=(1,2,3,4,5)$ and $\rho_1=(2,3,1,4,5)$ only. The corresponding Chan-Paton factors, with the indices appropriately relabeled, are
\begin{equation}\label{chanpat}
t(1,2,3,4,5)= (T^{a_5})^{\alpha_4}_{\alpha_1}\delta^{\alpha_3}_{\alpha_2}\qquad\qquad
t(2,3,1,4,5)= (T^{a_5})^{\alpha_3}_{\alpha_2}\delta^{\alpha_4}_{\alpha_1}.
\end{equation}
The leading terms in the low energy expansions of the corresponding partial amplitudes  are written in Eqs. (\ref{FINAL1}) and (\ref{FINAL2}), respectively, where the momentum and group indices require relabeling $1\to 5, ~2\to 4, ~ 3\to 2, ~4\to 3, ~5\to 1$. Inside the terms proportional to $g_a^3$, one can easily identify the contributions due to $t$-channel exchanges of virtual gluons. The Chan-Paton factors (\ref{chanpat}) reveal that in addition to gluons, there are also massless $U(1)$ baryon number gauge bosons propagating in this channel.  However, a mass of order $M_{\rm string}$ is generated for the anomalous gauge boson, by the Green-Schwarz mechanism that operates beyond the disk level.  For phenomenological purposes, one could subtract these contributions by hand, but it is  really not necessary because one could also argue that they are suppressed by color factors of order ${\cal O}(1/N)$ with respect to gluons. A more serious problem appears in the terms proportional to $g_ag_b^2$: although they exhibit correct $u$-channel singularities due to charged electroweak bosons, they miss the $t$-channel poles normally associated to the neutral electroweak boson! In principle, this problem could be cured by considering an electroweak stack of $D$-branes with $Sp(2)$ instead of $U(2)$ gauge symmetry.
The lesson to be learned from this example is that the amplitudes under consideration require  careful handling prior to investigating their phenomenological consequences.

In Ref. \cite{Anchordoqui:2009mm}, we made some concrete proposals how to apply four-fermion amplitudes in the LHC studies of model-dependent features of dijet signals. A similar procedure could be applied to the processes involving additional gauge bosons, however we do not want to prejudice the reader, hence we limit ourselves to summarizing the most important properties of the results derived in Section 6. The amplitudes for the scattering processes involving quark electro-weak doublets only are written in Eqs. (\ref{FINALgqqqq1}) and (\ref{FINALgqqqq2}), with the latter one contributing only in the case of identical flavors. They are covered by case {\it (i)} in Section 6.2, where the properties of the respective $G$-functions are extensively discussed. The processes involving both quark doublets and quark singlets require three stacks and are covered by case {\it (ii)}.
They are still described by Eq. (\ref{FINALgqqqq1}), with the $G$-functions appropriately modified to accommodate the instanton effects {\em etc.}, see  Eq. (\ref{givenby2}). Finally, the processes involving not only quarks but also leptons, necessary for
the discussion of jets associated to Drell-Yan pairs, are covered by the four-stack case {\em (iii)}. Here only one ordering is picked up from Eq. (\ref{FINALgqqqq1}), and the $G$-functions receive further modifications, see  Eq. (\ref{givenby3}). In all cases, the Chan-Paton factors are very simple, therefore squaring these amplitudes and summing over the group indices do not present any problems.

\section{Summary}
The fundamental string mass scale is an arbitrary parameter. If it is in the range of few TeVs, vibrating string modes, i.e. Regge excitations predicted by critical superstring theory, will be observed at the LHC. In this paper, we extended our discussion of
the scattering processes with the best potential for revealing the effects of Regge excitations, to the case of five external particles. The corresponding amplitudes describe processes underlying three jet production, photon (or electroweak bosons) plus dijets, jets associated to Drell-Yan lepton pairs and many other processes that can be used not only to discover Regge excitations of quarks and gluons but also to determine their masses, spins, decay widths \cite{Anchordoqui:2008hi}, branching ratios and other properties as predicted by superstring theory.

Without any doubt, the most interesting property of full-fledged string disk amplitudes is their universality, i.e. their model-independence. The amplitudes involving  arbitrary number of gluons and at most two fermions do not depend on the compactification details. In this work, we exhibited a larger universality, between
the amplitudes involving one quark-antiquark pair plus a number of gluons and their purely gluonic counterparts obtained by replacing the quark-antiquark pair  by two gluons. This becomes particularly striking after collecting Eqs. (\ref{side1}), (\ref{side2}), (\ref{m5}) and (\ref{m5q}):
\begin{eqnarray}
{\cal M}_{\rho}\Big(g^+_1,g^-_2, \bigg\{{g^-_3, g^+_4\atop q^-_3,\bar q^+_4}\Big)&=& V^{(4)}(\hat s_i)\ {\cal M}_{\rho}^{QCD}\Big(g^+_1,g^-_2, \bigg\{{g^-_3, g^+_4\atop q^-_3,\bar q^+_4}\Big)\nonumber
\\ &&\\ \nonumber
{\cal M}_{\rho}\Big(g^+_1,g^-_2, g^-_3,\bigg\{{g^-_4, g^+_5\atop q^-_4,\bar q^+_5}\Big) &=&
\Big[{V^{(5)}(\hat s_i)}  -  2i\
 P^{(5)}(\hat s_i)\ \epsilon(1,2,3,4)\Big]\ {\cal M}_{\rho}^{QCD}\Big(g^+_1,g^-_2, g^-_3,\bigg\{{g^-_4, g^+_5\atop q^-_4,\bar q^+_5}\Big)\ , \end{eqnarray}
where $\rho$ refers to any partial amplitude. As explained in Subsections 2.4. and 4.4, 
we expect the relation between purely gluonic amplitudes and those 
involving one quark-antiquark pair to remain valid for $N$-point amplitudes. 
In Sections 5 and 7, we also explained how the recently derived relations between partial amplitudes can be used to express all these universal amplitudes in terms of $(N-3)!$ functions of kinematic variables.

If fundamental strings are discovered at the LHC, the landscape problem will be nullified, however a whole class of new, difficult problems will come into focus. With the present understanding of superstring theory, it is not clear what to expect beyond the string threshold. The name used for describing this energy domain is ``trans-Planckian,''  dating back to the times when the string mass was automatically assumed to be near the Planck mass.
It is possible that we will be exploring such trans-Planckian physics very soon.

\vskip1cm
\goodbreak
\centerline{\noindent{\bf Acknowledgments} }\vskip 2mm

We wish to thank Ignatios Antoniadis, Gia Dvali, Daniel Haertl, and Stefan Theisen
for useful discussions.
The research of T.R.T.\ is supported by the U.S.  National Science
Foundation Grants PHY-0600304, PHY-0757959 and by the Cluster of
Excellence ``Origin and Structure of the Universe'' in Munich,
Germany.  He is grateful to Arnold Sommerfeld Center for Theoretical
Physics at Ludwig--Maximilians--Universit\"at, and to
Max--Planck--Institut f\"ur Physik in M\"unchen, for their kind
hospitality. The diagrams have been created by the program JaxoDraw \cite{Jaxo}.
Any opinions, findings, and conclusions or
recommendations expressed in this material are those of the authors
and do not necessarily reflect the views of the National Science
Foundation.

\break
\section*{Appendix}
\appendix

\section{Material for computing parton amplitudes}
\label{appA}

In this Appendix we present the world--sheet CFT correlators relevant to the computation of the string amplitudes in Sections 4 and 6. 

For the scalar $\phi$ bosonizing the superghost system we have
\begin{subequations}
\begin{align}
\langle e^{-\phi(z_{1})} \, e^{-\phi(z_{2})/2} \, e^{-\phi(z_{3})/2} \rangle
\ \ &= \ \ z_{12}^{-1/2} \, z_{13}^{-1/2} \, z_{23}^{-1/4} \label{1,3a}\ , \\
\langle e^{-\phi(z_{1})/2} \, e^{-\phi(z_{2})/2} \, e^{-\phi(z_{3})/2} \, e^{-\phi(z_{4})/2} \rangle \ \ &= \ \ \frac{1}{(z_{12} \, z_{13} \, z_{14} \, z_{23} \, z_{24} \, z_{34})^{1/4}}\ , \label{1,3aa}
\end{align}
\end{subequations}
while for the bosonic space--time coordinate fields $X^{\mu}$
\begin{subequations}
\begin{align}
\langle e^{ik _{1}   X(z_{1})} \, ... \, e^{ik_n   X(z_{n})} \rangle \ \ &= \ \ \prod_{i,j=1 \atop {i<j}}^{n} \bigl| z_{ij} \bigr|^{2\al' k_i   k_j}\ ,  
\label{1,4a} \\
\Bigl\langle  \pa X^{\mu}(z_{A}) \, \prod_{i=1}^{n} e^{ik_i   X(z_{i})} \Bigr\rangle \ \ &= \ \  \left(-2i\al' \, \sum_{r=1}^{n} \frac{k_r^{\mu}}{z_{A} \, - \, z_{r}} \right) \, \Bigl\langle \prod_{i=1}^{n} e^{ik_i   X(z_{i})} \Bigr\rangle  \ ,\label{1,4b} \\
\Bigl\langle \pa X^{\mu}(z_{A}) \, \pa X^{\nu}(z_{B}) \, \prod_{i=1}^{n} e^{ik_i   X(z_{i})} \Bigr\rangle \ \ &= \ \ \left(-4\al'^{2} \, \sum_{r,s=1}^{n} \frac{k^{\mu}_r \, k_s^{\nu}}{z_{A,r} \, z_{B,s}} \ - \ 2\al' \; \frac{\eta^{\mu \nu}}{z_{AB}^{2}} \right) \ \Bigl\langle \prod_{i=1}^{n} e^{ik_i X(z_{i})} \Bigr\rangle\ , \label{1,4c}
\end{align}
\end{subequations}
and for the NS-R SCFT with the $\psi^{\mu}$, $S_{\al}$ and $S^{\dbe}$ fields:
\bea
&&\ds{\vev{\psi^{\mu}(z_{1}) \, S_{\al}(z_{2}) \, S_{\dbe}(z_{3})}=
(2\ z_{12} \, z_{13})^{-1/2} \; \si^{\mu}_{\al \dbe}\ ,}\\[4mm]
&&\ds{\vev{\psi^{\mu}(z_{1}) \, \psi^{\nu} (z_{2}) \, \psi^{\la}(z_{3}) \, S_{\al}(z_{4}) \, S_{\dbe}(z_{5})}=(2\ 
z_{14} \, z_{15} \, z_{24} \, z_{25} \, z_{34} \, z_{35})^{-1/2} } \\[4mm]
&&\hskip2.75cm\times\ds{ \left\{\ \frac{z_{45}}{2} \; (\si^{\mu} \, \bar{\si}^{\nu} \, \si^{\la})_{\al \dbe} \ + \ \eta^{\mu \nu} \, \si^{\la}_{\al \dbe} \; \frac{z_{14} \, z_{25}}{z_{12}} \ - \ \eta^{\mu \la} \, \si^{\nu}_{\al \dbe} \; \frac{z_{14} \, z_{35}}{z_{13}} \ + \ \eta^{\nu \la} \, \si^{\mu}_{\al \dbe} \; \frac{z_{24} \, z_{35}}{z_{23}}\ \right\}\ .} 
\label{1,5b}
\eea
Because of the term proportional to $\psi^{\mu} \psi^{\nu}$ in the
gluon vertex operator \req{1,1b} for the computation of the $gggq\bar q$ amplitude
we also need the  seven point function \cite{spin}:
\begin{align}
\langle &\psi^{\mu}(z_{1}) \, \psi^{\nu} (z_{2}) \, \psi^{\la}(z_{3}) \, \psi^{\rho}(z_{4}) \, \psi^{\tau}(z_{5}) \, S_{\al}(z_{6}) \, S_{\dbe}(z_{7}) \rangle \ \ = \ 
(2\ z_{16} \, z_{17} \, z_{26} \, z_{27} \, z_{36} \, z_{37} \, z_{46} \, z_{47} \, z_{56} \, z_{57} )^{-1/2} \notag \\
&\times \Biggl\{ \ \si^{\mu}_{\al \dbe} \, \left( \eta^{\nu \la} \, \eta^{\rho \tau} \; \frac{z_{26} \, z_{37} \, z_{46} \, z_{57}}{z_{23} \, z_{45}} \ + \ \eta^{\nu \tau} \, \eta^{\la \rho} \; \frac{z_{26} \, z_{57} \, z_{36} \, z_{47}}{z_{25} \, z_{34}} \ - \ \eta^{\nu \rho} \, \eta^{\la \tau} \; \frac{z_{26} \, z_{47} \, z_{36} \, z_{57}}{z_{24} \, z_{35}} \right) \Biggr. \notag \\
& \ \ \ \ \ - \ \si^{\nu}_{\al \dbe} \, \left( \eta^{\mu \la} \, \eta^{\rho \tau} \; \frac{z_{16} \, z_{37} \, z_{46} \, z_{57}}{z_{13} \, z_{45}} \ + \ \eta^{\mu \tau} \, \eta^{\la \rho} \; \frac{z_{16} \, z_{57} \, z_{36} \, z_{47}}{z_{15} \, z_{34}} \ - \ \eta^{\mu \rho} \, \eta^{\la \tau} \; \frac{z_{16} \, z_{47} \, z_{36} \, z_{57}}{z_{14} \, z_{35}} \right) \notag \\
& \ \ \ \ \ + \ \si^{\la}_{\al \dbe} \, \left( \eta^{\mu \nu} \, \eta^{\rho \tau} \; \frac{z_{16} \, z_{27} \, z_{46} \, z_{57}}{z_{12} \, z_{45}} \ + \ \eta^{\mu \tau} \, \eta^{\nu \rho} \; \frac{z_{16} \, z_{57} \, z_{26} \, z_{47}}{z_{15} \, z_{24}} \ - \ \eta^{\mu \rho} \, \eta^{\nu \tau} \; \frac{z_{16} \, z_{47} \, z_{26} \, z_{57}}{z_{14} \, z_{25}} \right) \notag \\
& \ \ \ \ \ - \ \si^{\rho}_{\al \dbe} \, \left( \eta^{\mu \nu} \, \eta^{\la \tau} \; \frac{z_{16} \, z_{27} \, z_{36} \, z_{57}}{z_{12} \, z_{35}} \ + \ \eta^{\mu \tau} \, \eta^{\nu \la} \; \frac{z_{16} \, z_{57} \, z_{26} \, z_{37}}{z_{15} \, z_{23}} \ - \ \eta^{\mu \la} \, \eta^{\nu \tau} \; \frac{z_{16} \, z_{37} \, z_{26} \, z_{57}}{z_{13} \, z_{25}} \right) \notag \\
& \ \ \ \ \ + \ \si^{\tau}_{\al \dbe} \, \left( \eta^{\mu \nu} \, \eta^{\la \rho} \; \frac{z_{16} \, z_{27} \, z_{36} \, z_{47}}{z_{12} \, z_{34}} \ + \ \eta^{\mu \rho} \, \eta^{\nu \la} \; \frac{z_{16} \, z_{47} \, z_{26} \, z_{37}}{z_{14} \, z_{23}} \ - \ \eta^{\mu \la} \, \eta^{\nu \rho} \; \frac{z_{16} \, z_{37} \, z_{26} \, z_{47}}{z_{13} \, z_{24}} \right) \notag \\
& \ \ \ \ \ \ \ \ + \ \frac{z_{67}}{2} \, \biggl( + \ \eta^{\mu \nu} \, (\si^{\la} \, \bar{\si}^{\rho} \, \si^{\tau})_{\al \dbe} \; \frac{z_{16} \, z_{27}}{z_{12}} \ - \ \eta^{\mu \la} \, (\si^{\nu} \, \bar{\si}^{\rho} \, \si^{\tau})_{\al \dbe} \; \frac{z_{16} \, z_{37}}{z_{13}} \biggr. \notag \\
& \ \ \ \ \ \ \ \ \ \ \ \ \ \ \ \ \ \ \ + \eta^{\mu \rho} \, (\si^{\nu} \, \bar{\si}^{\la} \, \si^{\tau})_{\al \dbe} \; \frac{z_{16} \, z_{47}}{z_{14}} \ - \ \eta^{\mu \tau} \, (\si^{\nu} \, \bar{\si}^{\la} \, \si^{\rho})_{\al \dbe} \; \frac{z_{16} \, z_{57}}{z_{15}} \notag \\
& \ \ \ \ \ \ \ \ \ \ \ \ \ \ \ \ \ \ \ + \eta^{\nu \la} \, (\si^{\mu} \, \bar{\si}^{\rho} \, \si^{\tau})_{\al \dbe} \; \frac{z_{26} \, z_{37}}{z_{23}} \ - \ \eta^{\nu \rho} \, (\si^{\mu} \, \bar{\si}^{\la} \, \si^{\tau})_{\al \dbe} \; \frac{z_{26} \, z_{47}}{z_{24}} \notag \\
& \ \ \ \ \ \ \ \ \ \ \ \ \ \ \ \ \ \ \ + \eta^{\nu \tau} \, (\si^{\mu} \, \bar{\si}^{\la} \, \si^{\rho})_{\al \dbe} \; \frac{z_{26} \, z_{57}}{z_{25}} \ + \ \eta^{\la \rho} \, (\si^{\mu} \, \bar{\si}^{\nu} \, \si^{\tau})_{\al \dbe} \; \frac{z_{36} \, z_{47}}{z_{34}} \notag \\
& \ \ \ \ \ \ \ \ \ \ \ \ \ \ \ \ \ \ \ \biggl. \, - \ \eta^{\la \tau} \, (\si^{\mu} \, \bar{\si}^{\nu} \, \si^{\rho})_{\al \dbe} \; \frac{z_{36} \, z_{57}}{z_{35}} \ + \ \eta^{\rho \tau} \, (\si^{\mu} \, \bar{\si}^{\nu} \, \si^{\la})_{\al \dbe} \; \frac{z_{46} \, z_{57}}{z_{45}} \biggr) + \left( \frac{z_{67}}{2} \right)^{2} \, (\si^{\mu} \, \bar{\si}^{\nu} \, \si^{\la} \, \bar{\si}^{\rho} \, \si^{\tau})_{\al \dbe} \Biggr\}.
\label{1,6}
\end{align}
This correlator is derived thoroughly in a separate paper \cite{spin}.
The computation of the $gq\bar q q\bar q$ amplitude requires correlation functions involving four spin fields:
\beq
\langle S_{\al}(z_{1}) \, S_{\dbe}(z_{2}) \, S_{\ga}(z_{3}) \, S_{\dde}(z_{4}) \rangle= \frac{\vep_{\al \ga} \, \vep_{\dbe \dde}}{(z_{13} \, z_{24})^{1/2}}\ .\label{4fA1}
\eeq
In addition, in the presence of one  fermionic pair $\psi^{\mu} \psi^{\nu}$, the following correlator is derived in \cite{spin}:
\bea \label{4fA2}
&&\ds{\langle \psi^{\mu}(z_{1}) \, \psi^{\nu}(z_{2}) \, S_{\al}(z_{3}) \, S_{\dbe}(z_{4}) \, S_{\ga}(z_{5}) \, S_{\dde}(z_{6}) \rangle =
\frac{- 1}{ ( z_{13} \, z_{14} \, z_{15} \, z_{16} \, z_{23} \, z_{24} \, z_{25} \, z_{26})^{1/2} \, (z_{35} \, z_{46} )^{1/2}} }\\[5mm]
&&\hskip3cm\times\ \ds{\Biggl\{ \frac{\eta^{\mu \nu}}{z_{12}} \; \vep_{\al \ga} \, \vep_{\dbe \dde} \, z_{13} \, z_{15} \, z_{24} \, z_{26} \ + \ \frac{z_{15}}{2} \; \bigl( \si^{\mu}_{\al \dbe} \, \si^{\nu}_{\ga \dde} \, z_{24} \, z_{36} \ - \ \si^{\mu}_{\al \dde} \, \si^{\nu}_{\ga \dbe} \, z_{26} \, z_{34} \bigr) \Biggr. }\\[5mm]
&&\hskip8cm \ds{ - \Biggl. \frac{z_{13}}{2} \; \bigl( \si^{\nu}_{\al \dbe} \, \si^{\mu}_{\ga \dde} \, z_{26} \, z_{45} \ + \ \si^{\nu}_{\al \dde} \, \si^{\mu}_{\ga \dbe} \, z_{24} \, z_{56} \bigr) \Biggr\}\ . }
\eea

\section{Hypergeometric function relations}
\label{appB}

Here we present  the  restrictions, which follow from requiring gauge invariance, 
on the hypergeometric integrals in equations \req{new2,101} and \req{2,101}.
Each of the three $\xi_i \to
k_i,\ i=1,2,3$ operations (with the remaining polarization vectors untouched) gives a
separate  block of relations.
For $\xi_1  \ra k_1$ we obtain
\beq\ba{ll}
&(1-s_{1}) \, {H}^\rho_{10}\ - \ (s_{4}-s_{1}-s_{2}) \, {H}^\rho_{1} \ - \ (s_{2} -s_{4}-s_{5}) \, {H}^\rho_{4} =0\ , \\[2mm]
&s_{1} \, {H}^\rho_{2}  \ + \ (s_{4}-s_{1}-s_{2}) \, {H}^\rho_{3}  \ + \ (s_{2}-s_{4}-s_{5}) \, {H}^\rho_{7}   = 0\ , \\[2mm]
&s_{1} \, {H}^\rho_{5}  \ + \ (s_{4}-s_{1}-s_{2}) \, {H}^\rho_{6}  \ + \ (s_{2}-s_{4}-s_{5}) \, {H}^\rho_{8}  =0\ ,
\ea \label{B1}
\eeq
for $\xi_2 \ra k_2$
\beq\ba{ll}
&(1-s_{1}) \, {H}^\rho_{10}  \ + \ s_{2} \, {H}^\rho_{2}  \ + \ (s_{5}-s_{2}-s_{3}) \, {H}^\rho_{5}   =0\ , \\[2mm]
&-s_{1}\, {H}^\rho_{1}  \ + \ s_{2} \, {H}^\rho_{3}  \ + \ (s_{5}-s_{2}-s_{3}) \, {H}^\rho_{6} = 0\ ,\\[2mm]
&-s_{1} \, {H}^\rho_{4}  \ + \ s_{2} \, {H}^\rho_{7}  \ + \ (s_{5}-s_{2}-s_{3}) \, {H}^\rho_{8}  = 0\ ,
\ea \label{B2}
\eeq
and finally for  $\xi_3  \ra  k_3:$
\beq\ba{ll}
&{H}^\rho_{4}  \ - \ {H}^\rho_{5}  \ + \ {H}^\rho_{8}  =0\ ,  \\[2mm]
&-s_{1} \, {H}^\rho_{5}  \ + \ s_{2} \, {H}^\rho_{7}  \ + \ (s_{1}+s_{5}-s_{2}-s_{3}) \, {H}^\rho_{8} = 0\ ,   \\[2mm]
&(s_{4}-s_{1}-s_{2}) \, {H}^\rho_{6}  \ + \ s_{2} \, {H}^\rho_{7}  \ + \ (s_{1}-s_{3}-s_{4}) \, {H}^\rho_{8}  = 0\ ,  \\[2mm]
&{H}^\rho_{10}   \ - \ (s_{2}-s_{4}-s_{5}) \, {H}^\rho_{5}  \ - \ s_{1} \, {H}^\rho_{11}  \ - \ (s_{4}-s_{1}-s_{2}) \, {H}^\rho_{9} =0\ , \\[2mm]
&s_{2} \, {H}^\rho_{2}  \ + \ (s_{1}-s_{3}-s_{4}) \, {H}^\rho_{5}  \ + \ (s_{4}-s_{1}-s_{2}) \, {H}^\rho_{9} = 0\ ,  \\[2mm]
%%%%%%%%%%%%%%%%%%%%%%%%%%%
&(s_{4}-s_{1}-s_{2}) \, {H}^\rho_{1}  \ + \ s_{2} \, {H}^\rho_{2}  \ + \ (s_{1}-s_{3}-s_{4}) \, {H}^\rho_{4}  \ - \ s_{2} \, {H}^\rho_{7} = 0 \ , \\[2mm]
&s_{1} \, {H}^\rho_{2}  \ + \ (s_{4}-s_{1}-s_{2}) \, {H}^\rho_{3}  \ + \ (s_{2}-s_{4}-s_{5}) \, {H}^\rho_{7} = 0 \ , \\[2mm]
&(s_{4}-s_{1}-s_{2}) \, {H}^\rho_{6}  \ + \ s_{2} \, {H}^\rho_{7}  \ + \ (s_{1}-s_{3}-s_{4}) \, {H}^\rho_{8} = 0 \ ,\\[2mm]
&-s_{2} \, {H}^\rho_{3}  \ + \ (s_{2}+s_{3}-s_{1}-s_{5}) \, {H}^\rho_{6}  \ + \ s_{1} \, {H}^\rho_{9}= 0\ ,\\[2mm]
&{H}^\rho_{10}  \ + \ {H}^\rho_{5}  \ - \ {H}^\rho_{11} = 0 \ .
\ea\label{B3}
\eeq
Note, that these relations have to hold for any group ordering $\rho$.
In the next Appendix \ref{appC} we  present a two--dimensional basis of functions, which 
solves the equations \req{B1}, \req{B2} and \req{B3}.

\section{Two--dimensional basis of hypergeometric functions}
\label{appC}

In the previous Appendix B we have listed  universal relations \req{B1}, \req{B2} 
and \req{B3}, which  hold for the integrals ${H}_{i}^{\rho}$ for any 
given group ordering~$\rho$. In fact, these equations 
allow to express the hypergeometric integrals ${H}_{i}^{\rho}$ 
in terms of two basis functions. In Eq. \req{2,8} for any ordering $\rho$
we have chosen the pair:
$$h^\rho_{1}=H_3^\rho\ \ \ ,\ \ \ h^\rho_{2}=H_7^\rho\ .$$ 
 Hence, in the  case $(i)$ we work with the basis \req{BASIS1}, while  in 
 case $(ii)$ with \req{BASIS2}.
For a given ordering $\rho$ all the other nine functions ${H}_{i}^{\rho}$ 
can be expressed as linear combination \req{coeff} of the basis \req{2,8}.
The coefficients $C_{1}^{i} $, $C_{2}^{i},\ i=1,\ldots,11$, which are 
determined by the equations \req{B1}, \req{B2} and \req{B3},  
are rational polynomials in the kinematic invariants \req{sijnodim} and   
do {\em not} depend on the permutation $\rho$.
In the following we display the relations \req{coeff} for all eleven 
function~${H}_{i}^{\rho}$:
$$\ba{lcl}
&&\ds{{H}^\rho_{1}=\frac{s_{2}  -  s_{4}}{s_{1}} \  h^\rho_{1} \ - \ 
\frac{s_{2} - s_{4} - s_{5}}{s_{1}} \,  h^\rho_{2}\ , }\\[3mm]
&&\ds{{H}^\rho_{2}=\frac{s_{1}+s_{2}-s_{4}}{s_{1}} \,  h^\rho_{1} \ - \ 
\frac{s_{2} - s_{4} - s_{5}}{s_{1}} \,  h^\rho_{2}\ ,}\\[3mm]
%%%%%%%%%%%%%%%%%%%%%%
&&\ds{{H}^\rho_{3}=h^\rho_{1}\ ,}\\[3mm]
&&\ds{{H}^\rho_{4}=\frac{s_{4} \, (s_{4}-s_{1}-s_{2})}{s_{1} \, 
(s_{1}-s_{3}-s_{4})} \,  h^\rho_{1} \ - \ \frac{s_{4}\, (s_{4}+s_{5}-s_{2}) \, - 
\, s_{1} \, (s_{4}+s_{5})}{s_{1} \, (s_{1}-s_{3}-s_{4})} \,  h^\rho_{2}\ ,}\\[5mm]
&&\ds{{H}^\rho_{5}=\frac{s_{4} \, (s_{1}+s_{2}-s_{4})}{s_{1}-s_{3}-s_{4}} \; 
\left\{ \frac{1 }{ s_{2}+s_{3}-s_{5}} - \frac{1}{s_{1} } \right\} \,  h^\rho_{1}
-\frac{s_{4}+s_{5}-s_{2}}{s_{1}-s_{3}-s_{4}} \; \left\{ \frac{s_{1}-s_{3}-s_{4}+s_{5}}{s_{2}+s_{3}-s_{5}} + \frac{s_{4}}{s_{1}} \right\} \,  h^\rho_{2} \ ,} \\[5mm]
%%%%%%%%%%%%%%%%%%%%%%%%
&&\ds{{H}^\rho_{6}=\frac{s_{4}}{s_{2}+s_{3}-s_{5}} \,  h^\rho_{1} \ - \ 
\frac{s_{4}+s_{5}-s_{2}}{s_{2}+s_{3}-s_{5}} \,  h^\rho_{2}\ ,}\\[3mm]
&&\ds{{H}^\rho_{7}=h^\rho_{2}\ ,}\\[3mm]
&&\ds{{H}^\rho_{8}=\frac{s_{4} \, ( s_{1}+s_{2}-s_{4})}{(s_{1}-s_{3}-s_{4}) \, 
(s_{2}+s_{3}-s_{5})} \,  h^\rho_{1}
-  \frac{s_{2} \, (s_{3} +2s_{4}) \ - \ s_{4} \, (s_{4} + s_{5}) \, 
+ \, s_{1} \, (s_{4}+s_{5}-s_{2})}{(s_{1}-s_{3}-s_{4}) \, (s_{2}+s_{3}-s_{5})} \,  
h^\rho_{2}\ ,}\\[5mm]
%%%%%%%%%%%%%%%%%%%%%
&&\ds{{H}^\rho_{9}=\left\{ \frac{s_{2}-s_{4}}{s_{1}} \; + \;
\frac{s_{4}}{s_{2}+s_{3}-s_{5}} \right\} \,  h^\rho_{1} \ -
\ (s_{4}+s_{5}-s_{2}) \, \left\{ \frac{1}{s_{2}+s_{3}-s_{5}} \ -
\ \frac{1}{s_{1} } \right\} \,  h^\rho_{2}\ ,}\\[5mm]
&&\ds{{H}^\rho_{10}=\frac{(s_{1}+s_{2}-s_{4}) \,
\bigl[ (s_{1}-s_{3}) (s_{2}-s_{4}) \, - \, s_{4}s_{5} \bigr] }{s_{1} \, (s_{1}-1) \, 
(s_{1} - s_{3} - s_{4})} \,   h^\rho_{1}}\\[5mm]
&&\hskip1cm\ds{- \ \frac{(s_{2}-s_{4}-s_{5}) \, \bigl[ (s_{1}-s_{3}) 
(s_{1}+s_{2}-s_{4}) \, + \, (s_{1}-s_{4})s_{5} \bigr] }{s_{1} \, (s_{1}-1) \, (s_{1} - s_{3} - s_{4})} \,  h^\rho_{2} \ ,}
\ea$$
\bea
&&\ds{{H}^\rho_{11}=\left\{\ \frac{(s_{1}+s_{2}-s_{4}) \, \bigl( s_{1}^{2} s_{4} \, 
+ \, s_{1} \, \bigl[ -s_{4} + (s_{2}-2s_{4})(s_{2}+s_{3}-s_{5}) \bigr] \bigr)}{s_{1} 
\, (s_{1}-1) \, (s_{1} - s_{3} - s_{4}) \, (s_{2} + s_{3}-s_{5})} \right.}\\[5mm]
&&\hskip1cm\ds{- \ \left. \frac{(s_{1}+s_{2}-s_{4}) \, 
\bigl[ s_{2} s_{3} + s_{4}(-1-s_{3}+s_{5}) \bigr]}{s_{1} \, (s_{1}-1) \, 
(s_{1} - s_{3} - s_{4})} \right\} \,  h^\rho_{1}}\\[5mm]
&&\hskip1cm\ds{ - \ \left\{ \frac{(s_{2}-s_{4}-s_{5}) \, 
\bigl[ -s_{1}^{2} + s_{1}(1+s_{2}+2s_{3}+s_{4}-2s_{5}) \bigr]}{ (s_{1}-1) 
\, (s_{1} - s_{3} - s_{4}) \, (s_{2} + s_{3}-s_{5})} 
\right. } \\[5mm]
&&\hskip1cm\ds{+ \frac{(s_{2}-s_{4}-s_{5}) \, \bigl[ s_{2}^{2} - s_{4} - 2s_{2}s_{4}-s_{3}(1+s_{3}+2s_{4})+s_{5}+2(s_{3}+s_{4})s_{5}-s_{5}^{2} \bigr]}{ (s_{1}-1) \, (s_{1} - s_{3} - s_{4}) \, (s_{2} + s_{3}-s_{5})}}\\[5mm]
&&\hskip1cm\ds{- \ \left. \frac{(s_{2}-s_{4}-s_{5}) \,
\bigl[ s_{2}s_{3}+s_{4}(-1-s_{3}+s_{5}) \bigr]}{s_{1} \, (s_{1}-1) \,
(s_{1} - s_{3} - s_{4})}\ \right\} \,  h^\rho_{2}\ .} 
\label{2,9}
\eea

We would like to stress that the identities given in Appendix \ref{appB} and
\ref{appC} -- in particular the coefficients $C_{1}^{i}$, $C_{2}^{i}$
in the relations \req{coeff} -- are
completely independent on the given ordering $\rho$.
Hence, these relations allow to simplify all six orderings in the same way.
With these relations it is possible to write the partial amplitudes 
\req{2,101} and \req{new2,101} in terms of the two--dimensional basis \req{2,8}.
Obviously, the coefficients $C_i^{10}$ and $C_i^{11}$ are much more 
tedious than the others and show a potential tachyon pole 
$\frac{1}{s_1 - 1}$. So it makes sense to work with the original
kinematics ${\cal X}^1$, ${\cal X}^2$ instead of the 
particular blocks ${\cal K}^{10} = {\cal X}^{1} -{\cal X}^{2}$ and 
${\cal K}^{11} = -s_{1} {\cal X}^{1} + {\cal X}^{2}$.
Eventually, the final result for the generic (color stripped) $gggq\bar q$ 
subamplitudes can be given as \req{FINAL} for both cases $(i)$ and $(ii)$.

The coefficients $C_1^i, C^i_2$ for $i=1,...,9$ can be read off 
from \req{coeff} and \req{2,9}, while the coefficients $D^{i}_j$ 
are fixed by the combinations ${\cal K}^{10} = {\cal X}^{1} -{\cal X}^{2}$ 
and ${\cal K}^{11} = -s_{1} {\cal X}^{1} + {\cal X}^{2}$:
\beq
\ba{ll}
\ds{D_{1}^{1}}&= \ds{\ \ C_{1}^{10} \, - \, s_{1} \, C_{1}^{11} \eq \frac{s_{1}+s_{2}-s_{4}}{s_{1}-s_{3}-s_{4}} \; \left\{ \frac{s_{2} \, s_{3} \, + \, s_{4} \, (s_{5}-s_{3})}{s_{1} } \ - \ \frac{s_{1} \, s_{4}}{ s_{2}+s_{3}-s_{5}} \ - \ 
\bigl( s_{2} - 2s_{4} \bigr) \right\}\ ,}\\[4mm]
\ds{D_{1}^{2} }&=\ds{ \ \ -C_{1}^{10} \, + \, C_{1}^{11} \eq \frac{s_{4} \, (s_{1} + s_{2}-s_{4})}{s_{1} - s_{3} - s_{4}}
\; \left\{ \frac{1}{s_{2}+ s_{3}-s_{5}} \ - \ \frac{1}{s_{1}} \right\}\ ,}\\[4mm]
\ds{D_{2}^{1}}&=\ds{ \ \ C_{2}^{10} \, - \, s_{1} \, C_{2}^{11} \eq \frac{s_{2}-s_{4}-s_{5}}{s_{1}-s_{3}-s_{4}} \; \left\{ \frac{s_{1} \, ( s_{1} - s_{2}-2s_{3} - s_{4} +2s_{5} ) }{s_{2} +s_{3}- s_{5}} \ - \ \bigl(s_{2}-s_{3}-2s_{4}+s_{5} \bigr) \right.\ ,}\\[4mm]
&\ds{\left. + \ \frac{  s_{2} \, s_{3} \, + \, s_{4} \,(s_{5}-s_{3}) }{s_{1} } \right\}
\ ,}\\[4mm]
\ds{D_{2}^{2}}&=\ds{ \ \ -C_{2}^{10} \, + \, C_{2}^{11} \eq \frac{s_{4}+s_{5}-s_{2}}{s_{1}-s_{3}-s_{4}} \; \left\{ \frac{ s_{1}-s_{3}-s_{4}+s_{5}}{s_{2}+s_{3}-s_{5}} \ + \ 
\frac{s_{4}}{s_{1}} \right\}\ .}
\ea \label{eff1}
\eeq
Finally, in the form \req{FINAL} the unphysical tachyon pole $\frac{1}{s_1-1}$ 
originating from the expressions \req{2,9} for ${H}^\rho_{10}$ and ${H}^\rho_{11}$
is eliminated  and the residual pole structure reflects the relevant 
SM propagators from the Yang Mills limit of this scattering process. 
Explicitly, with \req{eff1} the contribution of ${H}^\rho_{10}$ and ${H}^\rho_{11}$
to the subamplitude \req{FINAL} is given by
\begin{align}
{H}^\rho_{10}& \, {\cal K}^{10} \ + \ {H}^\rho_{11}\, {\cal K}^{11} \ \ \equiv \ \ D_{1}^{1} \, h^{\rho}_{1} \, {\cal X}^{1} \ + \ D_{1}^{2} \, h^{\rho}_{1} \, {\cal X}^{2} \ + \ D_{2}^{1} \, h^{\rho}_{2} \, {\cal X}^{1} \ + \ D_{2}^{2} \, h^{\rho}_{2} \, {\cal X}^{2} \notag \\
&= \ \ \; \frac{s_{1}+s_{2}-s_{4}}{s_{1}-s_{3}-s_{4}} \; \left\{ \frac{s_{2} \, s_{3} \, + \, s_{4} \, (s_{5}-s_{3})}{s_{1} } \ - \ \frac{s_{1} \, s_{4}}{ s_{2}+s_{3}-s_{5}} \ - \ \bigl( s_{2} - 2s_{4} \bigr) \right\} \,  h^\rho_{1} \, {\cal X}^{1} \notag \\
& \ \ \ \ \ \ \ \ \ \ + \ \frac{s_{4} \, (s_{1} + s_{2}-s_{4})}{s_{1} - s_{3} - s_{4}}
\; \left\{ \frac{1}{s_{2}+ s_{3}-s_{5}} \ - \ \frac{1}{s_{1}} \right\} \,
 h^\rho_{1} \, {\cal X}^{2} \notag \\
& \ \ \ \ \ \ \ \ \ \ + \ \frac{s_{2}-s_{4}-s_{5}}{s_{1}-s_{3}-s_{4}} \; \left\{ \frac{s_{1} \, ( s_{1} - s_{2}-2s_{3} - s_{4} +2s_{5} ) }{s_{2} +s_{3}- s_{5}} \ - \ \bigl(s_{2}-s_{3}-2s_{4}+s_{5} \bigr) \right.  \notag \\
& \ \ \ \ \ \ \ \ \ \ \ \ \ \ \ \ \ \ \ \ \ \ \ \ \ \left. + \ \frac{  s_{2} \, s_{3} \, + \, s_{4} \,(s_{5}-s_{3}) }{s_{1} } \right\} \,  h^\rho_{2} \, {\cal X}^{1} \notag \\
& \ \ \ \ \ \ \ \ \ \ + \ \frac{s_{4}+s_{5}-s_{2}}{s_{1}-s_{3}-s_{4}} \; \left\{ \frac{ s_{1}-s_{3}-s_{4}+s_{5}}{s_{2}+s_{3}-s_{5}} \ + \ \frac{s_{4}}{s_{1}} \right\} \,  h^\rho_{2} \, {\cal X}^{2} \ .
\label{2,10}
\end{align}

\section{Subamplitudes of $\bm{gggq\bar q}$  and low energy expansions}
\label{appD}

{}From the expressions \req{2,101} and
\req{new2,101}, it can be seen that one partial amplitude 
$A(1_\rho,2_\rho,3_\rho,4_\rho,5_\rho)$ is specified by eleven
integrals ${H}_i^{\rho}$
or hypergeometric functions multiplying some kinematics ${\cal K}^i$
from the list \req{1,11}.
At any rate, for each case $(i)$ and $(ii)$
it is sufficient to work out only one ordering. 
More precisely, all six orderings of case
$(i)$, depicted in Figure  \ref{gggqqall}, may be obtained from 
the ordering $\rho=(1,2,3,4,5)$
by respective permutations of gluons. The partial amplitude $A(1,2,3,4,5)$
is discussed in more detail in the following.
Furthermore, the two orderings of case $(ii)$, depicted in
Figure  \ref{gggqqall1}, may be obtained
from the ordering $\rho=(1,2,4,3,5)$  by respective permutations of gluons.
A detailed analysis of the partial amplitude $A(1,2,4,3,5)$
is presented in this Section.

\ \\
\noindent
\underline{\sl Subamplitude $A(1,2,3,4,5)$}
\ \\

In this Subsection we focus on the hypergeometric integrals and the $\ap$--expansions
of the partial amplitude $A(1,2,3,4,5)$.
According to \req{REG1} in the amplitude \req{2,101} 
the parameters $x = z_2$ and $y = z_1/z_2$ have to be integrated over the region
${\cal I}_{(12345)}=\{x,y\in \RR\ |\ -\infty<x<0\ ,\ 1<y<\infty\}$.
After the change of variables
\beq
x \ \ \to \ \ -\frac{x\ (1-y)}{1-x} \co y \ \ \to \ \ \frac{1}{xy} \ ,
\eeq
the region ${\cal I}_{(12345)}$ is mapped to the unit square 
and \req{2,101} can be cast into:
%%%%%%%%%%%%%%%%%%%%%%
\begin{align}
&A(1,2,3,4,5)\eq -4 \,  \al'^{2} \,
g_{\te{D}p_{a}}^{3} \ 
\int^{1} \limits_{0} \dd x \int^{1} \limits_{0} \dd
y \ x^{s_{2}} \,  y^{s_5} \, (1-x )^{s_{3}} \, (1-y )^{s_4} \,
( 1-xy)^{s_{1}-s_{3}-s_{4}} \notag \\
& \ \ \ \times \ \Biggl[ \
\frac{{\cal K}^{1}}{(1-x)\, (1-y)} \ + \ \frac{{\cal
   K}^{2}}{x\, (1-x)\, y\, (1-y)} \ + \ \frac{{\cal K}^{3}\, (1-xy)}{x\, (1-x)\, y\, (1-y)} \ + \ \frac{{\cal K}^{4}}{(1-x) \, (1-xy)}  \Biggr. \notag \\
& \ \ \ \ \ \ \ \ \ \ + \ \frac{{\cal K}^{5}}{(1-x)\, y\,  (1-xy)} \ + \ \frac{{\cal K}^{6}}{y \, (1-x)}\ + \ \frac{{\cal K}^{7}}{x\, (1-x)\, y} \ +
\ \frac{{\cal K}^{8}\, (1-y)}{(1-x) \, y\, (1-xy)} \notag \\
& \ \ \ \ \ \ \ \ \ \ \ \Biggl. + \  \frac{{\cal K}^{9}}{(1-x)\, y\, (1-y)} \ + \ \frac{{\cal K}^{10}}{(1-x)\, (1-y)\, (1-xy)} \ + \ \frac{{\cal K}^{11}}{(1-x)\, y\ (1-y)\, (1-xy)} \ \Biggr] \label{2,3} \\
&\equiv 4\, \al'^{2} \, g_{\te{D}p_{a}}^{3} \,
\sum_{i=1}^{11} {H}^{(12345)}_{i} \ {\cal K}^{i}\ .
\label{2,4}
\end{align}

According to the results of Appendix \ref{appC} all eleven hypergeometric 
integrals ${H}^{(12345)}_{i}$ may be expressed in terms of a two dimensional basis
\req{2,8}. For the present case the latter is represented by the two integrals
\req{BASIS1}:
\bea
\ds{h^{(12345)}_{1} }&=&\ds{-\int^1 \limits_{0} \dd x \int^{1} \limits_{0} \dd y \
x^{s_2-1} \,  y^{s_5-1} \, (1-x)^{s_3-1} \, (1-y)^{s_4-1} \,
(1-xy)^{s_{1}-s_{3}-s_{4}+1}\ , }\\
\ds{h^{(12345)}_{2} }&=&\ds{-\int^{1} \limits_{0} \dd x \int^{1} \limits_{0} \dd y \
x^{s_{2}-1} \,  y^{s_5-1} \, (1-x)^{s_3-1} \, (1-y)^{s_4} \
(1-xy)^{s_{1}-s_{3}-s_{4}} \ .}
\label{12345}
\eea
With this basis \req{12345} the amplitude 
\req{2,4} may be cast into the form \req{FINAL}, with the coefficients 
to be read off from Eqs. \req{2,9} and \req{eff1}.

The low--energy expansion of the amplitude \req{2,4} is determined by the 
$\ap$--expansion of the eleven functions ${H}^{(12345)}_{i}$. 
However, due to the relations \req{2,9} it is enough to determine the 
$\ap$--expansion for the basis \req{12345} only:
\begin{subequations}
\begin{align}
h^{(12345)}_1  \ \ &= \ \ -\fc{1}{s_2s_4} \, - \, \fc{1}{s_2s_5} \, - 
\, \fc{1}{s_3s_5} \ - \
\zeta(2)\, \left(2-\fc{s_1}{s_3}-\fc{s_1}{s_4}-\fc{s_4}{s_2}-\fc{s_2}{s_5}-
\fc{s_3}{s_5}-\fc{s_5}{s_2}\right) \ + \ {\cal O}(\ap)\ , \label{expa1}  \\
h^{(12345)}_2  \ \ &= \ \ -\fc{1}{s_2s_5} \, - \, \fc{1}{s_3s_5} \ - \ \zeta(2)\,
\left(1-\fc{s_1}{s_3}-\fc{s_4}{s_2}-\fc{s_2}{s_5}-\fc{s_3}{s_5}\right) \ + 
\ {\cal O}(\ap)\ .
\label{expa2}
\end{align}
\end{subequations}

\ \\
\noindent
\underline{\sl Subamplitude $A(1,2,4,3,5)$}
\ \\

In this Subsection we focus on the hypergeometric integrals and the $\ap$ expansions
of the partial amplitude $A(1,2,4,3,5)$.
According to \req{REG2} in the amplitude \req{new2,101} 
we have to integrate over the region
${\cal I}_{(12435)}=\{x,y\in \RR\ |\ -\infty<x<0\ ,\ 0<y<1\}$.
After the change of variables
\beq
x \ \ \to \ \ -\frac{1-y}{(1-x)\, y} \co y \ \ \to \ \ x \,y\ ,
\eeq
the region ${\cal I}_{(12435)}$ is mapped to the unit square 
and \req{new2,101} can be cast into:
%%%%%%%%%%%%%%%%%%%%%%
\begin{align}
&A(1,2,4,3,5)\eq -4\, \al'^{2} \,
g_{\te{D}p_{a}}^{2}\, g_{\te{D}p_{b}} \ 
\int^{1} \limits_{0} \dd x \int^{1} \limits_{0} \dd
y \ x^{-s_{2}-s_3+s_5} \,  y^{s_5} \, (1-x)^{s_{3}} \, (1-y)^{s_1-s_3-s_4} \,
(1-xy)^{s_{4}} \notag \\
& \ \ \times \ \Biggl[ \
\frac{{\cal K}^{1}}{(1-x)\, (1-xy)} \ + \ \frac{{\cal
   K}^{2}}{y\, (1-x)\, (1-xy)} \ + \ \frac{{\cal K}^{3}\, (1-y)}{y\, (1-x)\, (1-xy)}
\ + \ \frac{{\cal K}^{4}}{(1-x) \, (1-y)}  \Biggr. \notag \\
& \ \ \ \ \ \  + \ \frac{{\cal K}^{5}}{x\, (1-x)\, y\,  (1-y)} \ + \
\frac{{\cal K}^{6}}{x \, y \, (1-x)}\ + \ \frac{{\cal K}^{7}}{(1-x)\, y} \ +
\ \frac{{\cal K}^{8}\, (1-xy)}{x \, (1-x) \, y\, (1-y)} \notag \\
& \ \ \ \ \ \ \ \Biggl. + \  \frac{{\cal K}^{9}}{x \, (1-x)\, y\,
 (1-xy)} \ + \ \frac{{\cal K}^{10}}{(1-x)\, (1-y)\, (1-xy)} \ + \
\frac{{\cal K}^{11}}{x\, (1-x)\, y\ (1-y)\, (1-xy)} \ \Biggr] \label{2,3new} \\
&\equiv \ \
4\, \al'^{2} \, g_{\te{D}p_{a}}^{2}\, g_{\te{D}p_{b}} \, \sum_{i=1}^{11} {H}^{(12435)}_{i}\  
{\cal K}^{i}\ .
\label{2,4new}
\end{align}

According to the results of Appendix \ref{appC} all eleven hypergeometric 
integrals ${H}^{(12435)}_{i}$ may be expressed in terms of a two dimensional basis
\req{2,8}. For the present case the latter is represented by the two integrals
\req{BASIS2}:
\bea
\ds{h^{(12435)}_{1} }&=&\ds{-\int^1 \limits_{0} \dd x \int^{1} \limits_{0} \dd y \
x^{-s_{2}-s_3+s_5} \,  y^{s_5-1} \, (1-x)^{s_{3}-1} \, (1-y)^{s_1-s_3-s_4+1} \,
(1-xy)^{s_{4}-1}\ ,}\\
\ds{h^{(12435)}_{2} }&=&\ds{-\int^{1} \limits_{0} \dd x \int^{1} \limits_{0} \dd y \
x^{-s_{2}-s_3+s_5} \,  y^{s_5-1} \, (1-x)^{s_{3}-1} \, (1-y)^{s_1-s_3-s_4} \,
(1-xy)^{s_{4}} \ . }
\label{12435}
\eea
With this basis \req{12435} the amplitude 
\req{2,4new} may be cast into the form \req{FINAL}, with the coefficients 
to be read off from Eqs. \req{2,9} and \req{eff1}.

The low--energy expansion of the amplitude \req{2,4new} is determined by the 
$\ap$--expansion of the eleven functions ${H}^{(12435)}_{i}$. 
However, due to the relations \req{2,9} it is enough to determine the 
$\ap$--expansion for the basis \req{12435} only:
\begin{subequations}
\begin{align}
h^{(12435)}_1  \ \ &= \ \ -\fc{1}{s_3s_5} \ + \
\zeta(2)\, \lf(1+\fc{s_1}{s_3}-\fc{s_2}{s_5}-\fc{s_3}{s_5}\ri) \ + \ {\cal O}(\ap) \label{newexpa1}  \\
h^{(12435)}_2  \ \ &= \ \ -\fc{1}{s_3s_5} \ + \ \zeta(2)\,
\lf(\fc{s_1}{s_3}-\fc{s_2}{s_5}-\fc{s_3}{s_5}\ri) \ + \ {\cal O}(\ap) \label{newexpa2}
\end{align}
\end{subequations}

% \break

\end{document}